\documentclass[prd,twocolumn,superscriptaddress]{revtex4-2}
\usepackage[utf8]{inputenc}
\usepackage{amsmath,amssymb}
\usepackage{braket}
\usepackage{dsfont}
\usepackage[colorlinks=true,linkcolor=blue,urlcolor=blue,citecolor=blue,anchorcolor=blue]{hyperref}
\usepackage[capitalise]{cleveref}
\Crefname{section}{Sec.}{Secs.}
\crefrangelabelformat{equation}{(#3#1#4--#5#2#6)}
\usepackage{tikz}
\usepackage{multirow}
\usepackage{amsfonts}
\usepackage{ulem}
\usepackage{diagbox}
\usepackage{subfigure}
\usepackage{slashed} 
\usepackage{cancel}

\newcommand{\ie}{{i.e.}}
\newcommand{\eg}{{e.g.}}

\newcommand{\pii}{$\pi^0_i$} 
\newcommand{\pif}{$\pi^0_f$}
\newcommand{\piz}{$\pi^0$}

\newcommand{\tf}{t_{s}}
\newcommand{\tc}{t_{\rm ins}}
\newcommand{\xins}{\vec{x}_{\rm ins}}
\newcommand{\ti}{t_{\rm src}}
\newcommand{\xsrc}{\vec{x}_{\rm src}}

\newcommand{\OJ}{\mathcal{J}}
\newcommand{\OO}{\mathcal{O}}
\newcommand{\JN}{\OJ_N}
\newcommand{\JNb}{\OJ^\dagger_N}
\newcommand{\JNd}{\OJ^\dagger_N}
\newcommand{\Jpi}{\OJ_\pi}
\newcommand{\JNpi}{\OJ_{N\pi}}
\newcommand{\JNpib}{\OJ^\dagger_{N\pi}}
\newcommand{\JNpid}{\OJ^\dagger_{N\pi}}
\newcommand{\NN}{\braket{\JN\JNd}}
\newcommand{\NNpi}{\braket{\JN\JNpid}}
\newcommand{\NpiN}{\braket{\JNpi\JNd}}
\newcommand{\NpiNpi}{\braket{\JNpi\JNpid}}
\newcommand{\NJN}{\braket{\JN\OO\JNd}}
\newcommand{\NJNpi}{\braket{\JN\OO\JNpid}}
\newcommand{\NpiJN}{\braket{\JNpi\,\OO\,\JNd}}
\newcommand{\NpiJNpi}{\braket{\JNpi\,\OO\,\JNpid}}

\newcommand{\Cdpt}{C}

\newcommand{\Ctpt}{\Omega}

\newcommand{\vp}{\vec{p}}
\newcommand{\vpp}{\vec{p}^{\,\prime}}
\newcommand{\vq}{\vec{q}}

\newcommand{\pvec}{\vec{p}}
\newcommand{\pprime}{\prime\prime}

\usepackage{xparse} 
\NewDocumentCommand\Nf{mgg}{N\textsubscript{f}=#1\IfNoValueTF{#2}{}{+#2}\IfNoValueTF{#3}{}{+#3}}
\NewDocumentCommand\vol{mg}{#1\textsuperscript{3}\IfNoValueTF{#2}{}{×#2}}

\begin{document}
\title{Investigation of pion-nucleon contributions to nucleon matrix elements}
\newcommand{\UCY}{Department of Physics,
University of Cyprus, P.O. Box 20537, 1678 Nicosia, Cyprus}
\newcommand{\CYI}{Computation-based Science and Technology Research Center,
The Cyprus Institute, 20 Kavafi Str., Nicosia 2121, Cyprus}
\newcommand{\HISKP}{HISKP (Theory), Rheinische
Friedrich-Wilhelms-Universit{\"a}t Bonn, Nu{\ss}allee 14-16, 53115
Bonn, Germany}
\author{Constantia Alexandrou}
\affiliation{\UCY}
\affiliation{\CYI}
\author{Giannis Koutsou}
\affiliation{\CYI}
\author{Yan Li}
\affiliation{\UCY}
\author{Marcus Petschlies}
\affiliation{\HISKP}
\author{Ferenc Pittler}
\affiliation{\CYI}
\date{\today}

\begin{abstract}
We investigate contributions of excited states to nucleon matrix elements computed in lattice QCD by employing, in addition to the standard nucleon interpolating operator,   pion-nucleon ($\pi$-$N$) operators. We solve a generalized eigenvalue problem (GEVP) to obtain an optimal interpolating operator that minimizes overlap with the $\pi$-$N$ states. We derive a variant of the standard application of the GEVP method, which allows for constructing 3-point correlation functions using the optimized interpolating operator without requiring the computationally demanding combination that includes $\pi$-$N$  operators in both sink and source. We extract nucleon matrix elements   using two twisted mass fermion ensembles, one ensemble generated using pion mass of 346 MeV and  one ensemble tuned to reproduce the physical value of the pion mass. Especially, we determine the isoscalar and isovector scalar, pseudoscalar, vector, axial, and tensor matrix elements. We include results obtained using a range of kinematic setups, including momentum in the sink. Our results using this variational approach are compared with previous results obtained using the same ensembles and multi-state fits without GEVP improvement. We find that for the physical mass point ensemble, the improvement, in terms of suppression of excited states using this method, is most significant for the case of the matrix elements of the isovector axial and pseudoscalar currents. 
\end{abstract}

\maketitle

\section{Introduction}
The study of nucleon structure, starting directly from the QCD Lagrangian, remains as one of the primary goals of nuclear and particle physics. First principles calculations of nucleon structure quantities, starting from the simplest case of those associated with the local matrix elements, provide insights into a plethora of phenomenologically important quantities related to the distribution of quarks within the nucleon as well to searches for physics beyond the Standard Model (BSM).
In this work we investigate nucleon charges and form factors, the importance of which is summarised below.

The scalar matrix elements of the nucleon are connected to the so-called nucleon $\sigma$-terms, which measure the mass generated by quarks in the nucleon. The closely-related isovector scalar charge is proportional to the QCD contribution of the neutron-proton mass splitting, i.e. to the mass splitting in the absence of QED, while the individual flavor components of the scalar charge are relevant to direct-detection  searches of dark matter, which rely on elastic scattering of nuclei with weakly interacting massive particles. They are also relevant for possible scalar interactions in BSM theories.

The vector matrix elements yield the nucleon electromagnetic form factors that are well-studied  quantities experimentally for understanding the internal structure of protons and neutrons. The size of the proton is typically defined in terms of the charge radius, which is directly related to the slope of the electric Sachs form factor at zero momentum transfer, and the nucleon magnetic moment is similarly defined as the value of the magnetic Sachs form factor at zero momentum transfer. While proton form factors are accurately determined, neutron form factors but also strange form factors for both the proton and the neutron still have large uncertainties and extracting them from lattice QCD would provide valuable input. 

The axial matrix elements yield the axial and induced pseudoscalar form factors, which quantify the scattering of neutrinos off nucleons relevant for neutrino oscillation experiments. Beyond the nucleon axial charge, defined as the axial form factor at zero momentum transfer, and which is known to high precision from $\beta$-decay, the two axial form factors are less well known experimentally compared to the proton electromagnetic form factors.

Finally, the tensor matrix elements yield the tensor charge, which enters charge and parity (CP) violating processes. Tensor interactions, similar to the scalar case, can serve as probes for BSM physics. Furthermore, the tensor matrix elements are related to Mellin moments of the transversity parton distribution function (PDF), which are less precisely known than, for example, the moments of the unpolarized PDF.

Lattice QCD has made significant progress in the \textit{ab initio} computation of nucleon structure quantities, with a series of recent simulations available with quark masses tuned to their physical values. Indeed, the Flavor Lattice Averaging Group (FLAG), which traditionally reviewed mesonic lattice quantities, has since 2019~\cite{FlavourLatticeAveragingGroup:2019iem} included nucleon structure quantities in their summary tables, including  results obtained with physical or near-physical pion mass ensembles. Such ``physical point'' calculations eliminate the need for chiral extrapolations from heavier-than-physical quark masses, which has been a significant unquantified source of systematic uncertainty in nucleon structure calculations. However, a major issue that arises as statistical accuracy  increases is that of the extraction of nucleon matrix elements free of or with minimal contamination of excited states. Indeed, excited state contamination has been attributed to the discrepancies of  recent lattice QCD results with those from phenomenology and experiment, as for example, the axial form factors~\cite{RQCD:2019jai,Jang:2019vkm,Barca:2022uhi,Jang:2023zts} and the nucleon $\sigma$-terms~\cite{Gupta:2021ahb,Gupta:2023cvo}. Chiral perturbation theory also suggests that using ensembles with physical pion mass, matrix elements that include $\pi N$ initial or final states may be amplified in the axial~\cite{Bar:2018xyi}, pseudoscalar~\cite{Bar:2019gfx}, and electromagnetic~\cite{Bar:2021crj} form factors.

A common way to address these amplified excited state contributions is to explicitly model the excited state dependence with increasing Euclidean time via fits that include two or three states. This significantly increases resource requirements,  since as the time separation between the sink and source increases an exponentially increasing statistics is needed to keep statistical errors approximately constant. An alternative approach, which is the focus of this work, is to variationally optimize the nucleon interpolating operator, such that the overlap with the excited states is suppressed. A first study on such a variational approach was done in Ref.~\cite{Barca:2022uhi} using an ensemble with a pion mass of $m_\pi=429$~MeV.

In this work, we study more operators and use two gauge ensembles one of which at the physical pion mass. We use a variational basis that includes, apart from the standard nucleon interpolating operator, a $\pi-N$ interpolating operator and optimize the variational basis by solving a generalized eigenvalue problem  for the 2-point nucleon correlation function. 
We then proceed to compute nucleon 3-point correlation functions using the optimized interpolating operators and extract  local matrix elements with the scalar, vector, axial, and tensor current operators. We introduce a new method that allows us to avoid explicit calculation of the computationally demanding 3-point function with $\pi-N$  interpolating operators at source and sink times.  We study these quantities   using two twisted mass fermion gauge ensembles,  one with pion mass of 346~MeV and one  with light quark mass tuned to reproduce the physical pion mass. We compare the improvement obtained via this variational approach with existing results that have used multi-state fits for the same quantities and the same ensembles.

The remainder of this paper is organized as follows: in Sec.~\ref{sec:setup}, we provide details of the ensembles used and the 3-point correlation function computed; in Secs.~\ref{sec:Id} and~\ref{sec:gevp}, we provide details on our GEVP setup; in Sec.~\ref{sec:anff}, we present our results; and in Sec.~\ref{sec:conclusions}, we summarize our findings and give our conclusions.

\section{Simulation details}
\label{sec:setup}
We use two gauge ensembles generated by the Extended
Twisted Mass Collaboration (ETMC), one tuned to the physical pion mass and the other one with a heavier pion mass $m_\pi=346\,$MeV~\cite{ETM:2015ned,Alexandrou:2018egz}. 
The physical pion mass ensemble is simulated with a degenerate doublet
of light quarks (\Nf{2}), and the heavier pion mass
ensemble is simulated with additionally a strange and a charm quark
with masses tuned to approximately their physical values (\Nf{2}{1}{1}).
A clover term is included in both actions.
The parameters of the gauge ensembles are given in \cref{tab:ens}.

\begin{table*}[!ht]
    \caption{Parameters of the ensembles used in this work, including
      the ensemble name (first column), the quark flavors simulated in
      the sea (second column), the dimensions of the ensembles in
      lattice units (third column), the lattice spacing (fourth
      column), spatial extent (fifth column), pion mass (sixth
      column), the dimensionless product of the pion mass with the
      spatial extent (seventh column), and the nucleon mass (eighth
      column).  Further details are given in
Refs.~\cite{ETM:2015ned,Alexandrou:2018egz,ExtendedTwistedMass:2021gbo}.}\label{tab:ens}
    \centering \renewcommand\arraystretch{1.5}
    \begin{ruledtabular}
    \begin{tabular}{cccccccc}
      Ensembles & Flavors (N\textsubscript{f}) & $N_L^3\times N_T$ & $a$ [fm] & $L$ [fm] & $m_\pi$ [MeV] & $m_\pi\,L$ & $m_N$ [MeV] \\ \hline
      cA211.53.24 & 2+1+1 & $24^3 \times 48$ & 0.0908 & 2.27 & 346  & 3.99 & 1193(18) \\
      cA2.09.48 & 2 & $48^3 \times 96$ & 0.0938 & 4.50 & 131  & 2.98 & 931(3)  \\
    \end{tabular}
    \end{ruledtabular}
\end{table*}

\subsection{2- and 3-point functions}
We use the single nucleon $N$ interpolating fields $\JN$ for $N = p$ (proton) and $n$ (neutron)
\begin{align}
  \OJ_{p}(\vec{x};t) &= \epsilon_{abc} \,  [u^{a}(\vec{x};t)^T \, \mathcal{C}\gamma_5 \,  d^b(\vec{x};t)]\, u^c(\vec{x};t) \,,
  \nonumber\\
  \OJ_{n}(\vec{x};t) &= \epsilon_{abc} \,  [d^{a}(\vec{x};t)^T \, \mathcal{C}\gamma_5 \,  u^b(\vec{x};t)]\, d^c(\vec{x};t) 
  \label{eq:interpolators-nucleon}
\end{align}
and the single pion interpolators for charged and neutral pion
\begin{align}
  \OJ_{\pi^+}(\vec{x};t) &= \bar{d}(\vec{x};t) \, i\gamma_5 \, u(\vec{x};t) \,,
  \nonumber \\
  \OJ_{\pi^0}(\vec{x};t) &=  \frac{1}{\sqrt{2}}\,\left( 
  \bar{u}(\vec{x};t) \, i\gamma_5 \, u(\vec{x};t) - \bar{d}(\vec{x};t) \, i\gamma_5 \, d(\vec{x};t) 
  \right)  \,.
  \label{eq:interpolators-pion}
\end{align}
The pion-nucleon $N\pi$ two-hadron interpolating field is generated from the product
\begin{align}
    \JNpi(\vec{x}_1,\vec{x}_2;t) &= \JN(\vec{x}_1;t) \,\, \Jpi(\vec{x}_2;t)
    \label{eq:interpolators-Npi}
\end{align}
with appropriate isospin combinations. \\

We apply Gaussian smearing \cite{Gusken:1989ad,Alexandrou:1992ti} to quark fields 
$\psi = u\,,\,\,d$
entering the interpolating fields of Eqs.~(\ref{eq:interpolators-nucleon}), (\ref{eq:interpolators-pion}), and (\ref{eq:interpolators-Npi})
\begin{align}
  \psi^a_{\rm smear}(\vec{x};t) &= \sum_{\vec{y}} F^{ab}(\vec{x},\vec{y}; t ) \,\, \psi^b(\vec{y},t) 
    \label{eq:smear}
\end{align}
with smearing kernel 
\begin{align}
    F&=(\mathds{1}+\alpha_G\,H)^{N_G} \,,
    \label{eq:hopping} \\
    H(\vec{x},\vec{y}; t )
    &=
    \sum_{i=1}^{3}\, \left[U_i(\vec{x}, t)\, \delta_{ \vec{x},\vec{y}-\hat{i}} + U^\dagger_i(\vec{x}-\hat{i}, t)\, \delta_{\vec{x},\vec{y}+\hat{i}} \right] \,,
    \nonumber
\end{align}
and parameters $\alpha_G=4.0$ and $N_G=50$, that are tuned in order to approximately give a smearing radius for the nucleon of 0.5 fm for the physical point ensemble \cite{Alexandrou:2018sjm,Alexandrou:2019ali} (ensemble cA2.09.48 in Tab. \ref{tab:ens}). The same parameters are also applied to the ensemble cA211.53.24.
For the links $U$ entering the hopping matrix $H$ in Eq. (\ref{eq:hopping}), we apply APE smearing \cite{APE:1987ehd} with parameters $n_{\rm APE}=50$ and $\alpha_{\rm APE}=0.5$ to reduce statistical errors due to ultraviolet fluctuations.

Given that we use single-nucleon and pion-nucleon interpolators, 
there are four possible types of 2-point functions depending on the combinations of $\JN$ and $\JNpi$ at the source ($\ti$) and sink ($\tf$) time slices,
namely
\begin{widetext}
\begin{align}
    \begin{bmatrix}
    \braket{\JN(\vec{x}_1;\tf)\,\, \JNb(\xsrc;\ti)} & \braket{\JN(\vec{x}_1;\tf) \,\, \JNpib(\xsrc,\vec{x}_2;\ti)}\\
    \braket{\JNpi(\vec{x}_1,\vec{x}_2;\tf) \,\, \JNb(\xsrc;\ti)} & \braket{\JNpi(\vec{x}_1,\vec{x}_2;\tf) \,\, \JNpib(\xsrc,\vec{x}_3;\ti)}
    \end{bmatrix} \,,
    \label{eq:2pt}
\end{align}
and four types of 3-point functions with an insertion operator $\OO$ at the insertion time slice $\tc$, namely,
\begin{align}
    \begin{bmatrix}
    \braket{\JN(\vec{x}_1;\tf)\,\, \OO(\xins;\tc)\,\, \JNb(\xsrc;\ti)} & \braket{\JN(\vec{x}_1;\tf)\,\, \OO(\xins;\tc)\,\, \JNpib(\xsrc,\vec{x}_2;\ti)}\\
    \braket{\JNpi(\vec{x}_1,\vec{x}_2;\tf)\,\, \OO(\xins;\tc) \,\, \JNb(\xsrc;\ti)} & \braket{\JNpi(\vec{x}_1,\vec{x}_2;\tf) \,\, \OO(\xins;\tc)\,\, \JNpib(\xsrc,\vec{x}_3;\ti)}
    \end{bmatrix} \,.
    \label{eq:3pt}
\end{align}
\end{widetext}
The expectation value $\braket{\cdots}$ in the above denotes the standard Feynman path-integral over the lattice action.
In most of the analysis in this work, we do not consider any effects of finite time extent of the lattice and thus 
understand $\braket{\cdots}$ as the vacuum expectation value $\braket{\Omega|\cdots|\Omega}$.

We perform Fourier transformations over every $\vec{x}$ argument in the 2-point and 3-point functions of Eqs.~(\ref{eq:2pt}) and (\ref{eq:3pt})
except the source point $\xsrc$. The latter is fixed at multiple randomly selected sites.
We project the pion-nucleon interpolators to isospin 1/2 of the nucleon state (cf. \cref{app:interfields,app:iso}). 
In addition, for each total momentum we project single-nucleon and pion-nucleon interpolators to those irreducible representations (irreps) of the lattice rotation groups
that contain total angular momentum $J = 1/2$. We refer to \cref{app:interfields} for details about the lattice irrep projection 
of the single- and two-hadron operators \cite{Morningstar:2013bda,Prelovsek:2016iyo}.

In the following we consider two reference frames: the center-of-mass frame with total momentum zero $\pvec = 0$
and with fixed irrep $G_{1g}$ and the first moving frame with total momentum 
$\pvec =\vec{1}_z = (2\pi/L)(0,0,1)$ with fixed irrep $G_1$. 
For the other five momenta with the same $|\vp|=2\pi/L$, they have been effectively taken into account via data symmetrization as explained in \cref{app:dsbs}.
Both these irreps are 2-dimensional, and we use $r \,\in\, \left\{ 0,\,1 \right\}$ to denote the two irrep rows.
Moreover, the isospin shall be fixed to that of the proton, $(I,\,I_3) = (1/2,\,+1/2)$.

With these specifications it becomes sufficient to label interpolators
simply as $\OJ^r_k(\vp;t)$ with explicit irrep row index $r$ and the index $k$ distinguishing the single- / two-hadron
interpolators together with 3-momentum distribution given the total momentum $\vp$.
In the center-of-mass frame at $\pvec = 0$ and with irrep $G_{1g}$ we thus have the range
\begin{align}
  k \, \in \, \left\{ 
  N \,, \,\,  N(1) \pi(1) \right\}\,,
  \label{eq:cmf-operators}
\end{align}
where the integer arguments in $N\pi$ give the nucleon and pion 3-momentum
$|\pvec_N|/(2\pi/L)$ and $|\pvec_{\pi}|/(2\pi/L)$, respectively, while
the single hadron momenta fulfill $\pvec = \pvec_{N} + \pvec_{\pi}$.

In the first moving frame at $\pvec= \vec{1}_z$ and irrep $G_1$ we
have three possible values for $k$,
\begin{align}
    k \, \in \, 
    \left\{ 
  N \,, \,\,  N(1)\pi(0) \,,\,\, N(0)\pi(1)
    \right\} 
    \label{eq:jkp1}\,.
\end{align}
From the projected interpolating fields, the matrix $C$ of 2-point functions is given by
\begin{align}
  \Cdpt_{jk}^{r}(\vp; t)=\braket{\OJ^{r}_j(\vp; t) \,\, [\OJ^{r}_k(\vp;0)]^\dagger} \,,
  \label{eq:C2pt}
\end{align}
where $r = 0,\,1$ is the irrep row index. 
We note, that with the above definition,  both in the center-of-mass frame 
and in the moving frame  we  have the 
relation of $C$ to the standard nucleon 2-point function $\Cdpt_{NN}$ with projection matrix $\Gamma_0=\frac{1}{4}(1+\gamma_4)$
via
\begin{align}\label{eq:C2ptNN}
  \Cdpt_{NN}(\vp;t)=\frac{1}{2}\sum_{r=0,1}\Cdpt^r_{NN}(\vp;t) \,.
\end{align}
More details are explained in \cref{app:23pt}.

Similarly, the 3-point functions are defined as
\begin{align}
  &\quad\Ctpt^{r',r}_{jk}(\OO;\vpp,\vp;\tf,\tc) \nonumber\\
  &=
  \braket{\OJ^{r'}_j(\vpp;\tf)\,\, \OO(\vec{q};\tc)\,\, [\OJ^{r}_k(\vp;0)]^\dagger} \,,
  \label{eq:C3pt}
\end{align}
with 3-momentum transfer $\vec{q}=\vp-\vpp$, and irrep rows $r',\,r$ for the sink and source operator, respectively.

For the insertion operator $\OO$, we consider the scalar $S(x)=\bar{\psi}(x)\psi(x)$, vector $V_\mu(x)=\bar{\psi}(x)\gamma_\mu \psi(x)$, pseudoscalar $P_5(x)=\bar{\psi}(x)\gamma_5 \psi(x)$, axial $A_\mu(x)=\bar{\psi}(x)\gamma_\mu\gamma_5 \psi(x)$, and tensor $T_{\mu\nu}(x)=\bar{\psi}(x)\sigma_{\mu\nu}\psi(x)$ currents. We will study both the isoscalar and isovector 
flavor combinations of these currents.
We renormalize the insertion operators with the renormalization constants given in \cite{Alexandrou:2020okk,Alexandrou:2019brg,Alexandrou:2015sea,ETMC:inpre}, where they are renormalized non-perturbatively using the RI-MOM massless scheme and converted to $\overline{\rm MS}$ at 2 GeV.

Similar with the 2-point function case, we have the relation of $\Ctpt$ to the standard nucleon 3-point function $\Ctpt_{NN}$ with projection matrix $\Gamma$ ($\Gamma=\Gamma_0$ for unpolarized projection and $\Gamma=\Gamma_k=i\gamma_5\gamma_k\Gamma_0$ for polarized projection) via
\begin{align}\label{eq:C3ptNN}
&\quad\Ctpt_{NN}(\Gamma,\OO;\vpp,\vp;\tf,\tc) \nonumber\\
&= \sum_{r^\prime,r=0,1} \Ctpt_{NN}^{r^\prime r}(\OO;\vpp,\vp;\tf,\tc) \,\hat{\Gamma}_{r^\prime r}(\vpp,\vp) \,,
\end{align}
where the relation between $\Gamma$ and $\hat{\Gamma}_{r^\prime r}(\vpp,\vp)$ is explained in \cref{app:23pt}.

\subsection{Topologies}

\begin{figure}[!ht]
    \centering
    \includegraphics[width=\columnwidth]{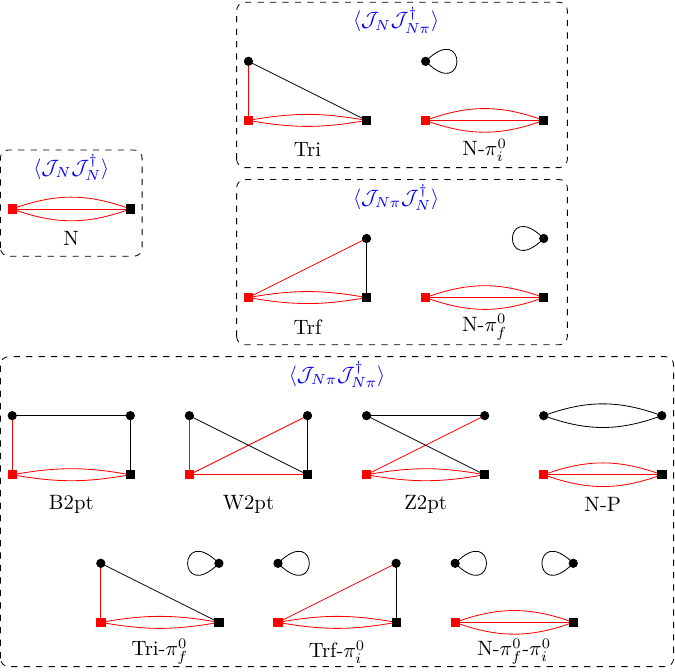}
    \caption{Topologies for the 2-point functions $\NN$, $\NNpi$, $\NpiN$, and $\NpiNpi$, where all arguments of the operators are suppressed. A name for each topology is given below it. 
    We denote the nucleon (pion) interpolating field by a square (circle). For each topology, the left (right) side represents the source $\ti$ (sink $\tf$) time slice, and the middle for the insertion time slice $\tc$.
    The red square denotes a nucleon interpolating field at $\ti$ and fixed source position $\xi$.
    Red  lines denote  point-to-all quark propagators 
    and black sequential or stochastic propagators.
    }
    \label{fig:diags_2pt}
\end{figure}

\begin{figure}[!ht]
    \centering
    \includegraphics[width=\columnwidth]{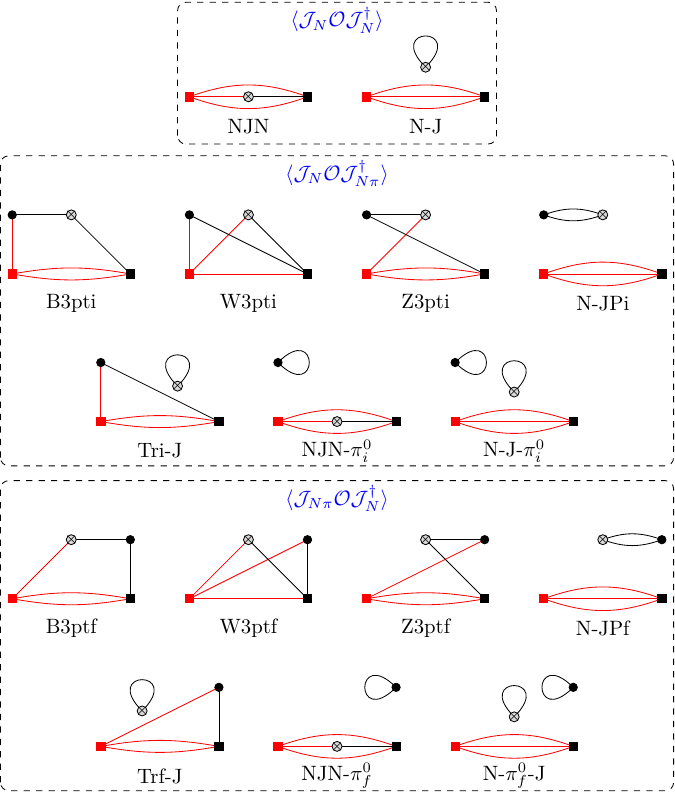}
    \caption{Topologies for the 3-point functions $\NJN$,
      $\NJNpi$, and $\NpiJN$. Circles with a cross inside denote the
      current. The rest of the notation is as in~\cref{fig:diags_2pt}.}
    \label{fig:diags_3pt_Id}
\end{figure}

\begin{figure}[!ht]
    \centering
    \includegraphics[width=\columnwidth]{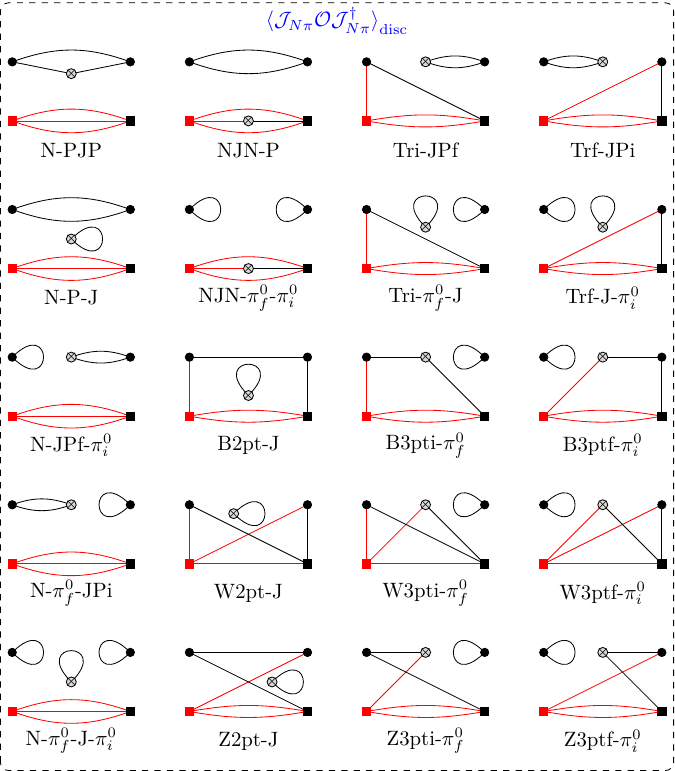}
    \caption{As in \cref{fig:diags_2pt} for disconnected topologies of $\NpiJNpi$.}
    \label{fig:diags_NpiJNpi_disc}
\end{figure}

\begin{figure}[!ht]
    \centering
    \includegraphics[width=\columnwidth]{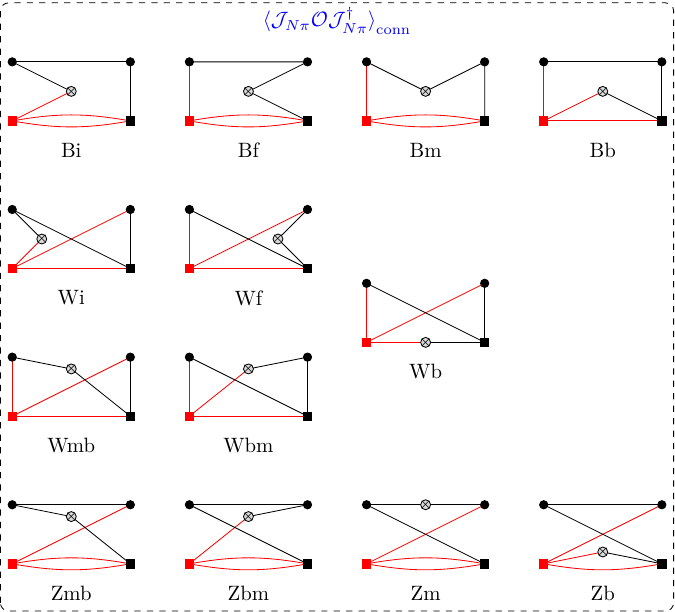}
    \caption{As in \cref{fig:diags_2pt} for fully quark-connected topologies of the 3-point function $\NpiJNpi$. These topologies are not included in the current work.}\label{fig:diags_NpiJNpi_conn}
\end{figure}

After performing the path integral over fermions, the 3-point functions can be  expressed as  a sum of products of quark propagators, each of which corresponds to a unique Wick-contraction of quark fields. These, in turn, can be represented by quark flow diagrams. We identify different topologies
depending on the quark pairing. A topology does not distinguish the flavors of quarks nor the quarks in the same interpolating field. Diagrams with the same topology can be computed in the same way. The topologies for the 2-point and 3-point functions 
in Eqs. (\ref{eq:2pt}) and (\ref{eq:3pt}) are summarized in \cref{fig:diags_2pt,fig:diags_3pt_Id,fig:diags_NpiJNpi_conn,fig:diags_NpiJNpi_disc}.
In the diagrams we suppress all space-time arguments.


In Figs. \ref{fig:diags_2pt} through \ref{fig:diags_NpiJNpi_disc} we consider topologies that include neutral pion loops.
These loops give non-zero contributions in the twisted mass fermion formulation, since isospin symmetry is broken at non-zero lattice spacing $a$.
They are found to be important for the restoration of isospin symmetry, as discussed in \cref{app:iso}.

For the quark-disconnected diagrams, we list, as the diagram name, the quark-connected sub-diagrams separated by hyphens.
We use the following scheme to denote the various sub-diagrams:
 `\pii', `\pif', and `J', denote a neutral pion loop at source time $\ti$, a neutral pion loop at
 sink time $\tf$, and a quark loop generated by the current operator $\OO$ at $\tc$, respectively. 
 A pion propagator from $\ti$ to $\tf$  is denoted by `P' and  a pion interpolator at $\ti$ ($\tf$) contracted with the current at $\tc$ 
 by `JPi' (`JPf'). We name the connected sub-diagram from contraction of a pion at $\ti$ with the current operator insertion and a pion at $\tf$ as `PJP'.
Each fully quark-connected topology is also assigned a name
as shown below each diagram in \cref{fig:diags_2pt,fig:diags_3pt_Id,fig:diags_NpiJNpi_conn,fig:diags_NpiJNpi_disc}.
We also have the connected triangle diagrams `Tri' and `Trf' depending on whether the vertices for the two-particle operator of the triangle are at $\ti$ or $\tf$, respectively, as well as the nucleon 2-point function `N'.

In computing these diagrams, we fix $\xsrc$ and $\ti$. In practice $N_{\text{src}}$ random lattice sites are used for $(\ti,\,\xsrc)$ to 
increase statistics.
Only for the purpose of discussion we take without loss of generality $\xsrc=\vec{0}$ and $\ti=0$. 

\begin{table}[!ht]
    \caption{Statistics used in this work for each class of topology as given in the second column and for the two gauge ensembles. We indicate the number of configurations by $N_{\text{cfg}}$, and the
      number of source points $(\ti,\,\xsrc)$ per gauge configuration by $N_{\text{src}}$,
    the number of stochastic time-slice sources by $N_{\text{stoc}}$, and the
    number of stochastic sources used in the one-end-trick approach by  $N_{\text{oet}}$. 
    For the 3-point functions with tensor insertions (marked with ``(tensor)'' in the table),
    the  statistics is less and indicated separately in the table.
    For convenience, we refer to `\pif' and `\pii' collectively as `\piz', to `JPf' and `JPi' collectively as `JP', to `X2pt', `X3pti', `X3ptf' collectively as `X' for X=B,W,Z. 
    }\label{tab:statistics}
    \centering
    \renewcommand\arraystretch{1.5}
    \begin{ruledtabular}
    \begin{tabular}{cccc}
        Number & Subset & cA211.530.24 & cA2.09.48 \\ \hline
        $N_{\text{cfg}}$ & all cases & 2467 & 1228 \\ \hline
        $N_{\text{src}}$ & N & 121 & 272 \\
         & NJN & 16 & 8 \\
         & NJN (tensor) & 4 & 1 \\
         & Tri & 96 & 16 \\
         & B, W, Z & 9 & 8 \\ 
         & BWZ3pt (tensor) & 3 & 4 \\  \hline
        $N_{\text{stoc}}$ & \piz & 200 & 100 \\
        & J & 200 & 400 \\
        & B, W & 12 & 12 \\ \hline
        $N_{\text{oet}}$ & P, JP, Z & 1 & 1 \\
    \end{tabular}
    \end{ruledtabular}
\end{table}

In what follows we briefly illustrate the method we used for the computation of diagrams for the different topologies
in Figs. \ref{fig:diags_2pt}, \ref{fig:diags_3pt_Id} and \ref{fig:diags_NpiJNpi_disc}.
The statistics we achieved in the following computations is given in \cref{tab:statistics}.

We build quark contractions from a combination of point-to-all, sequential and stochastic timeslice-to-all propagators and follow
the method detailed in Ref.~\cite{Alexandrou:2023elk}. In the figures red quark lines denote point-to-all quark propagators, while
black lines represent sequential or stochastic quark propagators.

To compute the quark loops arising from neutral pion ``\pif'', ``\pii'' and the current operators ``J'' we use stochastic timeslice sources. 
Also the P, JPi and JPf connected sub-diagrams are computed with stochastic timeslices and additionally we employ the one-end-trick. 

We note that exchange symmetry of source and sink (charge
conjugation, $\gamma_5$-hermiticity, time reversal) can be used to
transform several diagram pairs into each other. We used these
symmetry relations to obtain ``Trf'' from ``Tri'', ``X3ptf'' from
``X3pti'' for X = B, W, Z, and ``JPf'' from ``JPi''.

We note that the quark-disconnected 3-point function diagrams in
Fig.~\ref{fig:diags_NpiJNpi_disc} are straightforwardly obtained from
the sub-diagrams in Figs.~\ref{fig:diags_2pt}
and~\ref{fig:diags_3pt_Id} with the addition of pion and current
operator loops. We thus compute all the quark-disconnected diagrams
for the 3-point functions with two-hadron $N\pi$ interpolators simultaneously at
both source and sink. The fully connected diagrams in
Fig.~\ref{fig:diags_NpiJNpi_conn} bear a significantly higher
computational cost and are omitted in the present work.

\section{Optimal combination of 3-point functions using GEVP}\label{sec:Id}
With the 2-point functions defined in \cref{eq:C2pt}, the GEVP reads:
\begin{align}\label{eq:GEVP}
    &\quad \sum_{k}\Cdpt^r_{jk}(\vp;t)\,v^r_{nk}(\vp;t,t_0) \nonumber\\
    &= \lambda_n(\vp;t,t_0)\sum_{k}\Cdpt^r_{jk}(\vp;t_0)\,v^r_{nk}(\vp;t,t_0) \,.
\end{align}
We solve the GEVP per irrep row $r$ to account for the parity symmetry breaking in twisted-mass formulation as explained in \cref{app:pbreak}.

In this work the sum on index $k$ runs over 2 interpolating fields in the center-of-mass frame with $\vp=\vec{0}$, and 3 for the moving
frame with $|\vp| = 2\pi/L$ (cf. \cref{eq:cmf-operators,eq:jkp1}). \\

As the salient concept of this section we introduce three different variants of 3-point functions, two of which make use of the eigenpairs obtained 
from the GEVP. We henceforth assume the model situation, that the operator set entering the GEVP is a basis, i.e. that there are
as many independent operators as there are states, which contribute significantly. 
Under the model assumption, eigenvalues in \cref{eq:GEVP} give eigenenergies of the system, and the eigenvectors no longer depend on time, \ie,
\begin{align}\label{eq:lvmodel}
    \lambda_n(\vp;t,t_0)&=e^{-E_n(\vp)(t-t_0)} \,,\nonumber\\
    v^r_{nk}(\vp;t,t_0)&=v^r_{nk}(\vp)\,.
\end{align}

The following discussion is valid for any correlation matrix $\Cdpt$ and associated matrix of 3-point functions $\Ctpt$, so for brevity 
we can leave out the irrep row indexing $r,\,r'$.

The 2-point function has the spectral representation
\begin{align}
  \Cdpt_{jk}(\vp;t)
  &= 
  \sum\limits_{n}\,\braket{\Omega\,|\,\OJ_j(\vp;t)\,|\,n(\pvec)}\,\braket{n(\pvec)\,|\,\OJ_k(\vp;0)^{\dagger}\,|\,\Omega}
  \nonumber \\
  &= \sum\limits_n \, Z^*_{jn}(\vp) \, Z_{kn}(\vp)  \, e^{-E_n(\vp)t} \,,
\end{align}
with the sum over all contributing states $|n(\vp)\rangle$. In particular the state $n = 0$ is the 
nucleon state.
The GEVP is solved by eigenvectors $v_m$, which fulfill
the matrix equation
\begin{align}
  \sum\limits_{k} \, v_{mk}(\vp)\,Z_{kn}(\vp) &= A_n(\vp)\,\delta_{mn} \,,
\label{eq:viz}
\end{align}
where $A_n(\vp)$ depends on the normalization of states $|n(\pvec)\rangle$ and those of the lattice interpolators $\OJ_k$. \\

We normalize states as  $\braket{n(\pvec)\,|\,m(\pvec)} = \delta_{mn}$ 
and choose the normalization of the eigenvectors to satisfy $(v^{-1})_{N0} = 1$, such that
\begin{align}\label{eq:nmlz}
  A_0(\pvec) &= Z_{N0}(\pvec) \,.
\end{align}

With the eigenvectors $v_n$ we obtain the state-projected or ``GEVP-improved'' nucleon interpolator
\begin{align}
  \tilde{\OJ}^\dagger_N(\vp;t) &= \sum\limits_{k} \,v_{0k}(\vp) \, \OJ_{k}^{\dagger}(\vp;t) 
  \label{eq:Jtilde}
\end{align}
and our choice of eigenvector normalization ensures, that
\begin{align}
  \braket{N(\vp)\,|\, \tilde{\OJ}_N^\dagger(\vp;t)\,|\,\Omega} &= \braket{N(\vp)\,|\,\OJ_N^\dagger(\vp;t)\,|\,\Omega} \,.
\end{align}
We will focus on the contribution from the GEVP ground state $n = 0$.
Then this convention allows direct comparison of 3-point functions with ordinary source and sink
interpolators $\OJ_N$ and those built with GEVP-improved ones $\tilde{\OJ}_N$. \\

The 3-point functions defined in \cref{eq:C3pt} have the corresponding spectral expansion:
\begin{align}
    &\Ctpt_{jk}(\OO;\vpp,\vp;\tf,\tc)=\sum_{m,n} Z^{*}_{jm}(\vpp)\,Z_{kn}(\vp)
    \nonumber\\
    &\times\braket{m(\vpp)|\OO|n(\vp)} e^{-E_{m}(\vpp)(\tf-\tc)}e^{-E_{n}(\vp)\tc} \,.
    \label{eq:3pt-spectral}
\end{align}
In particular we define $I_0$ 
as the 3-point function with $j=k=N$ as a special case of Eq. (\ref{eq:3pt-spectral}):
\begin{align}
  I_0 = I_0(\OO;\vpp,\vp;\tf,\tc) &:=\Ctpt_{NN}(\OO;\vpp,\vp;\tf,\tc) \,.
  \label{eq:I0}
\end{align}
We use the spectral decomposition of the 3-point function 
to distinguish 3 contributions: first from the nucleon ground states
$n = m = 0$
with our target matrix element $\braket{N(\vpp)\,|\,\OO\,|\,N(\vp)}$, second
all contributions with nucleon ground state at sink and excited states $n > 0$ at source
or vice versa, and third from matrix elements with excited states at both source and sink
\begin{align}
  I_0 &= A_0(\vpp)^*\,A_0(\vp)
  \nonumber\\
  &\qquad \times\braket{N(\vpp)\,|\,\OO\,|\,N(\vp)} \, e^{-E_{N}(\vpp)(\tf-\tc)}\, e^{-E_{N}(\vp)\tc}
      \nonumber \\
    & \nonumber \\
  &+  \sum_{n>0} \, Z^{*}_{N 0}(\vpp)\,Z_{N n}(\vp) 
  \nonumber \\
  & \qquad \times \braket{N(\vpp)\,|\,\OO\,|\,n(\vp)} \, e^{-E_{N}(\vpp)(\tf-\tc)}\, e^{-E_{n}(\vp)\tc}
  \nonumber \\
  &+  \sum_{n>0} \, Z^{*}_{N n}(\vpp)\,Z_{N 0}(\vp) 
  \nonumber \\
  & \qquad \times \braket{n(\vpp)\,|\,\OO\,|\,N(\vp)} \, e^{-E_{n}(\vpp)(\tf-\tc)}\, e^{-E_{N}(\vp)\tc}
  \nonumber \\
  \nonumber \\
  &+  \sum_{n>0,m>0} \, Z^{*}_{N m}(\vpp)\,Z_{N n}(\vp)
  \nonumber \\
  & \qquad \times \braket{m(\vpp)\,|\,\OO\,|\,n(\vp)} \, e^{-E_{m}(\vpp)(\tf-\tc)}\, e^{-E_{n}(\vp)\tc} \,.
  \label{eq:I0-2}
\end{align}

One can remove the contamination by excited states $n>0$ or $m  > 0$ in $I_0$ entirely by using the GEVP-improved operators $\tilde{\OJ}_N$ 
defined in \cref{eq:Jtilde}
when projecting with the eigenvector $v_0$ for the GEVP ground state. We define $I$ as the corresponding fully GEVP-improved 
3-point function
\begin{align} 
  I &:=\sum\limits_{j,k}\, v^{*}_{0j}(\vpp)\,v_{0k}(\vp)\,\Ctpt_{jk}(\OO;\vpp,\vp;\tf,\tc)
  \nonumber \\
  &= A_0(\vpp)^*\,A_0(\vp)
  \nonumber \\
  &\qquad \times \braket{N(\vpp)\,|\,\OO\,|\,N(\vp)} \, e^{-E_{N}(\vpp)(\tf-\tc)}\, e^{-E_{N}(\vp)\tc}
  \label{eq:I-gevp-imp}
\end{align}

However, in practice computing the 3-point functions $\Ctpt_{jk}$ with $j, k \neq N$, i.e. 2-hadron $N\pi$ interpolators
at source and sink, is very costly. Thus, using the fully GEVP-improved $I$ is not straightforward.

Therefore, we propose a modification of the exact ground state projection in Eq. (\ref{eq:I-gevp-imp}).
The goal is to remove only the leading contamination by terms with matrix elements of nucleon ground state and any one excited state, i.e. 
to remove all terms with matrix elements $\braket{N(\vpp)\,|\,\OO\,|\,n(\vp)}$ and $\braket{n(\vpp)\,|\,\OO\,|\,N(\vp)}$. \\

In \cref{app:Idproof} we show that this is achieved with the linear combination
\begin{align}
  I_d &:= \sum\limits_{j,k} \, d_{jk} \,\,v^{*}_{0j}(\vpp)\,v_{0k}(\vp)\,\Ctpt_{jk}(\OO;\vpp,\vp;\tf,\tc) \,,
  \label{eq:Id}
\end{align}
where the 3-point function weights $d_{jk}$ are given by

\begin{align}
  d_{jk}  &=  \left\{
    \begin{matrix}
      1-W^{*}(\vpp)\,W(\vp)  \,, & j = k = N \,; \\
      & \\
      1+W^{*}(\vpp) \,, & j = N \,,\\
      & k \ne N \,; \\
      & \\
      1+W(\vp)  \,, & j \ne N \,, \\
      & k = N \,; \\
      & \\
      0 \,, & \mathrm{else}
    \end{matrix}
  \right.
  \label{eq:dij}
\end{align}
and  $W(\vp)$ is computed from the eigenvectors of the GEVP ground state as
\begin{align}
  W(\vp) &= \frac{1}{v_{0N}(\vp)\,[v^{-1}]_{N0}(\vp)}-1 \,.
  \label{eq:Wfactor}
\end{align}
Here we note $[v^{-1}]_{N0}=1$ if the normalization convention chosen previously is applied.
This modification $I_d$ will not require the computation of any 3-point function with $N\pi$ interpolators at both source and sink. \\

Further in \cref{app:Idproof}, we show that $I_d$ has the decomposition
\begin{align}
   I_d &= A^*_0(\vpp)\,A_0(\vp)
  \nonumber\\
  &\qquad \times\braket{N(\vpp)\,|\,\OO\,|\,N(\vp)} \, e^{-E_{N}(\vpp)(\tf-\tc)}\, e^{-E_{N}(\vp)\tc}
      \nonumber \\
    & \nonumber \\
  & - \sum_{n>0,m>0} \, Z^{*}_{Nm}(\vpp)\,Z_{N n}(\vp)
  \nonumber \\
  & \qquad \times \braket{m(\vpp)\,|\,\OO\,|\,n(\vp)} \, e^{-E_{m}(\vpp)(\tf-\tc)}\, e^{-E_{n}(\vp)\tc} \,,
  \label{eq:Id-2}
\end{align}
under the model assumption  that there are
as many independent interpolators as there are states, which contribute significantly.

This partial removal of excited state effects is useful, since for relatively large $\tc$ and $\tf-\tc$, one has
the enhanced exponential suppression of the purely excited state effects
\begin{align}\label{eq:hier}
    e^{-E_{N}(\vpp)(\tf-\tc)}e^{-E_{n}(\vp)\tc} \gg e^{-E_{m}(\vpp)(\tf-\tc)}e^{-E_{n}(\vp)\tc} \,,\nonumber\\
    e^{-E_{m}(\vpp)(\tf-\tc)}e^{-E_{N}(\vp)\tc} \gg e^{-E_{m}(\vpp)(\tf-\tc)}e^{-E_{n}(\vp)\tc} \,,
    \nonumber
\end{align}
for all $n,\,m > 0$. Thus in $I_d$ the dominant excited contamination is removed in the asymptotic regime.
We note, that in \cref{eq:Id-2} the purely excited state contamination has the opposite sign compared to 
$I_0$ in \cref{eq:I0-2}. This enables $I_d$ to indicate the significance of such contamination.
Our interpretation is, that in the absence of a significant change between $I_0$ and $I_d$, 
contamination by GEVP states above the ground state is likely insignificant. Consequently,
a further computation of 3-point functions with 2-hadron $N\pi$ interpolators at both source and sink 
to achieve the fully GEVP state projected $I$ will likely not lead to further significant improvement.

\section{Determination of GEVP eigenvectors}
\label{sec:gevp}
In practice, one has to solve the GEVP in \cref{eq:GEVP} with a finite number of interpolating fields.
The solution returned by the incomplete GEVP is different from the eigenmodes returned by the complete GEVP (i.e. the model situation with a basis of operators).
However, one is able to approach the lowest few eigenmodes by taking the asymptotic limit because the higher modes are suppressed in  such limit, \ie, one has
\begin{align}
    \lambda_n(\vp;t,t_0) &\xrightarrow{t,t_0\to\infty} e^{-E_n(\vp)(t-t_0)} \,,\nonumber\\
    v_{nk}^r(\vp;t,t_0) &\xrightarrow{t,t_0\to\infty} v_{nk}^r(\vp) \,.
\end{align}
We note here the difference to \cref{eq:lvmodel} is that \cref{eq:lvmodel} holds only under the model assumption.

\begin{figure*}[!ht]
    \centering
    \includegraphics[width=\textwidth]{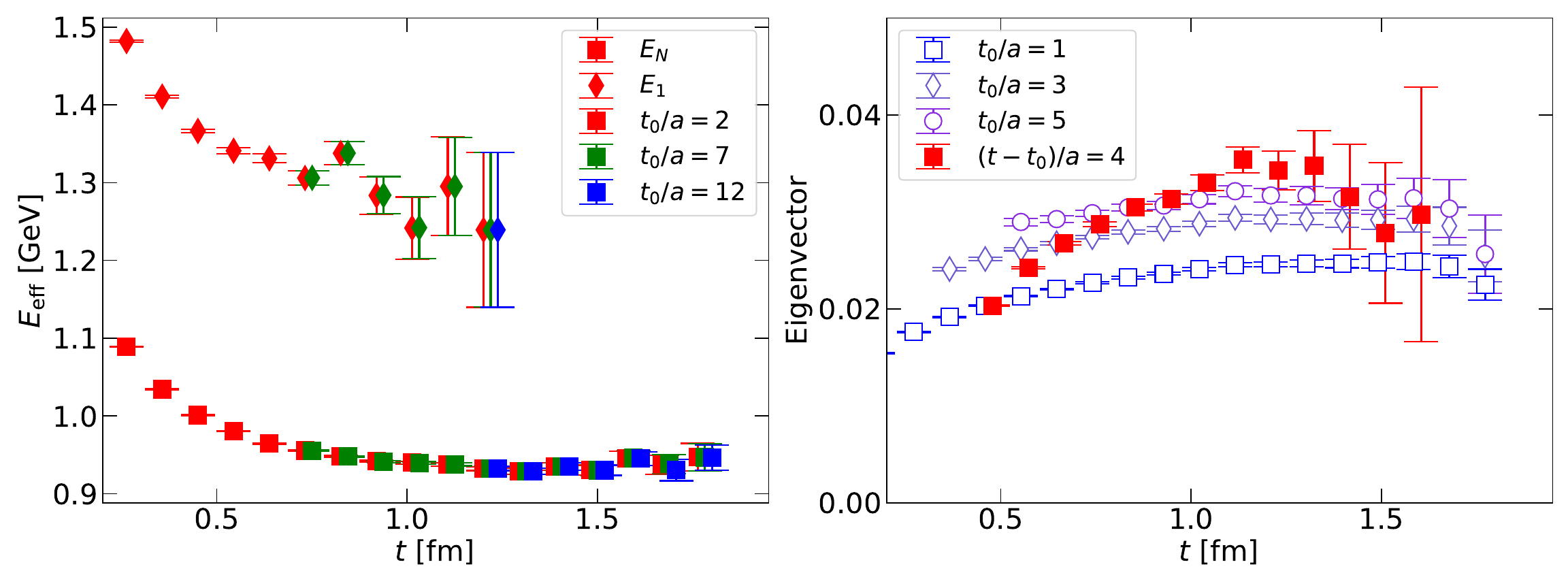}
    \caption{Ensemble cA2.09.48. We show an illustrative example of the $t_0$-dependence on effective energies (left) and eigenvectors (right) for  zero total momentum. 
 The eigenvector component used is $|v_{0,N(1)\pi(1)}/v_{0,N}|$.}\label{fig:t0dep}
\end{figure*}
\begin{figure}[!ht]
    \centering
    \includegraphics[width=\columnwidth]{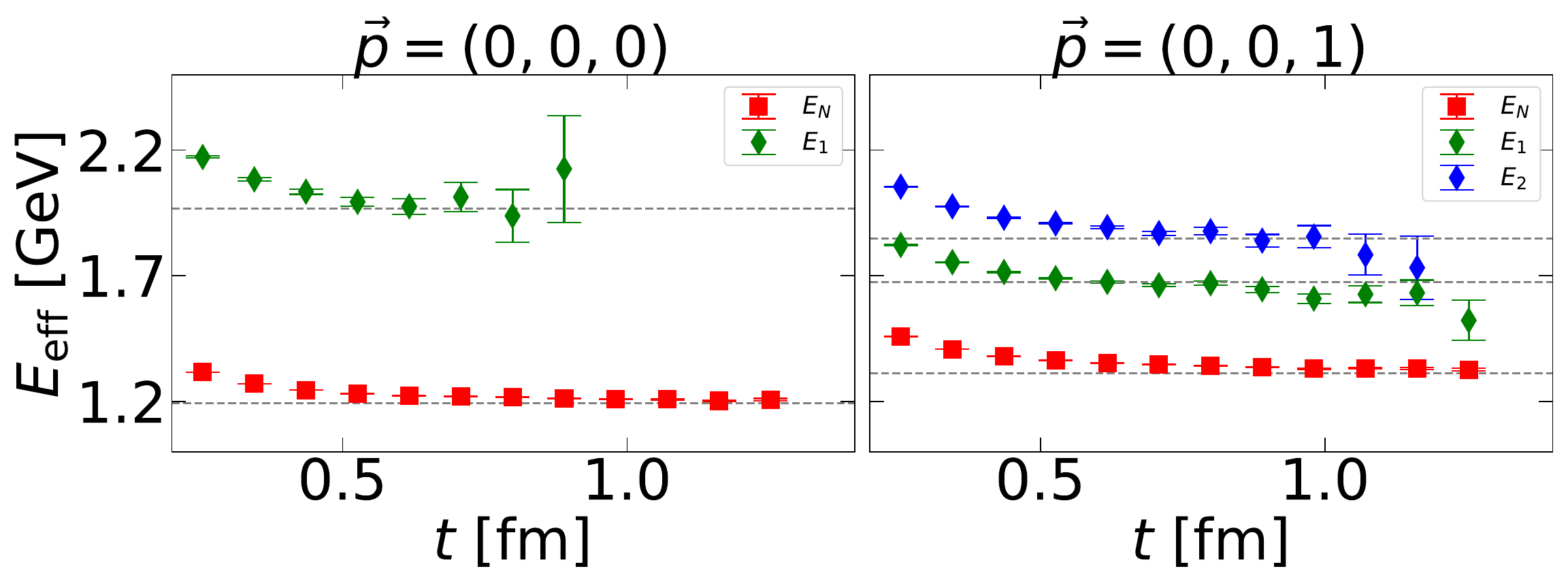}
    \caption{Ensemble cA211.530.24. Effective eigenenergies from GEVP performed with fixed $t_0/a=2$. Horizontal lines are non-interacting energy levels.
    }\label{fig:evalues24}
\end{figure}
\begin{figure}[!ht]
    \centering
    \includegraphics[width=\columnwidth]{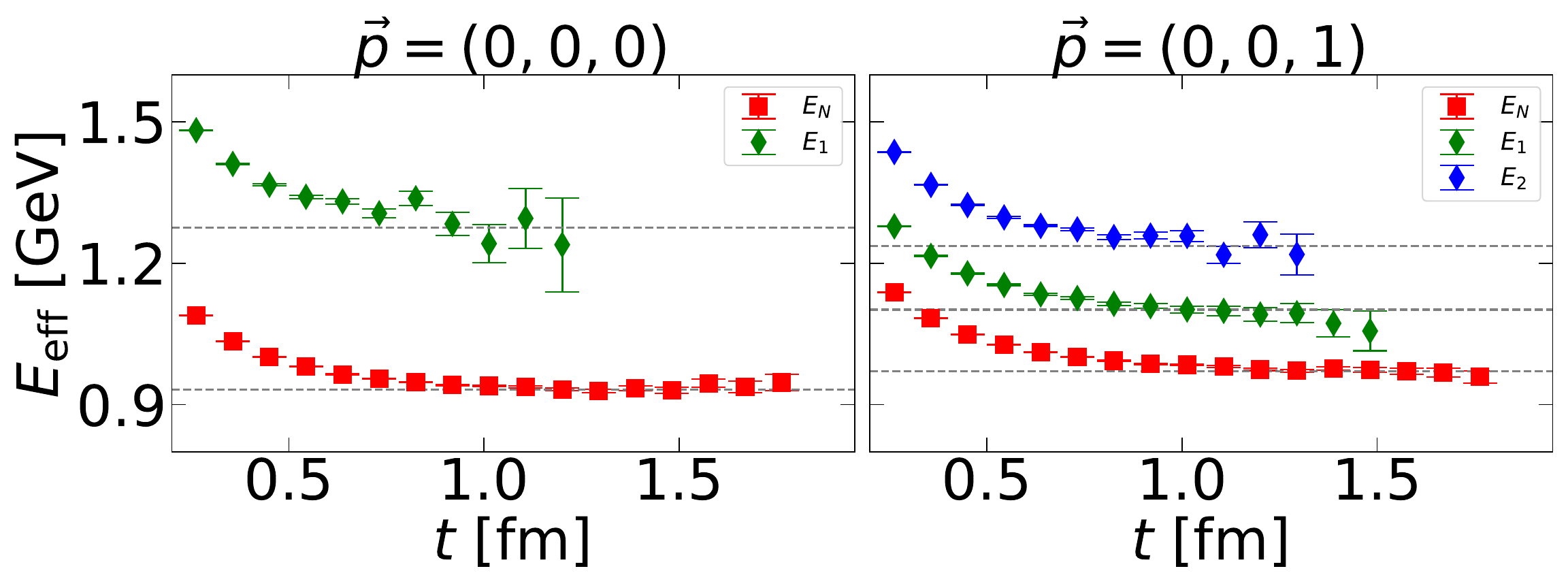}
    \caption{Ensemble cA2.09.48. Notation is the same as in \cref{fig:evalues24}.
    }\label{fig:evalues48}
\end{figure}

In the extraction of the eigenenergies, one can use the effective energy 
\begin{align}
    E^{\mathrm{eff}}_n(\vp;t,t_0)=\text{log}\frac{\lambda_n(\vp;t,t_0)}{\lambda_n(\vp;t+1,t_0)} \xrightarrow{t\to\infty} E_n(\vp) \,,
\end{align}
where one does not need the $t_0\to\infty$ limit as  proved in Ref.~\cite{Luscher:1990ck}. In Ref.~\cite{Blossier:2009kd}, it was  shown that  using $t_0\sim t/2$ one achieves a faster rate of convergence.
In practice, a fixed and relatively small $t_0$ is widely used, because in many cases $E^{\mathrm{eff}}_n(\vp;t,t_0)$ has very weak dependence on $t_0$, as illustrated in the left panel of \cref{fig:t0dep}, where points at same $t$ and different $t_0$, illustrated with different colors, almost overlap with each other.
Since almost no $t_0$-dependence is observed for the eigenenergies, we apply GEVP with fixed $t_0=2a$. The results are presented in \cref{fig:evalues24,fig:evalues48}.
In the extraction of the eigenvectors, however, we observe strong $t_0$-dependence, as illustrated in the right panel of \cref{fig:t0dep}.
Varying $t_0$, the results at large $t$ converge to a different constant value with small errors. 
In order to reliably extract the eigenvectors, we choose to fix $t-t_0=4a$. 
We note that any value of fixed $t-t_0$ can give the correct result  for $t$ and $t_0$  sufficiently large, since $t$ and $t_0$ will be increased at the same time to achieve the double asymptotic limit $t,\,t_0 \to \infty$ \cite{Blossier:2009kd}, and we choose $t-t_0=4a$ since it gives results with relatively less errors.
With the choice $t-t_0=4a$, the eigenvectors do converge at around $t=1.1\,$fm within  the larger errors. 
In particular, they converge to a value that the eigenvectors also approach as a function of $t$ as one increases 
$t_0$ from $t_0/a = 1$ to $t_0/a = 5$. 
Within this asymptotic regime, in what follows, we will perform plateau fits to determine the eigenvectors.

\begin{figure}[!ht]
    \centering
    \includegraphics[width=\columnwidth]{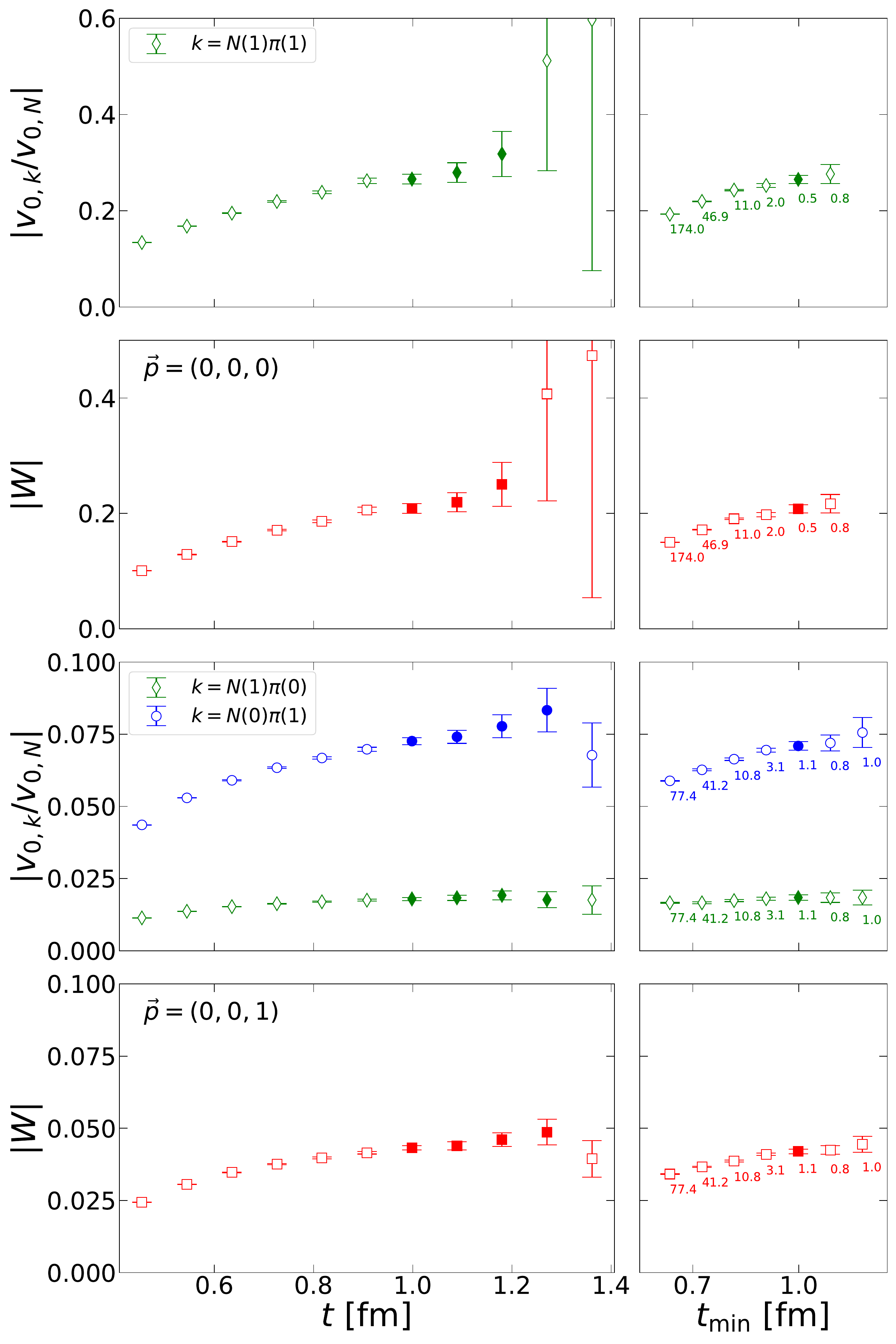}
    \caption{Ensemble cA211.530.24. Left panels: $|v_{0,k}/v_{0,N}|$ and $W$ extracted from GEVP at different $t$ with $t-t_0=4\,a$, for different total momenta $\vp$. 
    Right panels:  results of joint plateau fits as a function of the fit range parameter $t_{\mathrm{min}}$ ($t_{\mathrm{max}}$  chosen to make the relative errors of all fitted data smaller than 0.2). 
    The numbers below the points are the reduced $\chi^2$.
    The filled points show the most probable fit determined according to the Akaike Information Criterion (AIC).
    }\label{fig:2ptGEVP_24}
\end{figure}

\begin{figure}[!ht]
    \centering
    \includegraphics[width=\columnwidth]{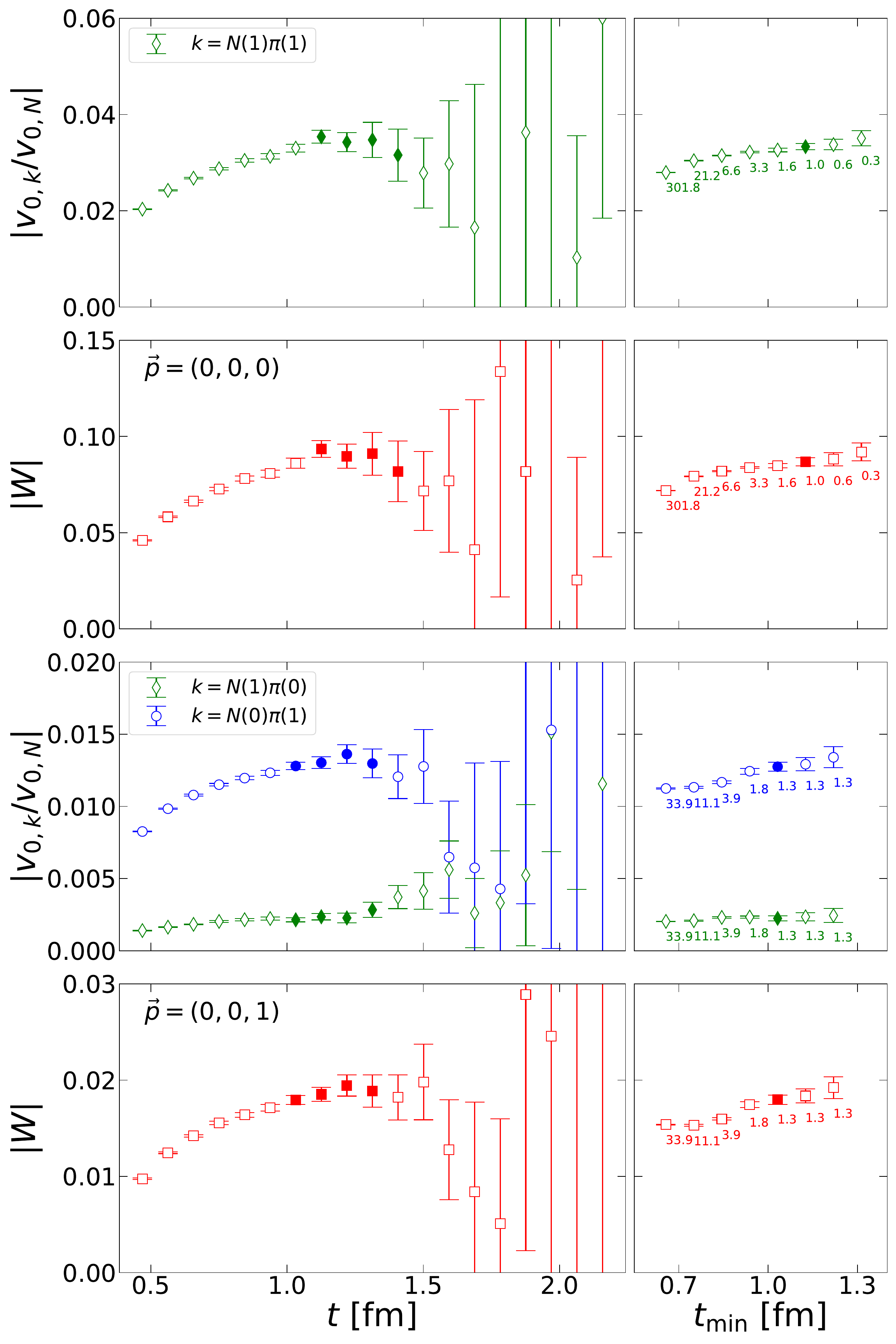}
    \caption{Ensemble cA2.09.48. the notation is the same as in \cref{fig:2ptGEVP_24}.}\label{fig:2ptGEVP_48}
\end{figure}

In \cref{fig:2ptGEVP_48,fig:2ptGEVP_24}, we show the quantities $|v_{0k}/v_{0N}|$ and $|W|$ (defined in \cref{eq:Wfactor}) determined from GEVP as a function of $t$ with $t-t_0$ fixed to be $4\,a$. 
We perform joint plateau fits that fit to $|v_{0,k}/v_{0,N}|$ and $|W|$ together for each momentum and ensemble.
As explained in \cref{app:pbreak}, in the case of $\vp=\vec{1}_z$, both real and imaginary parts of the these quantities are present and cannot be easily absorbed into redefinition of operators as a consequence of the broken parity of the twisted-mass action. 
We consider both the real and imaginary parts of these quantities, and use separate parameters in the fits.
We also use the Akaike Information Criterion (AIC) \cite{Jay:2020jkz,Neil:2022joj} to determine the most probable fits, that will be used for further analysis.

\begin{figure*}[!ht]
    \centering
    \includegraphics[width=\textwidth]{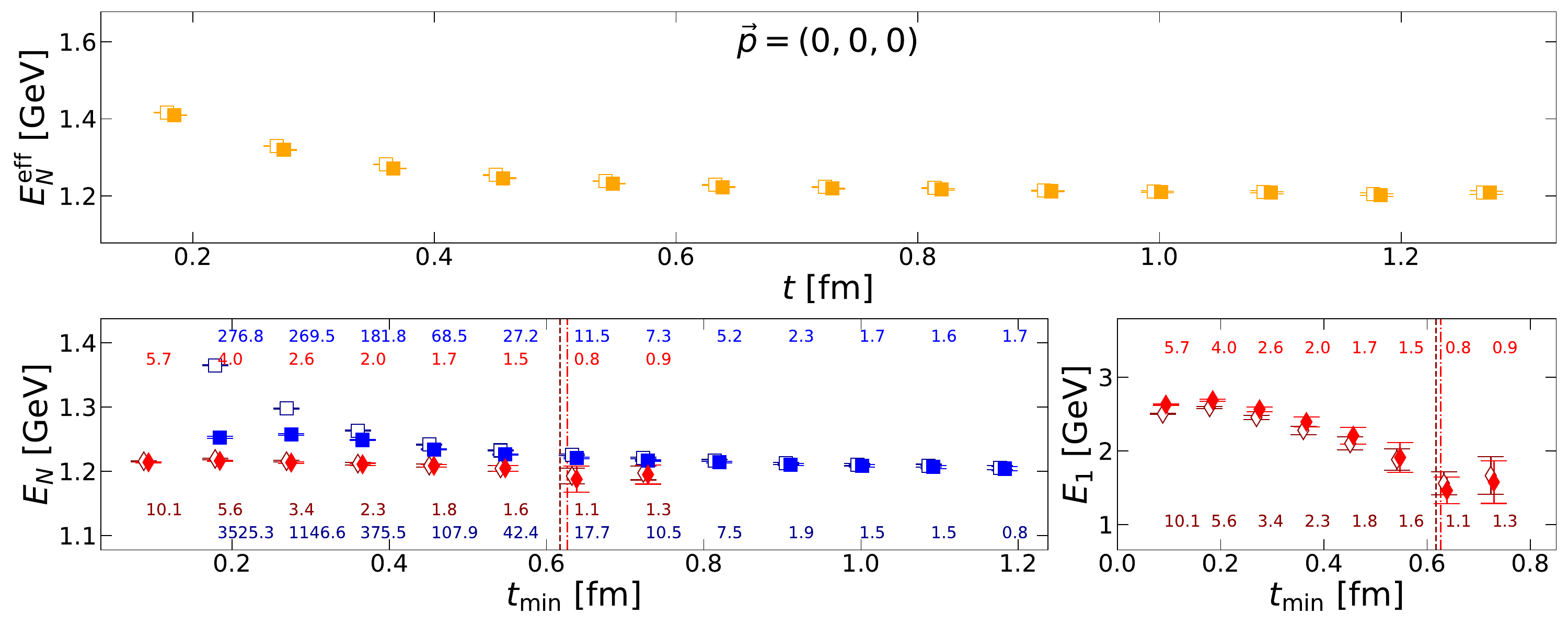}
    \includegraphics[width=\textwidth]{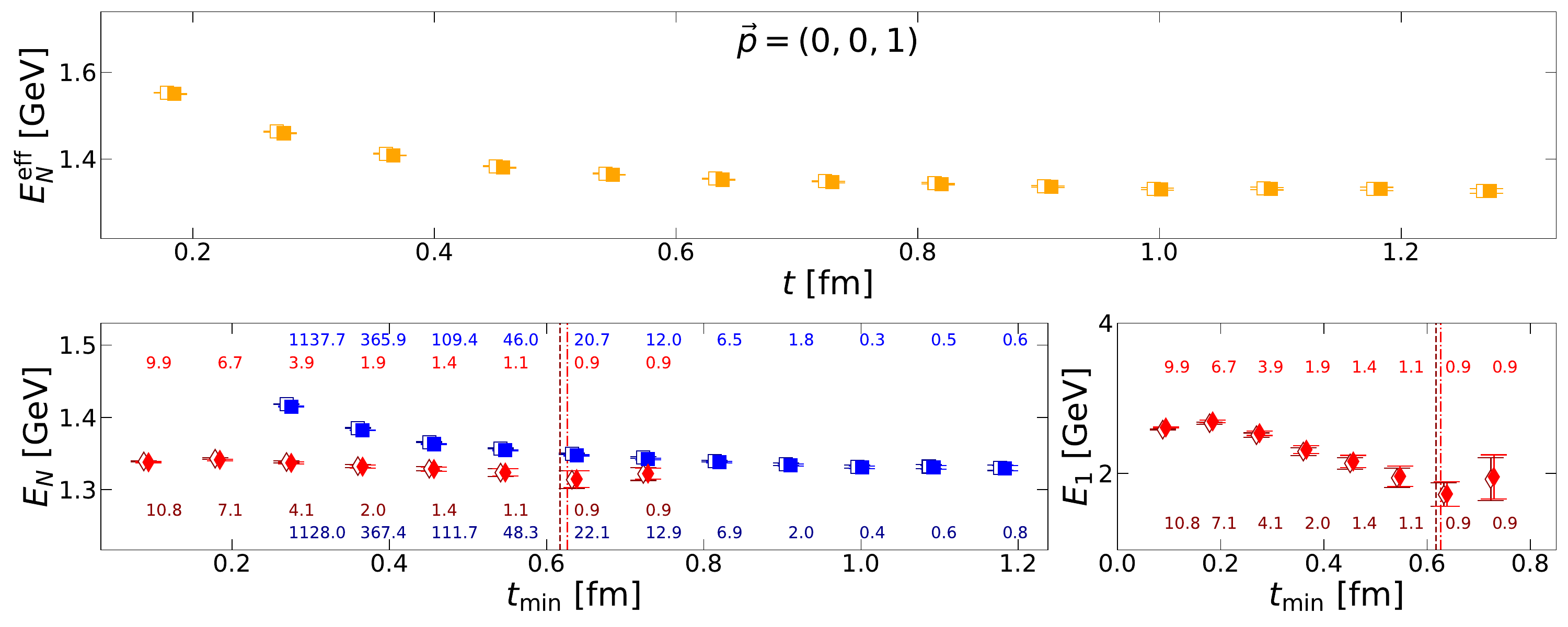}
    \caption{Ensemble  cA211.530.24. 
    Results for the effective energies and the fits to them for total momenta $\vp=(0,0,0)$ (top three panels) and $\vp=(0,0,1)\,2\pi/L$ (bottom three panels).
    The meaning of the three panels for a given total momentum is as follows.
    Upper panel: We show the nucleon effective energy $E_N^{\mathrm{eff}}(t)$ as a function of $t$. 
    Lower left panel: We show the convergence of the extracted value of $E_N$ as a function of the lowest time $t_{\mathrm{min}}$ used in the fit when we include one-state in the fit (squares), when we include two-states (diamonds).
    Lower right panel: The same as the lower left panel but for the values extracted for the mass of the first excited state.
    Open points are results from the usual nucleon 2-point functions. Filled points are results from the improved 2-point functions using GEVP eigenvectors extracted in \cref{fig:2ptGEVP_24}.
    The numbers in the lower panels are the reduced $\chi^2$ for different cases with the corresponding $t_{\mathrm{min}}$. The numbers on the bottom are for the case without GEVP improvement, and those on the top are for the case with GEVP improvement.
    Vertical lines (dashed lines without, dash-dotted lines with GEVP improvement) indicate the $t_{\mathrm{min}}$ of the most probable two-state fit determined according to the AIC.
    }\label{fig:2ptFit_24}
\end{figure*}

\begin{figure*}[!ht]
    \centering
    \includegraphics[width=\textwidth]{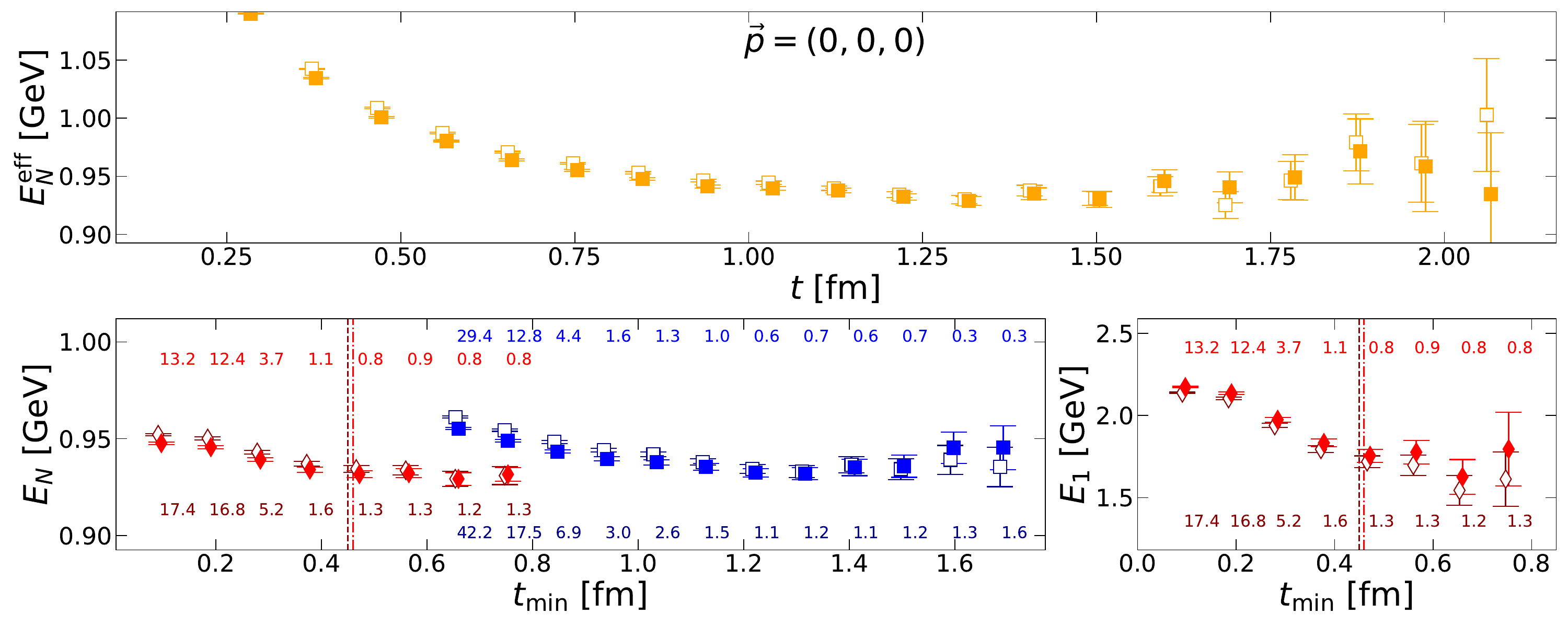}
    \includegraphics[width=\textwidth]{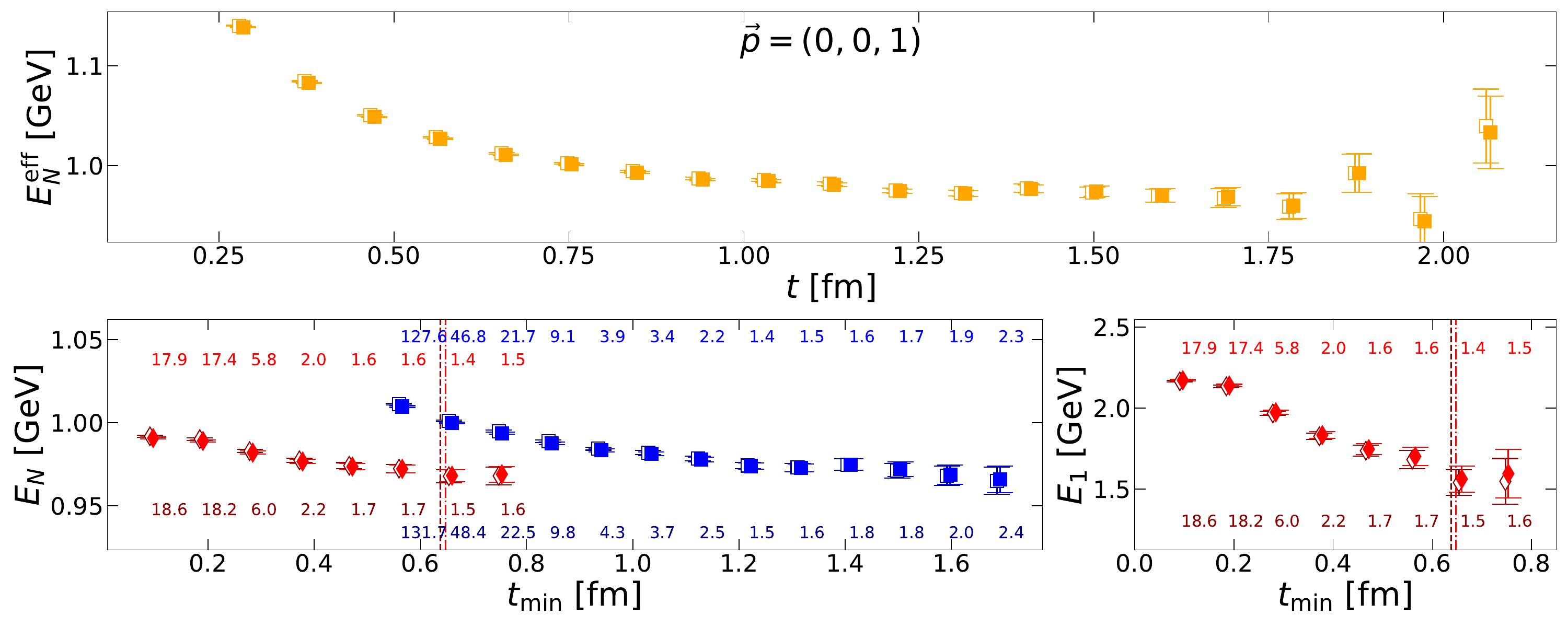}
    \caption{Ensemble  cA2.09.48. As in \cref{fig:2ptFit_24}. The GEVP eigenvectors used are extracted in \cref{fig:2ptGEVP_48}}\label{fig:2ptFit_48}
\end{figure*}

We use the eigenvectors extracted from  most probable fits using the Akaike information criterion (AIC) to improve the nucleon 2-point functions $\Cdpt_{NN}(\vp;t)$ defined in \cref{eq:C2ptNN} by replacing  $\JN^r$ with the GEVP improved $\tilde{\OJ}_N^r$:
\begin{align}\label{eq:ONt}
    \JN^r(\vp;t) \to \tilde{\OJ}_N^r(\vp;t) =\sum_{k} v_{0k}^r(\vp) \OJ_k^r(\vp;t)  \,,
\end{align}
where  $k$ sums over all the interpolating fields in the GEVP basis.
We show the effective energies of the nucleon
\begin{align}
    E_N^{\mathrm{eff}}(\vp;t)=\text{log}\frac{\Cdpt_{NN}(\vp;t)}{\Cdpt_{NN}(\vp;t+1a)} \,,
\end{align}
as a function of $t$ in \cref{fig:2ptFit_48,fig:2ptFit_24} without (open points) and with (filled points) GEVP improvement.
We perform one- and two-state fits to the effective energies. We use  AIC to determine the most probable two-state fit indicated with vertical lines.  As can be seen in \cref{fig:2ptFit_24}, there is an improvement in the extraction of the ground state energy for the heavier pion  mass ensemble suppressing excited states at earlier times. However, such an improvement is not observed for the  ensemble with the physical pion mass shown in \cref{fig:2ptFit_48}, which is in agreement with Ref.~\cite{Bar:2015zwa}.
The datasets used in the most probable fit will be used for joint fits with 3-point functions.

\section{Analysis of nucleon form factors}\label{sec:anff}
With the standard nucleon 2-point and 3-point functions defined in \cref{eq:C2ptNN,eq:C3ptNN}, one can construct ratios among them that converge to the desired nucleon matrix elements asymptotically.
In this work, we construct the following ratio 
\begin{align}\label{eq:ratio0}
    &\quad R^{\Gamma}_\OO(\vpp,\vp;\tf,\tc) =\frac{\Ctpt_{NN}(\Gamma,\OO;\vpp,\vp;\tf,\tc)}{\sqrt{\Cdpt_{NN}(\vpp;\tf)\Cdpt_{NN}(\vp;\tf)}} \nonumber\\
    &\qquad \times e^{-\left(E^{\rm eff}_{N}(\vpp;\tf,t_0)-E^{\rm eff}_{N}(\vp;\tf,t_0)\right)\,(\tc-\frac{\tf}{2})} \,, \nonumber\\
    &\xrightarrow[\tf-\tc\to\infty]{\tc\to\infty} \sum_{s^\prime, s} \tilde{\Gamma}_{s s^\prime} \braket{N(p^\prime,s^\prime)|\OO|N(p,s)} \equiv \Pi^{\Gamma}_\OO(p^\prime,p) \,,
\end{align}
where $\tilde{\Gamma}$ is the spin-projection matrix corresponding to $\Gamma$. The asymptotic limit is discussed in more details in \cref{app:deNME}.
We use the effective energies determined by taking $t_0/a=2$ as shown in \cref{fig:evalues24,fig:evalues48}.

\begin{figure*}[!ht]
    \centering
    \includegraphics[width=\textwidth]{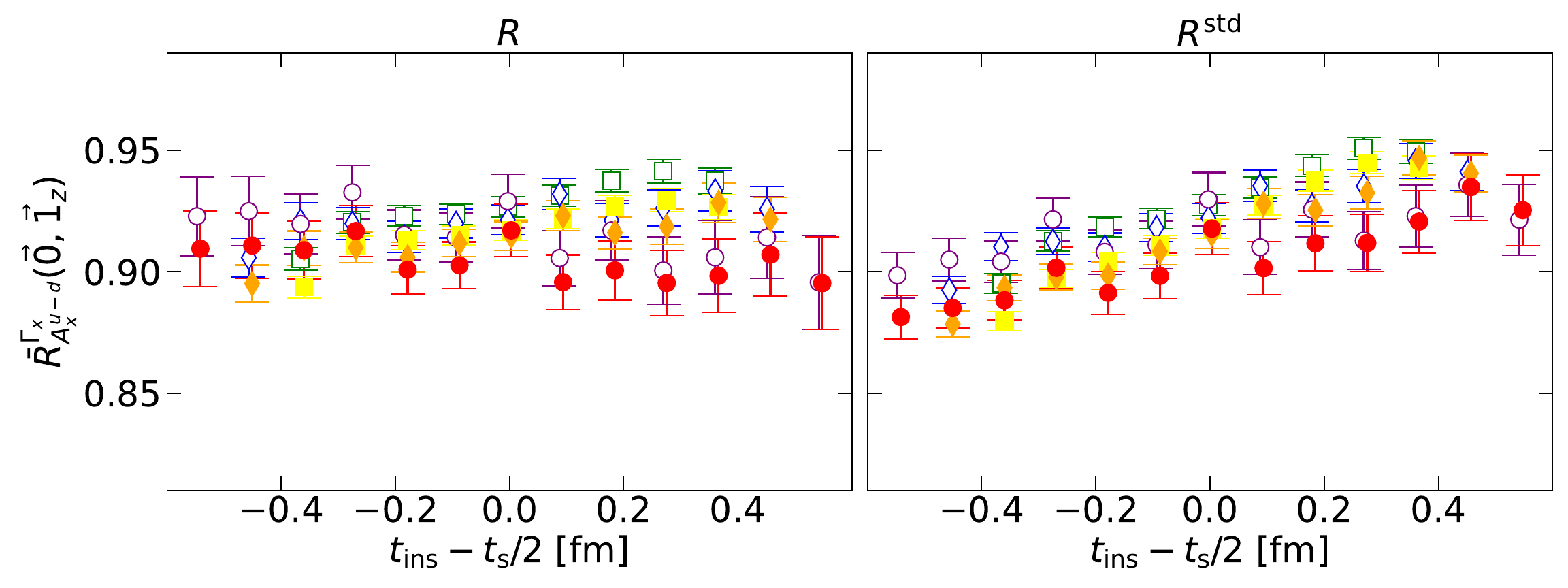}
    \caption{Ensemble cA211.530.24.  A comparison of the results using the ratios $R$ and $R^{\rm std}$ for the example case defined in \cref{eq:expRstd}. 
    Open and filled points refer to results without and with GEVP improvement, respectively.
    Squares, diamonds and circles  refer to $t_s/a=10,12$ and $14$, respectively.}\label{fig:3pt_raw_GA_01_24_R_Rstd}
\end{figure*}
A standard ratio  widely used in previous works is given by
\begin{align}\label{eq:ratio_std}
    & [R^{\text{std}}]^\Gamma_\OO(\vpp,\vp;\tf,\tc) = \frac{\Ctpt_{NN}(\Gamma,\OO;\vpp,\vp;\tf,\tc)}{\sqrt{\Cdpt_{NN}(\vpp;\tf)\Cdpt_{NN}(\vp;\tf)}} \nonumber\\
    &\qquad \times \sqrt{\frac{\Cdpt_{NN}(\vp;\tf-\tc)\Cdpt_{NN}(\vpp;\tc)}{\Cdpt_{NN}(\vpp;\tf-\tc)\Cdpt_{NN}(\vp;\tc)}} \nonumber\\
    &\xrightarrow[\tf-\tc\to\infty]{\tc\to\infty} \Pi^{\Gamma}_\OO(p^\prime,p) \,.
\end{align} 
Apart from  the fact that  two ratios are  exactly equal at the middle points $\tc-\tf/2=0$, $R^{\text{std}}$ usually has a smaller error due to using   correlated terms, but it has additional excited state contaminations from the 2-point functions at early time slices.
In \cref{fig:3pt_raw_GA_01_24_R_Rstd}, we present a comparison of the results using these two ratios for the example case (we suppress the time arguments)
\begin{align}\label{eq:expRstd}
    \bar{R}^{\Gamma_{x}}_{A_{x}^{u-d}}(\vec{0},\vec{1}_z) &= \mathcal{C}\left[i(E_N+m_N)\right]^{-1} R^{\Gamma_{x}}_{A_{x}^{u-d}}(\vec{0},\vec{1}_z) \nonumber\\ 
    &\hspace{2cm}\to G_A^{u-d}(Q_1^2) \,,
\end{align}
where $G_A^{u-d}(Q_1^2)$ is the isovector axial form factor at one unit of momentum transfer.
As expected, $R$ is noisier but tends to be flatter than $R^{\rm std}$, although within errors the two ratios give consistent results, as expected.
We note, that we use  a particular ratio  $R$  only for visualization purposes, while in the fits that we  perform, we will fit directly to 2-point and 3-point functions rather than to the ratio. 

Nucleon matrix elements $\Pi^\Gamma_\OO(p^\prime,p)$ can be expressed as linear combinations of nucleon form factors. 
These decompositions are discussed in more detail in \cref{app:deNME}.
Using the decompositions, we can recombine the ratios $R$ into new ratios $\bar{R}$, which then converge to individual nucleon form factors.
We will present these new ratios in \cref{sec:GEVPvi}, without and with GEVP improvement.

\subsection{Comparison among different GEVP projected 3-point functions} \label{sec:comtpf}\
In this section, we will compare results using $I_d$ with other alternative linear combinations of 3-point functions, which 
we label as $I^\prime$ and $I^{\pprime}$.
To illustrate these alternatives we introduce $I_{jk}$, the GEVP ground-state projected contribution from 3-point function $\Omega_{jk}$
\begin{align}
	I_{jk} &:= v^{*}_{0j}(\vpp)\,v_{0k}(\vp)\,\Ctpt_{jk}(\OO;\vpp,\vp;\tf,\tc) \,.
  \label{eq:Id}
\end{align} 
We restrict the detailed definitions to the case of just two operators, which we label as $N$ and $N\pi$.
This applies directly to the center-of-mass frame, where $N\pi$ corresponds to $N(1)\pi(1)$.
For the moving frames with two $N\pi$ interpolators ($N(1)\pi(0)$ and $N(0)\pi(1)$) the definitions are analogous.
We can then write
\begin{align}
	I_d &= d_{N,N}\,I_{N,N} + d_{N,N\pi}\,I_{N,N\pi} + d_{N\pi,N}\,I_{N\pi,N} \,.
	\label{eq:Id-sum} 
\end{align}
In Ref.~\cite{Barca:2022uhi}, a 3-point function without including the weights $d_{jk}$
\begin{align}
	I^\prime &= I_{N,N} + I_{N,N\pi} + I_{N\pi,N}
	\label{eq:Ip-sum}
\end{align}
is used, i.e. the fully GEVP-projected $I$ without contributions from 3-point functions $\NpiJNpi$ with $N\pi$ interpolator
at both source and sink.
Since we are computing all the disconnected diagrams for the latter 3-point functions, we define in addition the following combination of 3-point functions:
\begin{align}
    I^{\prime\prime}&=I_{N,N} + I_{N,N\pi} + I_{N\pi,N} + I^{\text{disc}}_{N\pi,N\pi} \,,
    \label{eq:Ipp-sum}
\end{align}
where the last term includes only all the diagrams with quark-disconnected topology.
To the three different 3-point functions $I_d$, $I^\prime$ and $I^{\prime\prime}$
in Eqs. (\ref{eq:Id-sum})-(\ref{eq:Ipp-sum}) correspond the ratios $R_d$, $R^\prime$, and $R^{\prime\prime}$, respectively.

\begin{figure*}[!ht]
    \centering
    \includegraphics[width=\textwidth]{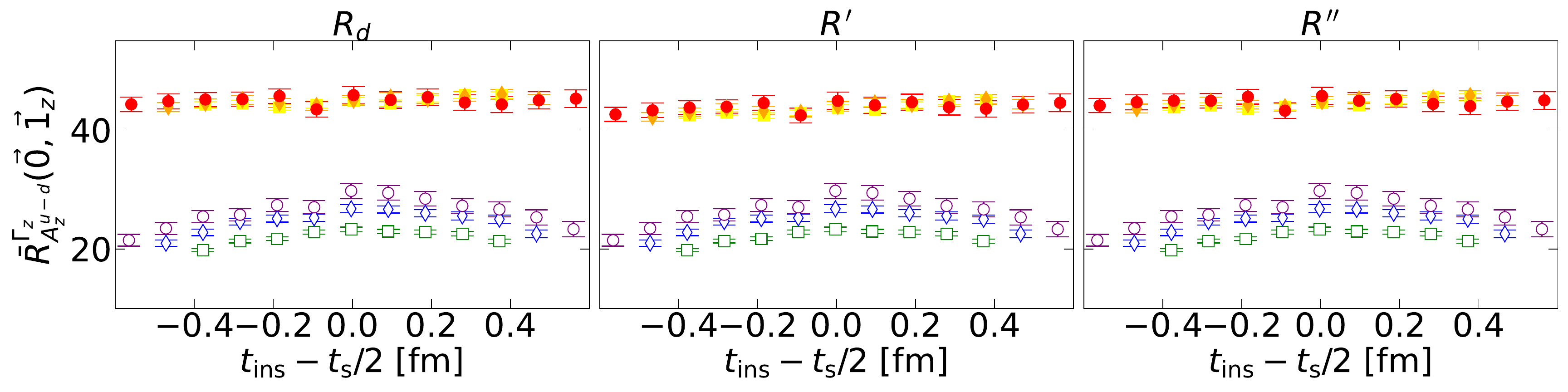}
    \caption{Ensemble  cA2.09.48. A comparison among results when using the  ratios $R_d$, $R^\prime$, and $R^{\prime\prime}$ for the example case defined in \cref{eq:expIdppp}.
    Open and filled points refer to results without and with GEVP improvement, respectively. The open symbols show the same results across the three panels.
    Square, diamond and circle points refer to $t_s/a=10,12$ and $14$ respectively.
    }\label{fig:3pt_raw_GP_01_zz_48_Id_Ip_Ipp}
\end{figure*}
In \cref{fig:3pt_raw_GP_01_zz_48_Id_Ip_Ipp}, we show the comparison among the results extracted for the three ratios (filled points) for the example case (we suppress the time arguments)
\begin{align}\label{eq:expIdppp}
    \bar{R}^{\Gamma_{z}}_{A_{z}^{u-d}}(\vec{0},\vec{1}_z) &=\mathcal{C}\left[-i\frac{p_1^2}{2m_N}\right]^{-1}{R}^{\Gamma_{z}}_{A^{u-d}_{z}}(\vec{0},\vec{1}_z)  \nonumber\\
     &\hspace{-1.5cm} -\left[\frac{-2m_N(E_N+m_N)}{p_1^2}\right]\bar{R}^{\Gamma_{x}}_{A^{u-d}_{x}}(\vec{0},\vec{1}_z) \to G_P^{u-d}(Q_1^2) \,,
\end{align}
where $G_P^{u-d}(Q_1^2)$ is the isovector induced pseudoscalar form factor at one unit of momentum transfer.
GEVP-improved results using $R_d$, $R'$ and $R''$ are compatible with each other.
Although $R_d$ is expected to be less contaminated than $R^\prime$, as explained in \cref{sec:Id}, the observed values of $W(\vec{p})$ shown in \cref{fig:2ptGEVP_48} are small and they do not make a statistically significant improvement.
In Ref.~\cite{Barca:2022uhi}, the use of $R^\prime$ that neglects the contribution from $\NpiJNpi$ is based on  chiral perturbation theory.
 $R_d$, proposed in this work, however, only relies on the hierarchical condition \cref{eq:hier} and holds without requiring the use of chiral perturbation theory.

\subsection{Application of GEVP to nucleon matrix elements of various  currents}\label{sec:GEVPvi}
The ratios of 3-point to 2-point functions given  in \cref{eq:ratio0}  converge to the desired nucleon matrix elements asymptotically.
Since nucleon matrix elements decompose into linear combinations of nucleon form factors, as discussed in \cref{app:deNME}, we can recombine ratios into new ratios denoted by $\bar{R}$ that converge directly to a desired nucleon form factor. The construction of the new ratios $\bar{R}$ will be presented below. 
2-point and 3-point functions related by symmetries will be averaged over. In what follows we will, thus, discuss ratios  $\bar{R}$ for which these symmetries are taken into account in the averages. 
Various symmetry transformations have been used as explained in more detail in \cref{app:dsbs}.
They include the complex conjugation that exchanges the source and sink for the nucleon interpolating field, 
which symmetrizes or anti-symmetrizes the ratio at $\tc=t$ and $\tc=\tf-t$ 
when the source and sink carry the same momentum, \ie, for $Q^2=(p^\prime-p)^2=0$ (cf. \cref{eq:ds_cc}).

We will present results extracted from  $\bar{R}$ with and without GEVP improvement.
$\bar{R}$ with GEVP improvement is derived from ratio $R_d$, while the case
without GEVP improvement is derived from $R^{\mathrm{std}}$.

Both isoscalar ($u+d$) and isovector ($u-d$) cases will be presented.
In the case of the isoscalar, we include the disconnected  quark contributions. We also study the disconnected contributions to the isovector although they will go to zero in the continuum limit. For the axial and pseudoscalar matrix elements, these contributions can be sizeable and account for the large cut-off effects seen in these quantities, as explained in \cref{app:iso}.
We use $\vec{1}_k, k=x,y,z$ to represent one-unit of momentum in spatial direction $k$, \eg for $k=z$ it means $\vec{1}_z = \frac{2\pi}{L}(0,0,1)$.
We use $Q_1^2$ to represent the $Q^2$ value for momentum transfer of $\vec{1}_k$ in any of the three spatial directions.
Given the ensemble parameters in Tab. \ref{tab:ens}, this transfer corresponds to  $Q_1^2=0.285\,$GeV$^2$ for ensemble cA211.530.24 
and $Q_1^2=0.074\,$GeV$^2$ for cA2.09.48.

\begin{figure}[!ht]
    \centering
    \includegraphics[width=\columnwidth]{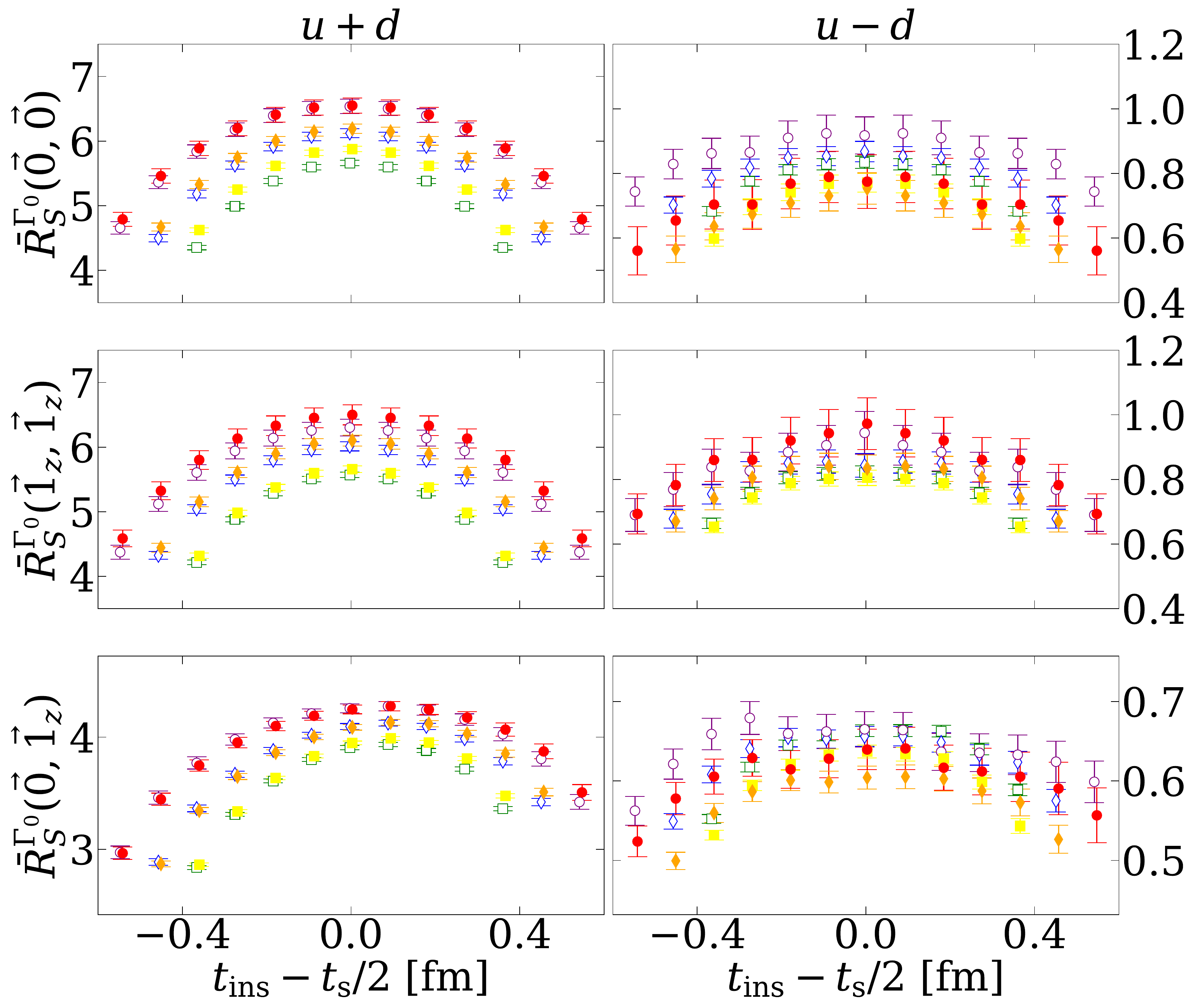}
    \caption{Ensemble cA211.530.24.
    We show results extracted from the ratios defined in \cref{eq:ratio_S}  the isoscalar (left) and isovector (right) cases.
    Both isoscalar and isovector cases include contributions from connected and disconnected topologies.
    From top to bottom, the results are  presented in the same order as the order of the  giving the ratios in \cref{eq:ratio_S}.
    Open and filled symbols refer to results without and with GEVP improvement, respectively.
    Squares, diamonds and circles denote results for $t_s=10a,12a$ and $14a$, respectively.
    }\label{fig:ratio_S_24}
\end{figure}
\begin{figure}[!ht]
    \centering
    \includegraphics[width=\columnwidth]{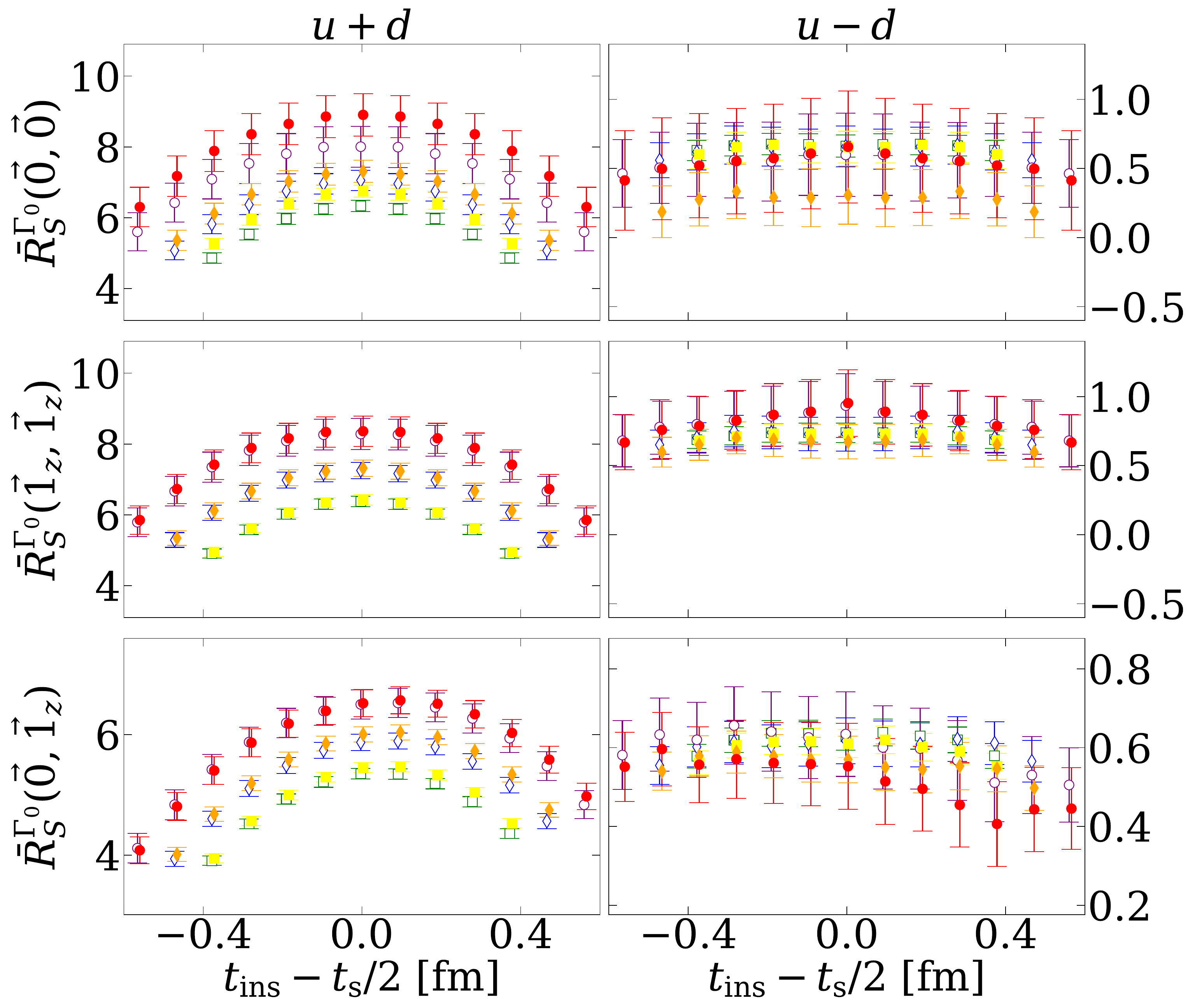}
    \caption{Ensemble cA2.09.48. The notation is the same as in \cref{fig:ratio_S_24}.}\label{fig:ratio_S_48}
\end{figure}
{\allowdisplaybreaks
To extract the matrix element of the scalar current, we construct the following ratios (we suppress the time arguments hereafter):
\begin{alignat}{2}\label{eq:ratio_S}
    \bar{R}^{\Gamma_0}_{S}(\vec{0},\vec{0})&=R^{\Gamma_0}_{S}(\vec{0},\vec{0}) \to g_S \,, \nonumber\\
    \bar{R}^{\Gamma_0}_{S}(\vec{1}_z,\vec{1}_z)&=\left[\frac{m_N}{E_N}\right]^{-1}R^{\Gamma_0}_{S}(\vec{1}_z,\vec{1}_z) \to g_S \,, \nonumber\\
    \bar{R}^{\Gamma_0}_S(\vec{0},\vec{1}_z)&=\mathcal{C}\left[E_N+m_N\right]^{-1} R^{\Gamma_0}_{S}(\vec{0},\vec{1}_z) \to G_S(Q_1^2) \,,
\end{alignat}
where the kinematic factor $\mathcal{C}$ is given by $\mathcal{C}=\sqrt{2E_N(E_N+m_N)}$.
The corresponding results are presented in \cref{fig:ratio_S_24,fig:ratio_S_48}.
}

\begin{figure}[!ht]
    \centering
    \includegraphics[width=\columnwidth]{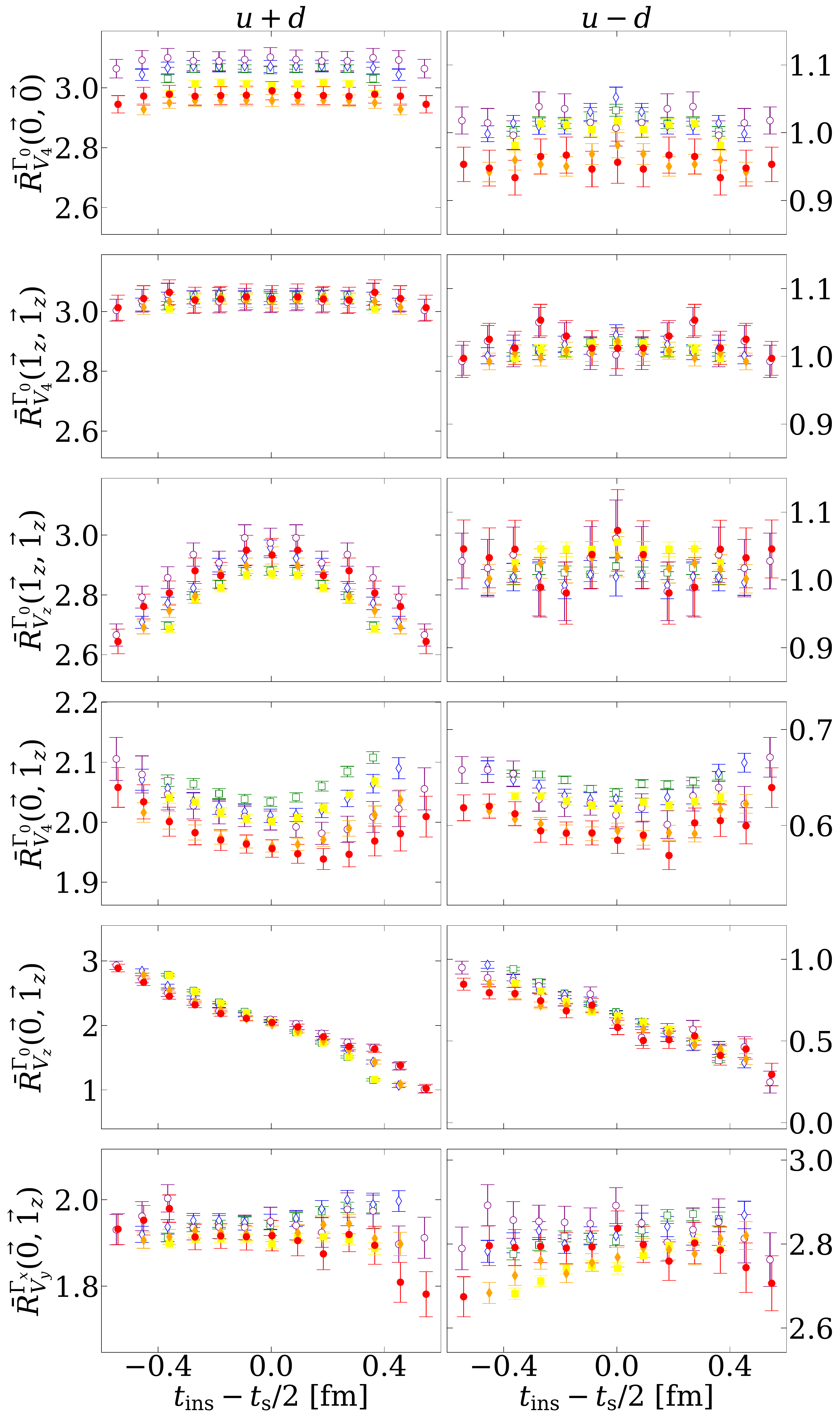}
    \caption{Ensemble  cA211.530.24. As in \cref{fig:ratio_S_24} for \cref{eq:ratio_V}.}\label{fig:ratio_V_24}
\end{figure}
\begin{figure}[!ht]
    \centering
    \includegraphics[width=\columnwidth]{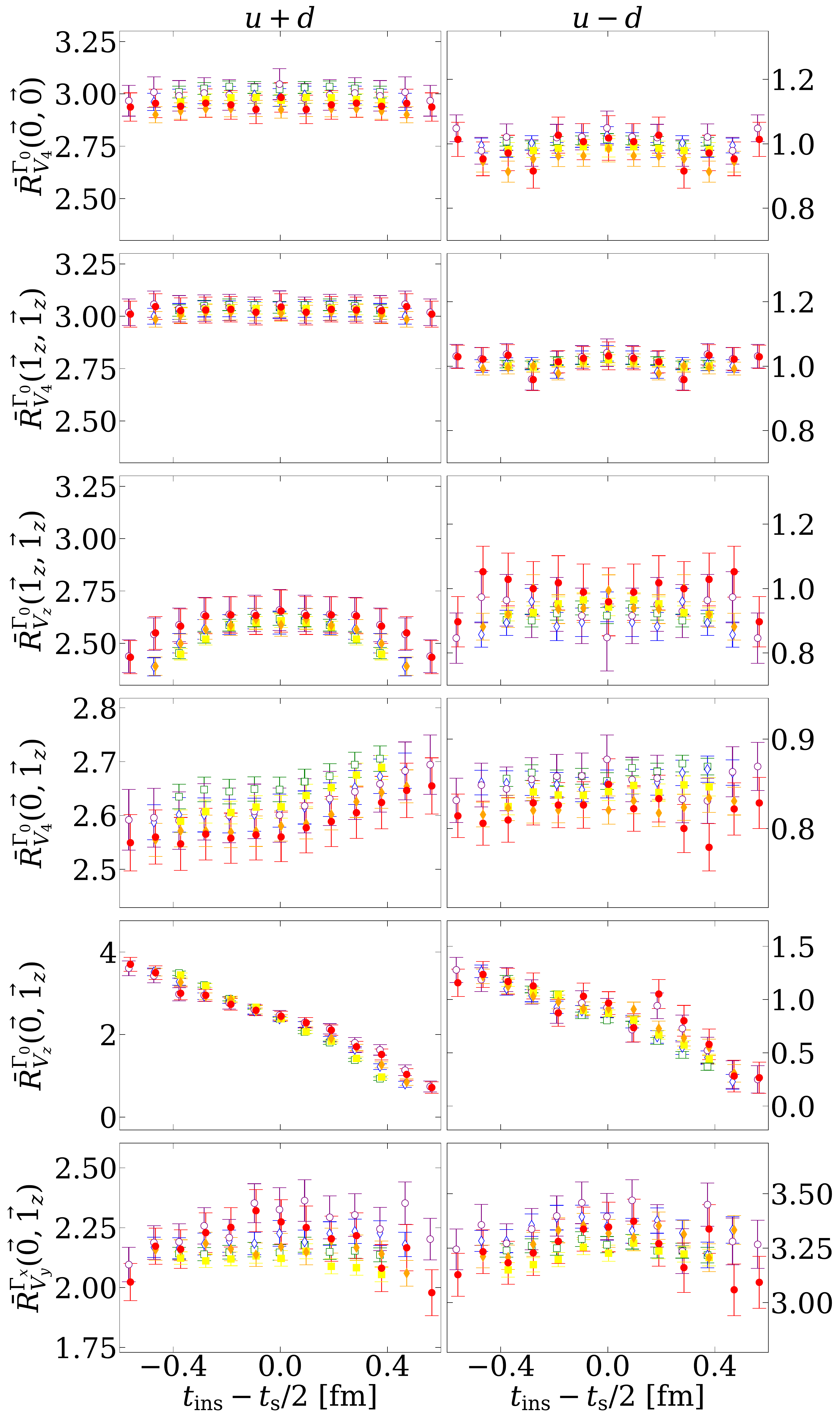}
    \caption{Ensemble  cA2.09.48. As in \cref{fig:ratio_S_24} for \cref{eq:ratio_V}.}\label{fig:ratio_V_48}
\end{figure}
{\allowdisplaybreaks
For extracting the matrix elements of the vector current and the electric and magnetic Sachs form factors $G_E(q^2)$ and $G_M(Q^2)$, respectively, we construct the following ratios:
\begin{alignat}{2}\label{eq:ratio_V}
    \bar{R}^{\Gamma_0}_{V_4}(\vec{0},\vec{0})&=R^{\Gamma_0}_{V_4}(\vec{0},\vec{0}) \to N_q \,,\nonumber\\
    \bar{R}^{\Gamma_0}_{V_4}(\vec{1}_z,\vec{1}_z)&=R^{\Gamma_0}_{V_4}(\vec{1}_z,\vec{1}_z) \to N_q \,,\nonumber\\
    \bar{R}^{\Gamma_0}_{V_z}(\vec{1}_z,\vec{1}_z)&=\left[-i\,\frac{p_1}{E_N}\right]^{-1}R^{\Gamma_0}_{V_z}(\vec{1}_z,\vec{1}_z) \to N_q \,, \nonumber\\
    \bar{R}^{\Gamma_0}_{V_4}(\vec{0},\vec{1}_z)&=\mathcal{C}\left[E_N+m_N\right]^{-1}{R}^{\Gamma_0}_{V_4}(\vec{0},\vec{1}_z) \to G_E(Q_1^2) \,,\nonumber\\
    \bar{R}^{\Gamma_0}_{V_z}(\vec{0},\vec{1}_z)&=\mathcal{C}\left[-i\,p_1\right]^{-1}{R}^{\Gamma_0}_{V_4}(\vec{0},\vec{1}_z) \to G_E(Q_1^2) \,,\nonumber\\
    \bar{R}^{\Gamma_x}_{V_y}(\vec{0},\vec{1}_z)&=\mathcal{C}\left[p_1\right]^{-1}{R}^{\Gamma_x}_{V_y}(\vec{0},\vec{1}_z) \to G_M(Q_1^2) \,,
\end{alignat}
where $N_q$ counts the number of quarks in the proton, i.e.  $N_q=N_u+N_d=3$ for the isoscalar case, and $N_q=N_u-N_d=1$ for the isovector case.  $E_N(p)=\sqrt{m_N^2+|\vec{p}^2}$ is the nucleon energy.
The results are presented in \cref{fig:ratio_V_24,fig:ratio_V_48}.
}

\begin{figure}[!ht]
    \centering
    \includegraphics[width=\columnwidth]{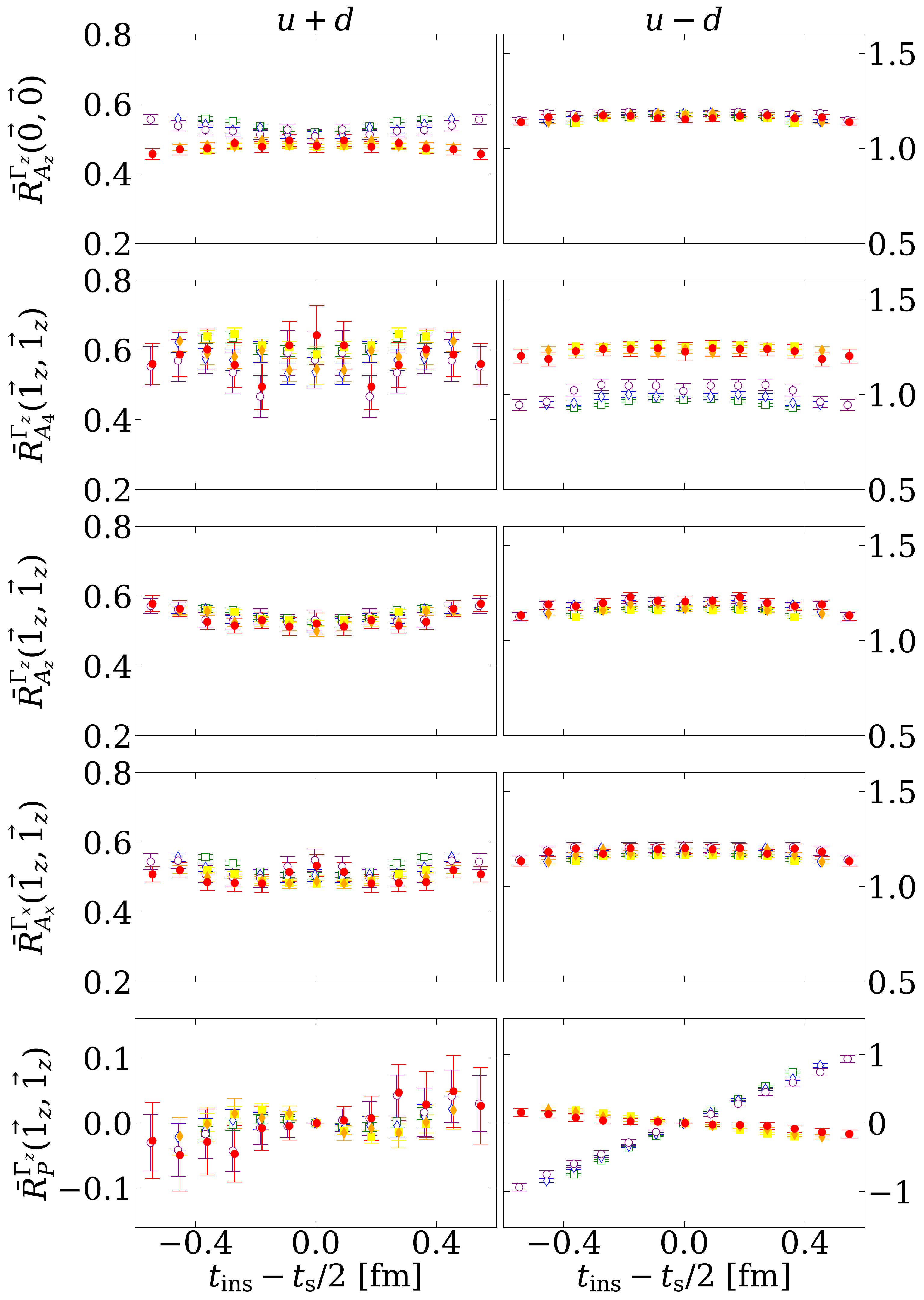}
    \caption{Ensemble  cA211.530.24. As in \cref{fig:ratio_S_24} for \cref{eq:ratio_AP0}.}\label{fig:ratio_AP0_24}
\end{figure}
\begin{figure}[!ht]
    \centering
    \includegraphics[width=\columnwidth]{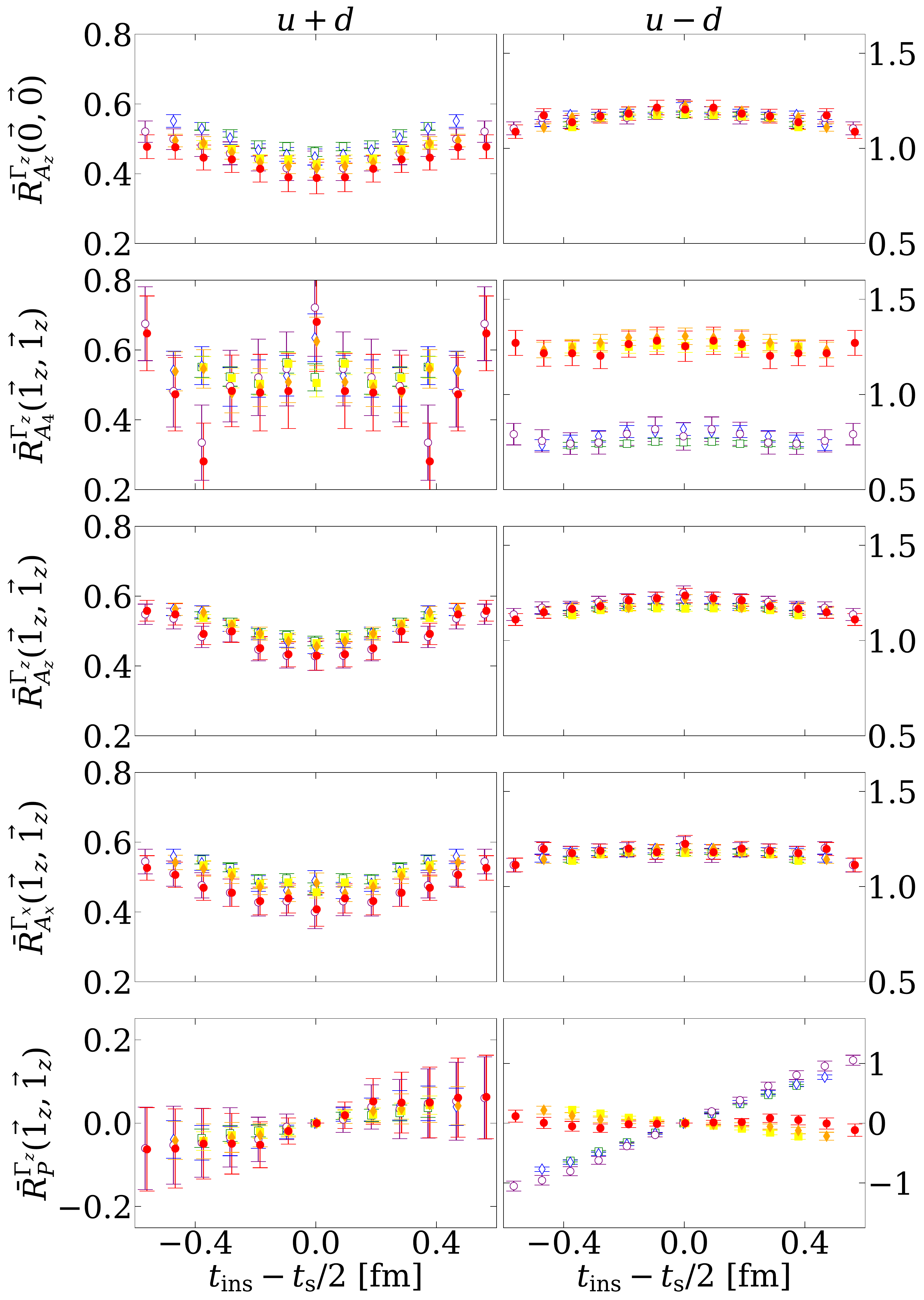}
    \caption{Ensemble  cA2.09.48. As in \cref{fig:ratio_S_24} for \cref{eq:ratio_AP0}.}\label{fig:ratio_AP0_48}
\end{figure}
{\allowdisplaybreaks
For the special case of studying the  pseudoscalar and axial form factors at $Q^2=0$, we consider the following ratios:
\begin{alignat}{2}\label{eq:ratio_AP0}
    \bar{R}^{\Gamma_z}_{A_z}(\vec{0},\vec{0}) &= -i\,R^{\Gamma_z}_{A_z}(\vec{0},\vec{0}) \to g_A \,, \nonumber\\
    \bar{R}^{\Gamma_z}_{A_4}(\vec{1}_z,\vec{1}_z) &= \left[-\frac{p_1}{E_N}\right]^{-1} R^{\Gamma_k}_{A_4}(\vec{1}_z,\vec{1}_z) \to g_A \,,\nonumber\\
    \bar{R}^{\Gamma_z}_{A_z}(\vec{1}_z,\vec{1}_z) &= \left[i\left( \frac{m_N}{E_N}+\frac{p_1^2}{E_N(E_N+m_N)} \right)\right]^{-1}  \nonumber\\
    &\hspace{2cm} \times R^{\Gamma_z}_{A_z}(\vec{1}_z,\vec{1}_z)\to g_A \,,\nonumber\\
    \bar{R}^{\Gamma_{x}}_{A_{x}}(\vec{1}_z,\vec{1}_z) &= \left[i\frac{m_N}{E_N}\right]^{-1} R^{\Gamma_{x}}_{A_{x}}(\vec{1}_z,\vec{1}_z) \to g_A \,,\nonumber\\
    \bar{R}^{\Gamma_z}_{P}(\vec{1}_z,\vec{1}_z) &= R^{\Gamma_z}_{P}(\vec{1}_z,\vec{1}_z) \to 0 \,.
\end{alignat}
The results are presented in \cref{fig:ratio_AP0_24,fig:ratio_AP0_48}.
}

\begin{figure}[!ht]
    \centering
    \includegraphics[width=\columnwidth]{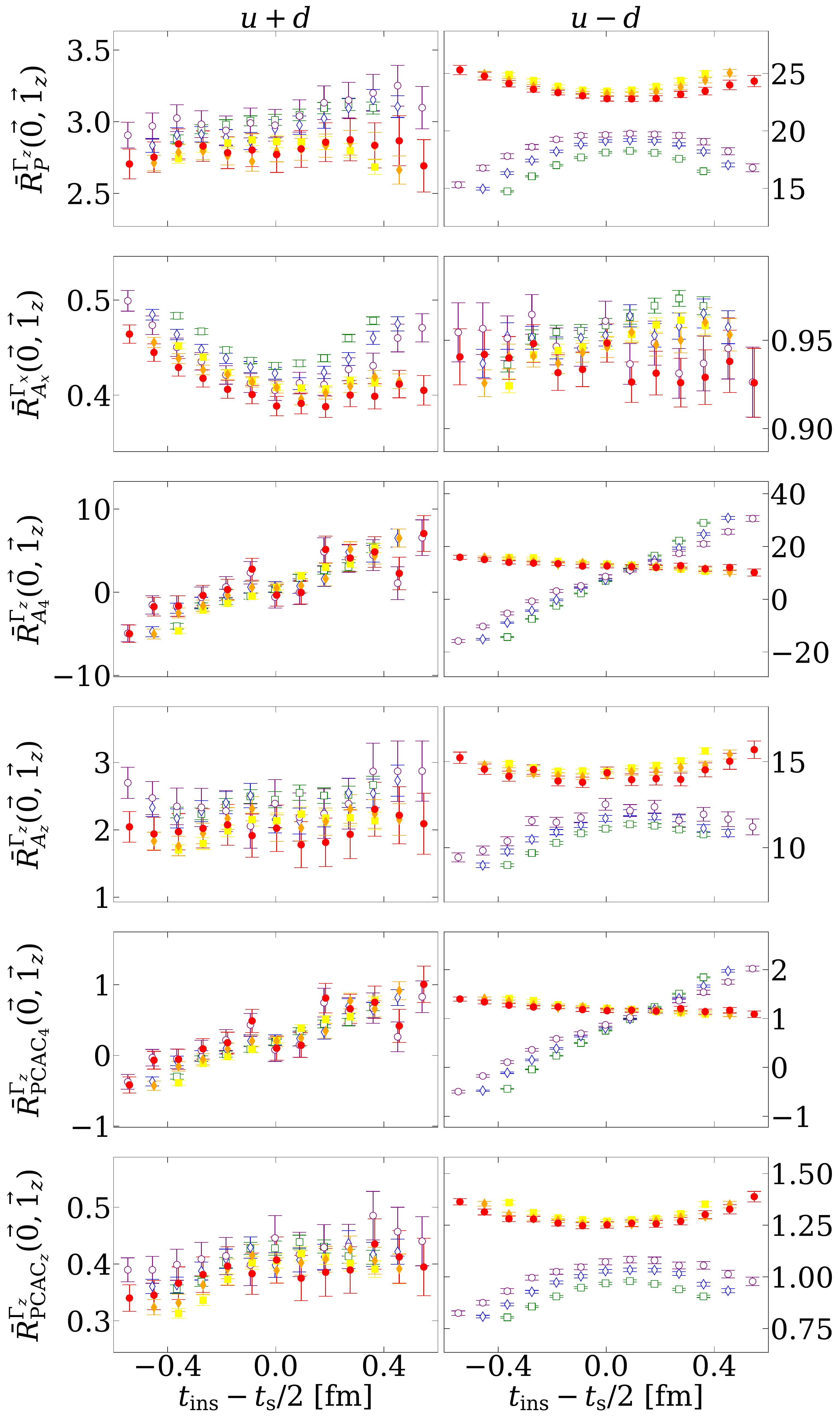}
    \caption{Ensemble  cA211.530.24. As in \cref{fig:ratio_S_24} for \cref{eq:ratio_AP1}.}\label{fig:ratio_AP1_24}
\end{figure}
\begin{figure}[!ht]
    \centering
    \includegraphics[width=\columnwidth]{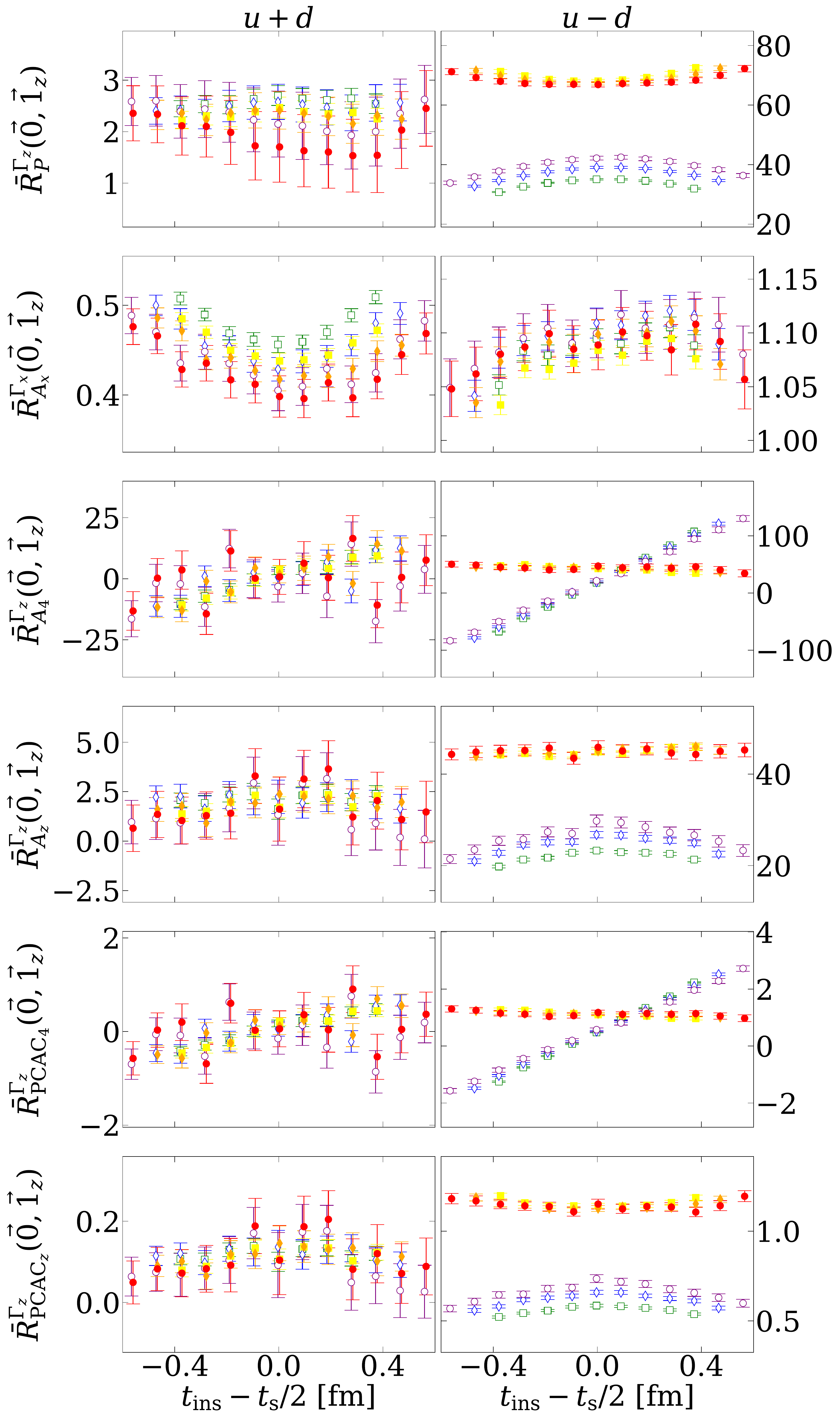}
    \caption{Ensemble  cA2.09.48. As in \cref{fig:ratio_S_24} for \cref{eq:ratio_AP1}.}\label{fig:ratio_AP1_48}
\end{figure}
{\allowdisplaybreaks
Similarly, for the matrix element with $Q^2=Q_1^2$ of the isovector pseudoscalar and axial currents  we consider the following ratios:
\begin{align}\label{eq:ratio_AP1}
    \bar{R}^{\Gamma_z}_{P}(\vec{0},\vec{1}_z) &= \mathcal{C}\left[-p_1\right]^{-1} R^{\Gamma_z}_{P}(\vec{0},\vec{1}_z) \hspace{0.5cm}\to  G_5(Q_1^2) \,, \nonumber\\
    \bar{R}^{\Gamma_{x}}_{A_{x}}(\vec{0},\vec{1}_z) &= \mathcal{C}\left[i(E_N+m_N)\right]^{-1} R^{\Gamma_{x}}_{A_{x}}(\vec{0},\vec{1}_z) \nonumber\\ 
    &\hspace{4cm}\to G_A(Q_1^2) \,, \nonumber\\ 
    \bar{R}^{\Gamma_{z}}_{A_{4}}(\vec{0},\vec{1}_z) &= \mathcal{C}\left[\frac{p_1(E_N-m_N)}{2m_N}\right]^{-1} {R}^{\Gamma_{z}}_{A_{4}}(\vec{0},\vec{1}_z) \nonumber\\
    &\hspace{0cm}-\left[ \frac{-2m_N}{E_N-m_N}\right] \bar{R}^{\Gamma_{x}}_{A_{x}}(\vec{0},\vec{1}_z)   \to G_P(Q_1^2) \,,\nonumber\\
    \bar{R}^{\Gamma_{z}}_{A_{z}}(\vec{0},\vec{1}_z) &=\mathcal{C}\left[-i\frac{p_1^2}{2m_N}\right]^{-1}{R}^{\Gamma_{z}}_{A_{z}}(\vec{0},\vec{1}_z)  \nonumber\\
     &\hspace{-1.5cm} -\left[\frac{-2m_N(E_N+m_N)}{p_1^2}\right]\bar{R}^{\Gamma_{x}}_{A_{x}}(\vec{0},\vec{1}_z) \to G_P(Q_1^2) \,,\nonumber\\
    \bar{R}^{\Gamma_{z}}_{\mathrm{PCAC}_4}(\vec{0},\vec{1}_z) &= \frac{\frac{m_q}{m_N}\bar{R}^{\Gamma_z}_{P}(\vec{0},\vec{1}_z) + \frac{Q^2}{4m_N^2} \bar{R}^{\Gamma_{z}}_{A_{4}}(\vec{0},\vec{1}_z)}{\bar{R}^{\Gamma_{x}}_{A_{x}}(\vec{0},\vec{1}_z)} \nonumber\\
    &\hspace{3.cm} \to r_{\mathrm{PCAC}}(Q_1^2)  \,,\nonumber\\
    \bar{R}^{\Gamma_{z}}_{\mathrm{PCAC}_z}(\vec{0},\vec{1}_z) &= \frac{\frac{m_q}{m_N}\bar{R}^{\Gamma_z}_{P}(\vec{0},\vec{1}_z) + \frac{Q^2}{4m_N^2}\bar{R}^{\Gamma_{z}}_{A_{z}}(\vec{0},\vec{1}_z)}{\bar{R}^{\Gamma_{x}}_{A_{x}}(\vec{0},\vec{1}_z)}  \nonumber\\
    &\hspace{3cm} \to r_{\mathrm{PCAC}}(Q_1^2) \,,
\end{align}
where $r_{\mathrm{PCAC}}$ is the PCAC ratio defined as
\begin{align}
    r_{\mathrm{PCAC}}(Q^2) = \frac{\frac{m_q}{mN}G_5(Q^2)+\frac{Q^2}{4m_N^2}G_P(Q^2)}{G_A(Q^2)} \,.
\end{align}
The ratio $r_{\mathrm{PCAC}}(Q^2)$   for the isovector case must be unity, as a consequence of the PCAC relation. For the isoscalar case, it is not $1$ because of the chiral anomaly \cite{Bell:1969ts,Alexandrou:2021wzv}.
The results are presented in \cref{fig:ratio_AP1_24,fig:ratio_AP1_48}.
}

\begin{figure}[!ht]
    \centering
    \includegraphics[width=\columnwidth]{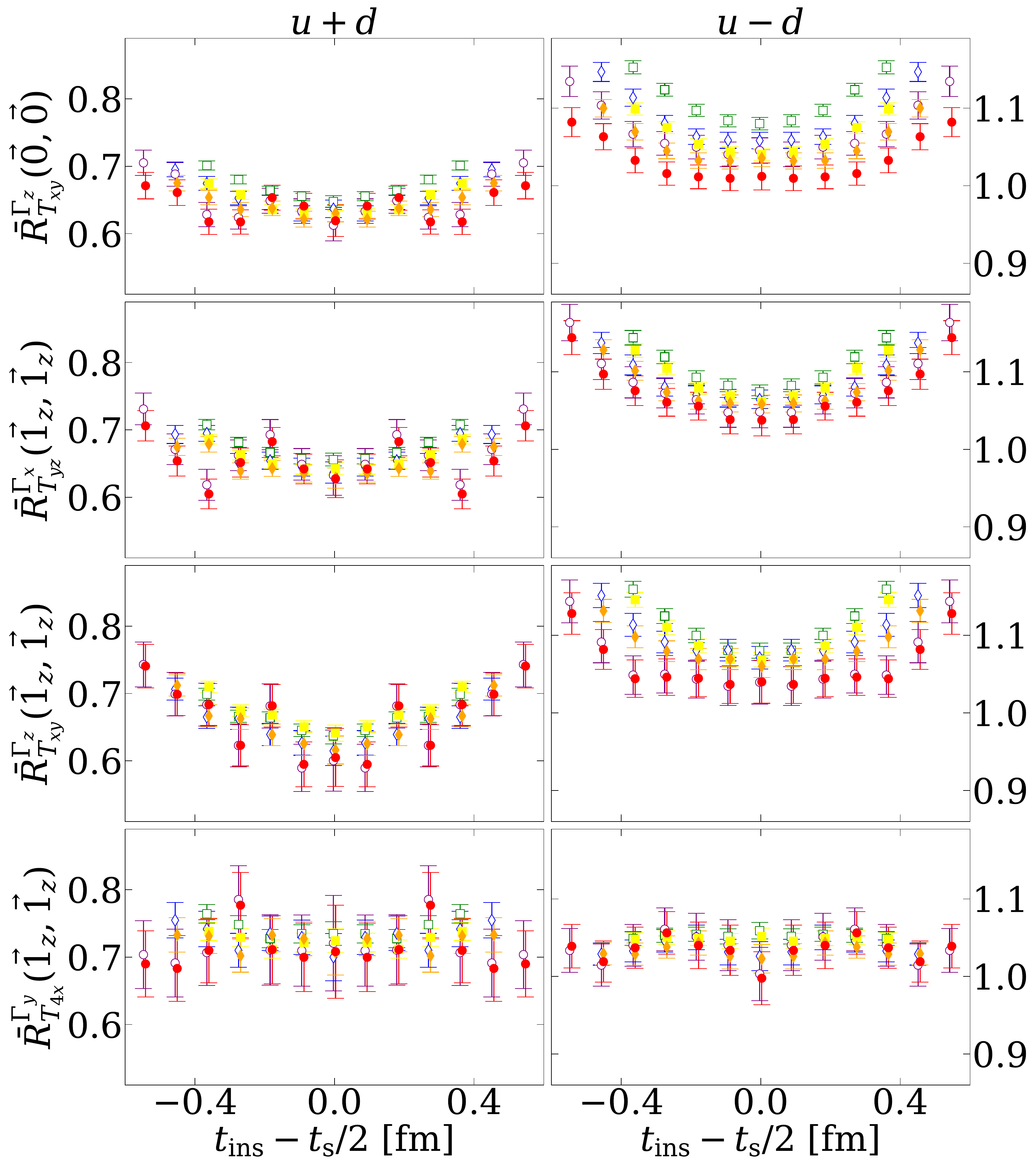}
    \caption{Ensemble  cA211.530.24. As in \cref{fig:ratio_S_24} for \cref{eq:ratio_T0}.}\label{fig:ratio_T0_24}
\end{figure}
\begin{figure}[!ht]
    \centering
    \includegraphics[width=\columnwidth]{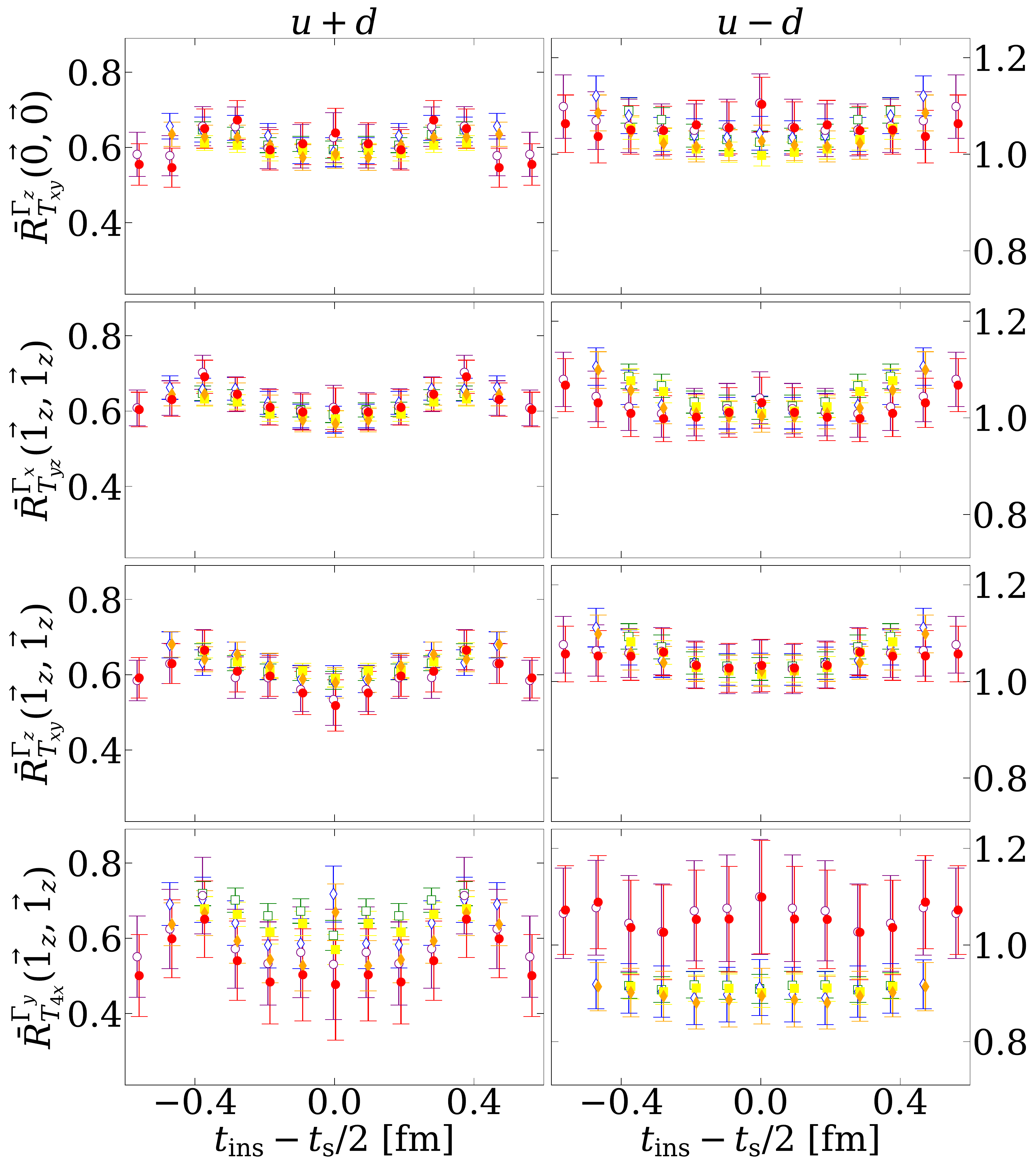}
    \caption{Ensemble  cA2.09.48. As in \cref{fig:ratio_S_24} for \cref{eq:ratio_T0}.}\label{fig:ratio_T0_48}
\end{figure}
{\allowdisplaybreaks
For the case of the tensor current for $Q^2=0$, we have the following ratios:
\begin{alignat}{2}\label{eq:ratio_T0}
    \bar{R}^{\Gamma_z}_{T_{xy}}(\vec{0},\vec{0})&=-i\,R^{\Gamma_z}_{T_{xy}}(\vec{0},\vec{0}) \to g_T \,, \nonumber\\
    \bar{R}^{\Gamma_x}_{T_{yz}}(\vec{1}_z,\vec{1}_z)&=-i\,R^{\Gamma_x}_{T_{yz}}(\vec{1}_z,\vec{1}_z) \to g_T \,, \nonumber\\
    \bar{R}^{\Gamma_z}_{T_{xy}}(\vec{1}_z,\vec{1}_z)&=-i\left[1-\frac{p_1^2}{E_N(E_N+m_N)}\right]^{-1}  \nonumber\\ 
    &\hspace{2cm}\times R^{\Gamma_z}_{T_{xy}}(\vec{1}_z,\vec{1}_z) \to g_T \,, \nonumber\\
    \bar{R}^{\Gamma_y}_{T_{4x}}(\vec{1}_z,\vec{1}_z)&=\left[\frac{-p_1}{E_N}\right]^{-1}R^{\Gamma_y}_{T_{4x}}(\vec{1}_z,\vec{1}_z) \to g_T \,.
\end{alignat}
The results are presented in \cref{fig:ratio_T0_24,fig:ratio_T0_48}.
}

\begin{figure}[!ht]
    \centering
    \includegraphics[width=\columnwidth]{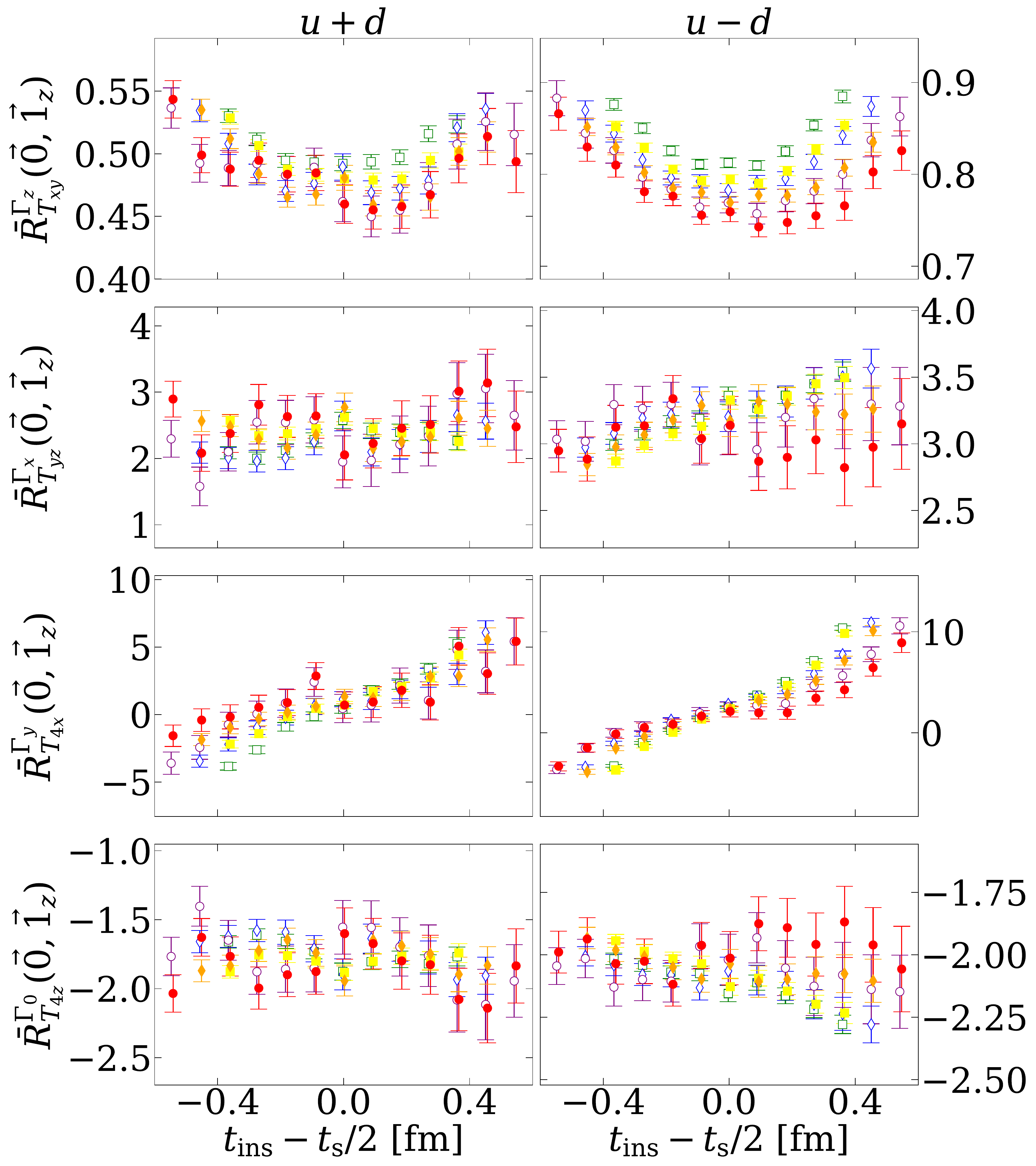}
    \caption{Ensemble  cA211.530.24. As in \cref{fig:ratio_S_24} for \cref{eq:ratio_T1}.}\label{fig:ratio_T1_24}
\end{figure}
\begin{figure}[!ht]
    \centering
    \includegraphics[width=\columnwidth]{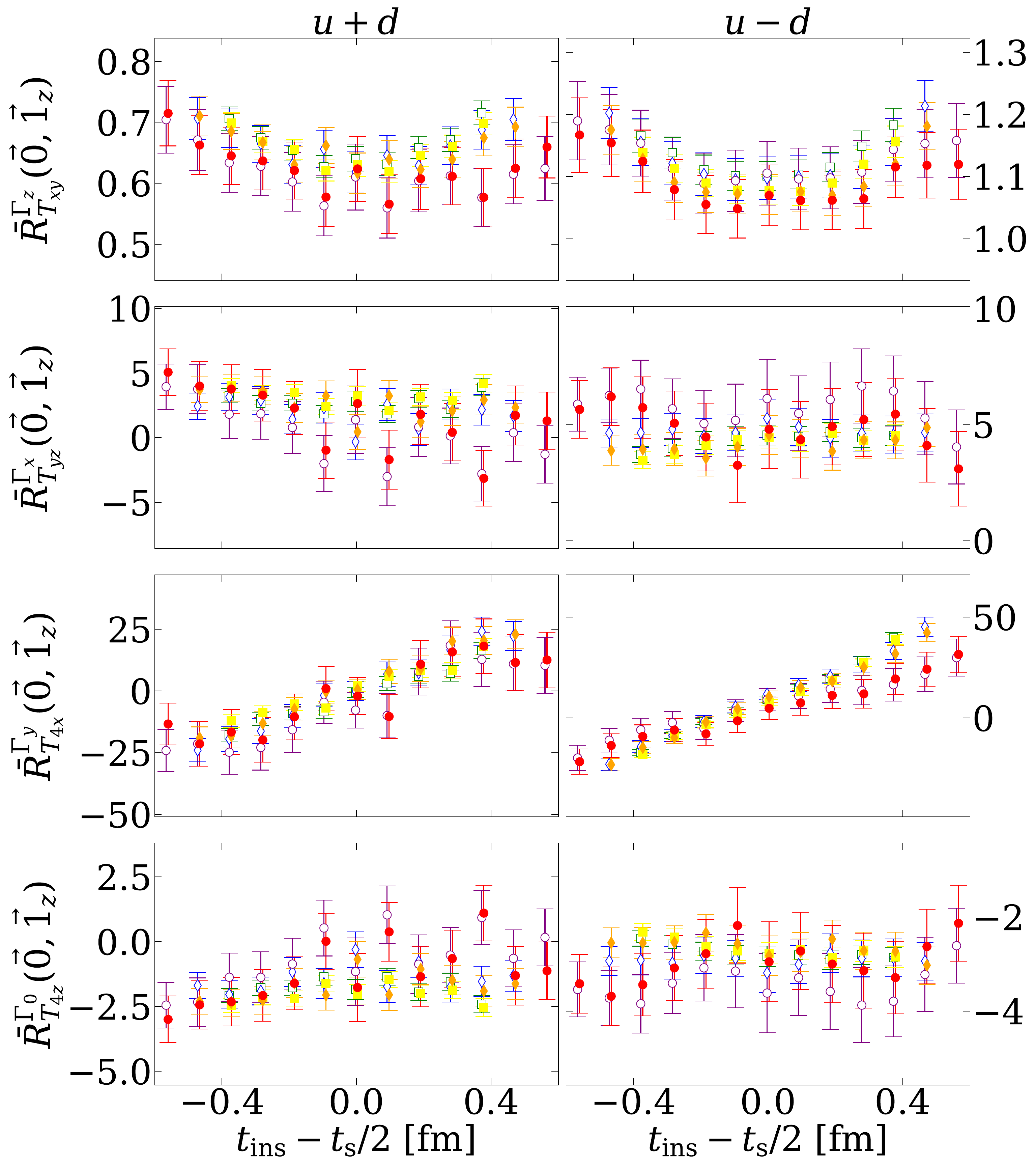}
    \caption{Ensemble  cA2.09.48. As in \cref{fig:ratio_S_24} for \cref{eq:ratio_T1}.}\label{fig:ratio_T1_48}
\end{figure}
{\allowdisplaybreaks
For the case of the tensor current for $Q^2=Q_1^2$, we have the following expressions:
\begin{alignat}{2}\label{eq:ratio_T1}
    \bar{R}^{\Gamma_z}_{T_{xy}}(\vec{0},\vec{1}_z)&=\mathcal{C}\left[2im_N(E_N+m_N)\right]^{-1}R^{\Gamma_z}_{T_{xy}}(\vec{0},\vec{1}_z)  \nonumber\\
    &  \hspace{4cm}\to A_{10}(Q_1^2) \,, \nonumber\\
    \bar{R}^{x}_{T_{yz}}(\vec{0},\vec{1}_z)&=\mathcal{C}\left[-i\,p_1^2\right]^{-1}R^{x}_{T_{yz}} \nonumber\\
    &\hspace{-1cm}- \left[\frac{2m_N(E_N+m_N)}{-p_1^2}\right] \bar{R}^{\Gamma_z}_{T_{xy}}(\vec{0},\vec{1}_z) \hspace{0cm}\to B_{10}(Q_1^2) \,,\nonumber\\
    \bar{R}^{y}_{T_{4x}}(\vec{0},\vec{1}_z)&=\mathcal{C}\left[p_1(E_N-m_N)\right]^{-1}R^{y}_{T_{4x}} \nonumber\\
    &\hspace{-1cm}-\left[\frac{-2m_N}{E_N-m_N}\right]\bar{R}^{\Gamma_z}_{T_{xy}}(\vec{0},\vec{1}_z) \hspace{0cm}\to B_{10}(Q_1^2) \,, \nonumber\\
    \bar{R}^{\Gamma_0}_{T_{4z}}(\vec{0},\vec{1}_z) &= \mathcal{C}\left[-2ip_1(E_N+m_N)\right]^{-1}R^{\Gamma_0}_{T_{4z}}(\vec{0},\vec{1}_z) \nonumber\\
    &\hspace{-2cm} - \left[\frac{m_N}{E_N+m_N} \right]\left(\bar{R}^{\Gamma_z}_{T_{xy}}(\vec{0},\vec{1}_z) +\bar{R}^{x}_{T_{yz}}(\vec{0},\vec{1}_z) \right) \hspace{0cm}\to \tilde{A}_{10}(Q_1^2) \,.
\end{alignat}
The results are presented in \cref{fig:ratio_T1_24,fig:ratio_T1_48}.
}

For all currents considered, we distinguish among three categories as far as the impact of GEVP  on the matrix elements  is concerned.
The first category includes ratios that do not change significantly after applying GEVP and it is observed  for the majority of currents considered including those that show large contamination from excited states and those which do not. 
In fact, for the ratios  that show significant contamination before using GEVP, we do not observe a flip of sign of the contamination after GEVP improvement, as suggested in \cref{eq:Id-2}.
Our conclusion is, therefore, that for these case the contamination observed is  unlikely to come from the lowest $N\pi$ states under consideration here. Furthermore,  including of the $\pi N$-$\pi N$ 3-point function in the construction of  optimized 3-point functions is unlikely to change this conclusion.
We note that this conclusion does not rely on the hierarchical condition \cref{eq:hier}, since it is based on the sign-flipping in \cref{eq:Id-2}. 

The second category includes a few cases without sharing obvious common characteristics.
For cases in this category, we observe mild or significant changes after using GEVP for the $m_\pi=346\,$MeV ensemble, but not for the physical point ensemble.
We note, there is no clear boundary between the first and second categories because of the ambiguity of significant changes.
This category can include, \eg, $\bar{R}^{\Gamma_0}_{S^{u-d}}(\vec{0},\vec{0})$ ([row 1,col 2] in \cref{fig:ratio_S_24}), $\bar{R}^{\Gamma_0}_{V_4^{u\pm d}}(\vec{0},\vec{0})$ ([row 1, col 1 and 2] in \cref{fig:ratio_V_24}), and $\bar{R}^{\Gamma_z}_{A_z^{u+d}}(\vec{0},\vec{0})$ ([row 1, col 1] in \cref{fig:ratio_AP0_24}).

The last category includes the isovector pseudoscalar and axial 3-point functions, where we observe significant effects of GEVP for both ensembles, as also observed in Ref.~\cite{Barca:2022uhi}.
In this category, the  common characteristic is that  the current has the same quantum number with the pion operator either at the sink or source.
In fact, for the two ensembles used in this work, we find that  
\begin{align}
\NJNpi \approx \NN \braket{\OO {\Jpi}^\dagger}
\end{align}
is well satisfied within the time slices under investigation and our normalization convention.
This  approximate equality can be explained from chiral perturbation theory by the fact that the right hand side is the leading order contribution \cite{Bar:2018xyi,Bar:2019gfx,Bar:2019igf} and it agrees with the findings 
of Ref.~\cite{Barca:2022uhi}.
However, this approximate quality is invalid for large times because the left- and right-hand sides have different asymptotic time-dependence: the left hand side  depends on the energy of a single nucleon and the right hand side  on the $N\pi$ non-interacting energy. 

\subsection{Analysis of the time-dependence of ratios for selective cases}
\begin{figure}[!ht]
    \centering
    \includegraphics[width=\columnwidth]{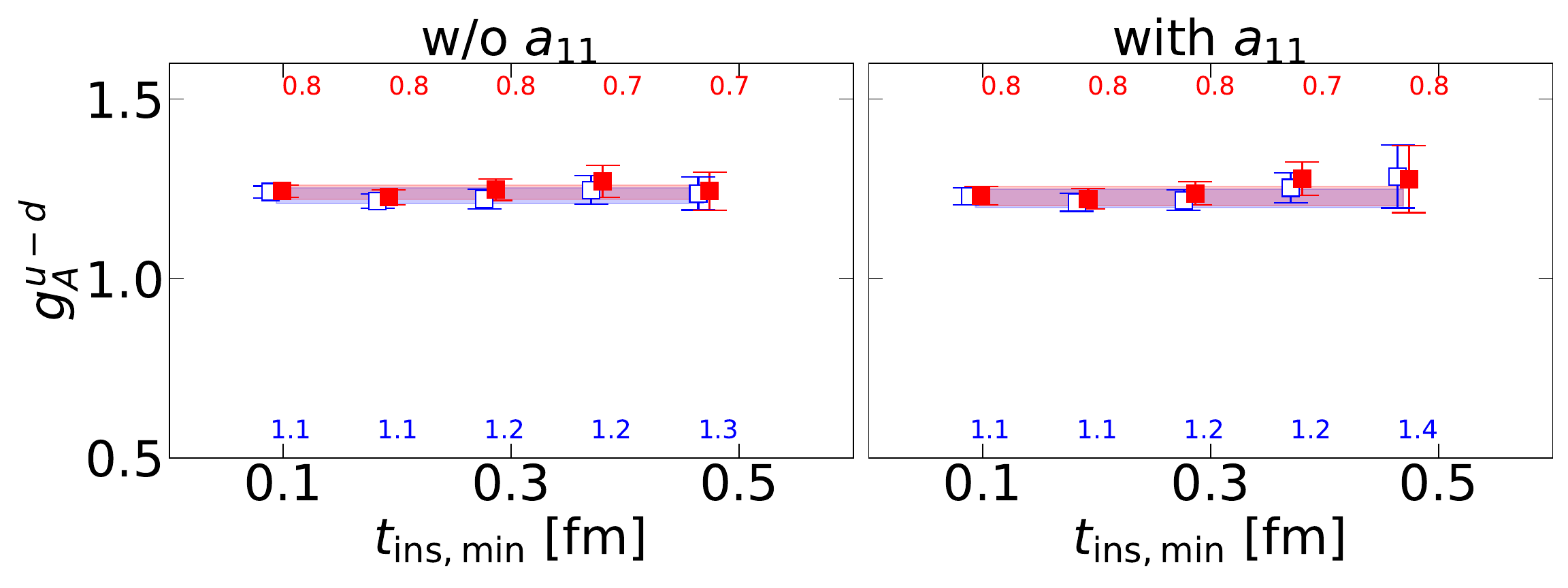}
    \caption{Ensemble  cA2.09.48. Values extracted from two-state fits to 2-point and 3-point functions used to define the ratio $\bar{R}^{\Gamma_z}_{A_z^{u-d}}(\vec{0},\vec{0})\to g^{u-d}_A$ in \cref{eq:ratio_AP0} without (left) and with (right) the $a_{11}$ term.
    }\label{fig:3pt_raw_sep_48_a11}
\end{figure}

In this section, we do further analysis for selected cases relevant for the nucleon sigma term, the isovector axial charge, and the isovector pseudoscalar, axial, induced pseudoscalar form factors. All cases in the third category will be covered.
We perform two-state fits to the 2-point and 3-point functions used to construct the ratios simultaneously with shared energies of the ground and first excited state.
The 2-point $\Cdpt$ and 3-point  $\Ctpt$ functions are parameterized as follows (we suppress all arguments other than time slices and momenta):
\begin{align}
    \Cdpt(\vp; t)&=c_0\,e^{-E_N(\vp)\,t}+c_1\,e^{-E_{1}(\vp)\,t} \,,\nonumber\\
    \Ctpt(\vpp,\vp;\tf,\tc)&=a_{00}\,e^{-E_{N}(\vpp)(\tf-\tc)}e^{-E_{N}(\vp)\tc}\nonumber\\
    &+a_{01}\,e^{-E_{N}(\vpp)(\tf-\tc)}e^{-E_{1}(\vp)\tc} \nonumber\\
    &+ a_{10}\,e^{-E_{1}(\vpp)(\tf-\tc)}e^{-E_{N}(\vp)\tc} \,.
\end{align}
Here we do not include the $a_{11}$ term, because it is more suppressed in terms of the exponential factors. Such a term does not impact the central values but  makes the errors large as illustrated in \cref{fig:3pt_raw_sep_48_a11} for the isovector axial charge.
We do not suggest to use such a term even if it could improve the errors. 
Because in terms of the exponential factors, its contribution is more suppressed than that of higher excited states. Including them may misaddress the effects from higher excited states, and mislead one with a statistically precise but overfitted results.

When $Q^2=0$, we have the same momentum at source and sink $\vpp=\vp$.
In addition, we have $E_{N}(\vpp)=E_{N}(\vp)$, $E_{1}(\vpp)=E_{1}(\vp)$, and $a_{01}=\pm a_{10}$.
So the independent parameters are $c_0,E_N(\vp),c_1,E_1(\vp),a_{00},a_{01}$, which are six in total.
We  fit to results obtained for $\tc\leq \tf/2$ only, because we make use of the symmetry to symmetrized or anti-symmetrized the 3-point functions.
For $Q^2\neq0$, we use two 2-point functions with the corresponding momentum values.
The independent parameters are $c_{0}^\prime,E_{N}(\vpp),c_{1}^\prime,E_{1}(\vpp)$ for the sink 2-point function, $c_{0},E_{N}(\vp),c_{1},E_{1}(\vp)$ for the source 2-point function, and additionally $a_{00},a_{01},a_{10}$ for the 3-point function, which are eleven in total.

 For the 2-point functions, we include points with $t\geq t_{\mathrm{min}}$ with $t_{\mathrm{min}}$ taken from the most probable two-state fits (indicated with vertical lines) in  \cref{fig:2ptFit_24,fig:2ptFit_48}.
We fit the  3-point functions in the internal  $t_{\mathrm{ins,min}}\leq\tc\leq \tf-t_{\mathrm{ins,min}}$ and  vary  $t_{\mathrm{ins,min}}$  from $a$ to $5a$.
We perform the model average of all these fits using AIC.

In addition, in Ref.~\cite{Alexandrou:2023qbg}, the authors analysed twisted mass fermion ensembles at the physical point with three different lattice spacings, namely  $a=0.07957(13)$ fm, $a=0.06821(13)$ fm, and $a=0.05692(12)$ fm. 
They studied the isovector pseudoscalar, axial, and induced pseudoscalar form factors at all three lattice spacings and did the continuum extrapolations.
We will take the continuum extrapolated results from Ref.~\cite{Alexandrou:2023qbg} to compare with our results from the physical point ensemble used here, which has  a larger lattice spacing.
%
Such a comparison is only meaningful if cut-off effects are small. 
As we will show, including the  quark disconnected contributions will help reduce the cut-off effects.
In Ref.~\cite{Alexandrou:2023qbg}, it was found that the pseudoscalar and induced pseudoscalar form factors suffer from large discretization artefacts.
In this work, including the  quark disconnected  contributions from, we observe that these cut-off effects are significantly reduced. 

\begin{figure}[!ht]
    \centering
    \includegraphics[width=\columnwidth]{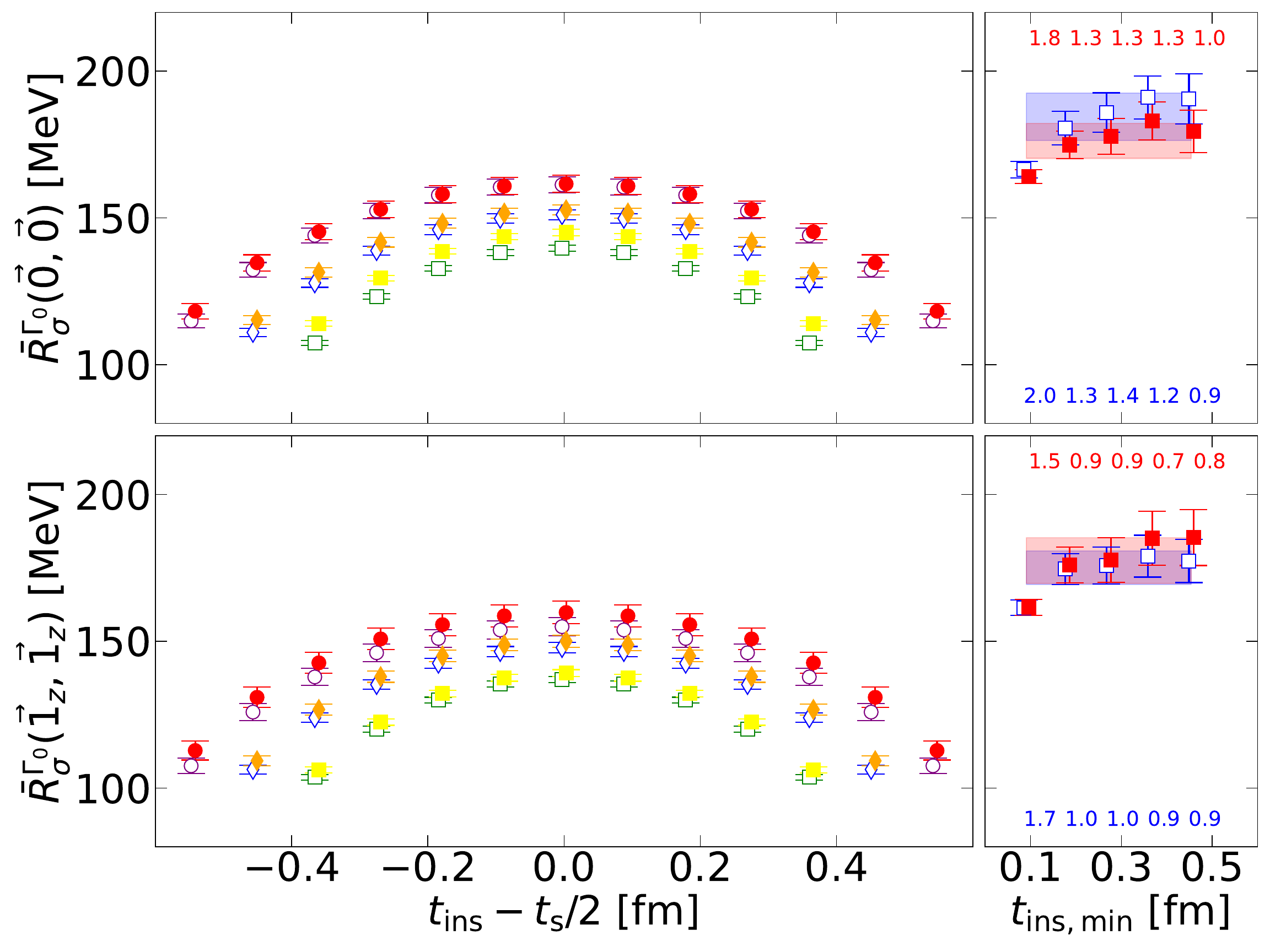}
    \caption{Ensemble  cA211.530.24. 
    The panels in the first column show results 
    for the ratio that yields the nucleon $\sigma_{\pi N}$, as defined in  \cref{eq:rbar_S}.
    Each panel in the second column shows results using the two-state fit to the 2-point and 3-point functions used to construct the ratio which gives the data shown in the left panel.
    Open and filled points refer to results without and with GEVP improvement, respectively.
    The blue (without GEVP) and red (with GEVP) bands are the results from the model average of the two-state fit.
    Below and above each point in the right panel we quote the  reduced $\chi^2$ without and with GEVP improvement, respectively.
    }\label{fig:3pt_raw_sep_S_24}
\end{figure}
\begin{figure}[!ht]
    \centering
    \includegraphics[width=\columnwidth]{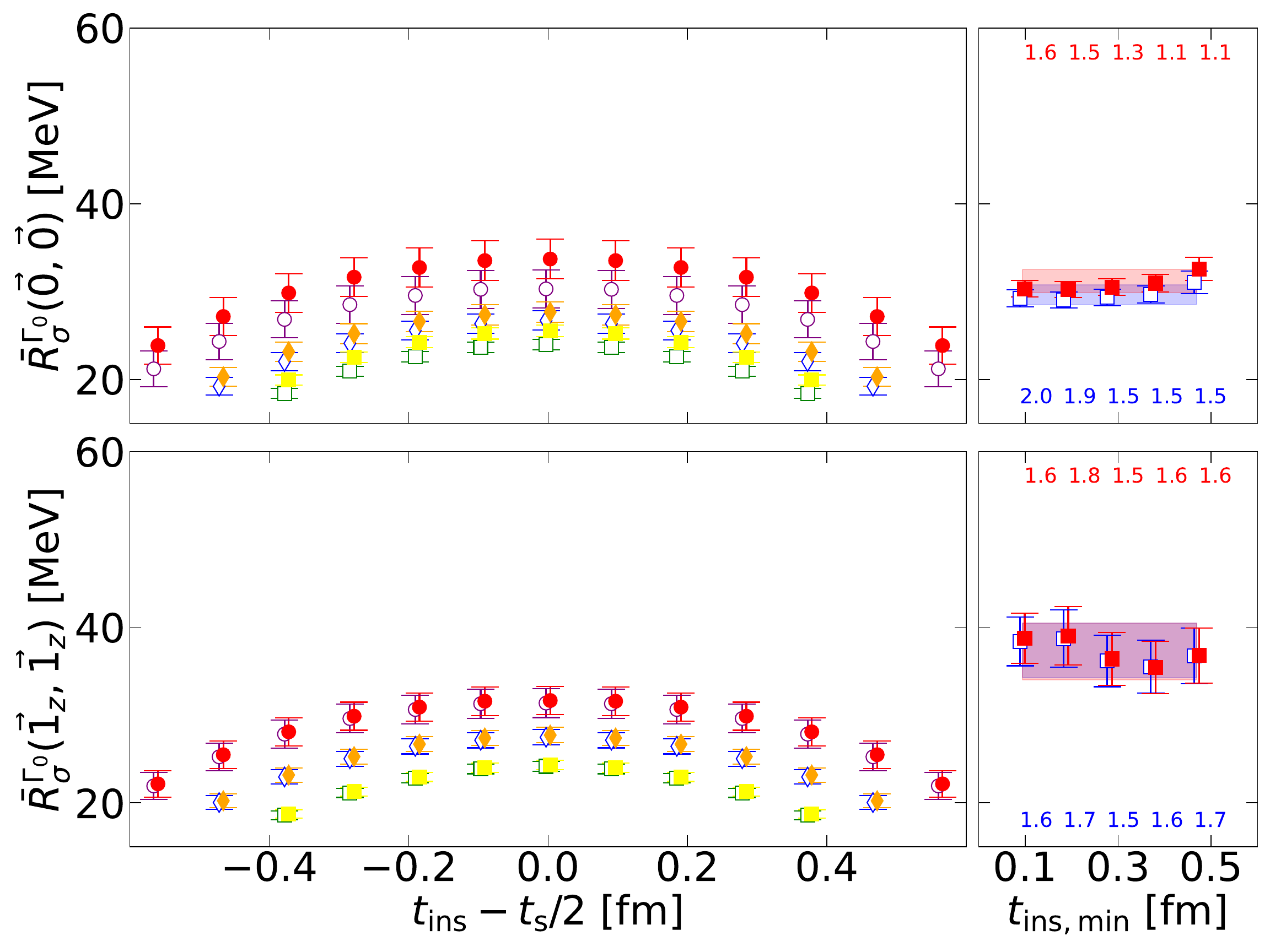}
    \caption{Ensemble  cA2.09.48. As in \cref{fig:3pt_raw_sep_S_24}}\label{fig:3pt_raw_sep_S_48}
\end{figure}

The first selected case that we discuss concerns  the nucleon sigma term  $\sigma_{\pi N}=m_q\,g^{u+d}_S$, that is renormalization independent.
It can be extracted  from   \cref{eq:ratio_S}. In particular,  we examine the following ratio:
\begin{alignat}{2}\label{eq:rbar_S}
    \bar{R}^{\Gamma_0}_{\sigma}(\vec{p},\vec{p})&=m_q\,\bar{R}^{\Gamma_0}_{S^{u+d}}(\vec{p},\vec{p}) &&\to \sigma_{\pi N} \,,
\end{alignat}
with $\vp=\vec{0}$ and $\vec{1}_z$.
The results for these two ratios and their two-state fits are presented in \cref{fig:3pt_raw_sep_S_24,fig:3pt_raw_sep_S_48}.
After applying GEVP, no significant changes are observed  nor a flip of sign of the contamination, as suggested in \cref{eq:Id-2}.
This reinforces our previous conclusion that the contamination observed in the nucleon $\sigma$ term case is unlikely to come from the lowest $N\pi$ states under consideration.
The analysis in Ref.~\cite{Gupta:2021ahb} suggests that this case should receive significant contaminations from $N\pi$ and $N\pi\pi$ states. Our finding does not support large contamination from the lowest  $N\pi$ states.

\begin{figure}[!ht]
    \centering
    \includegraphics[width=\columnwidth]{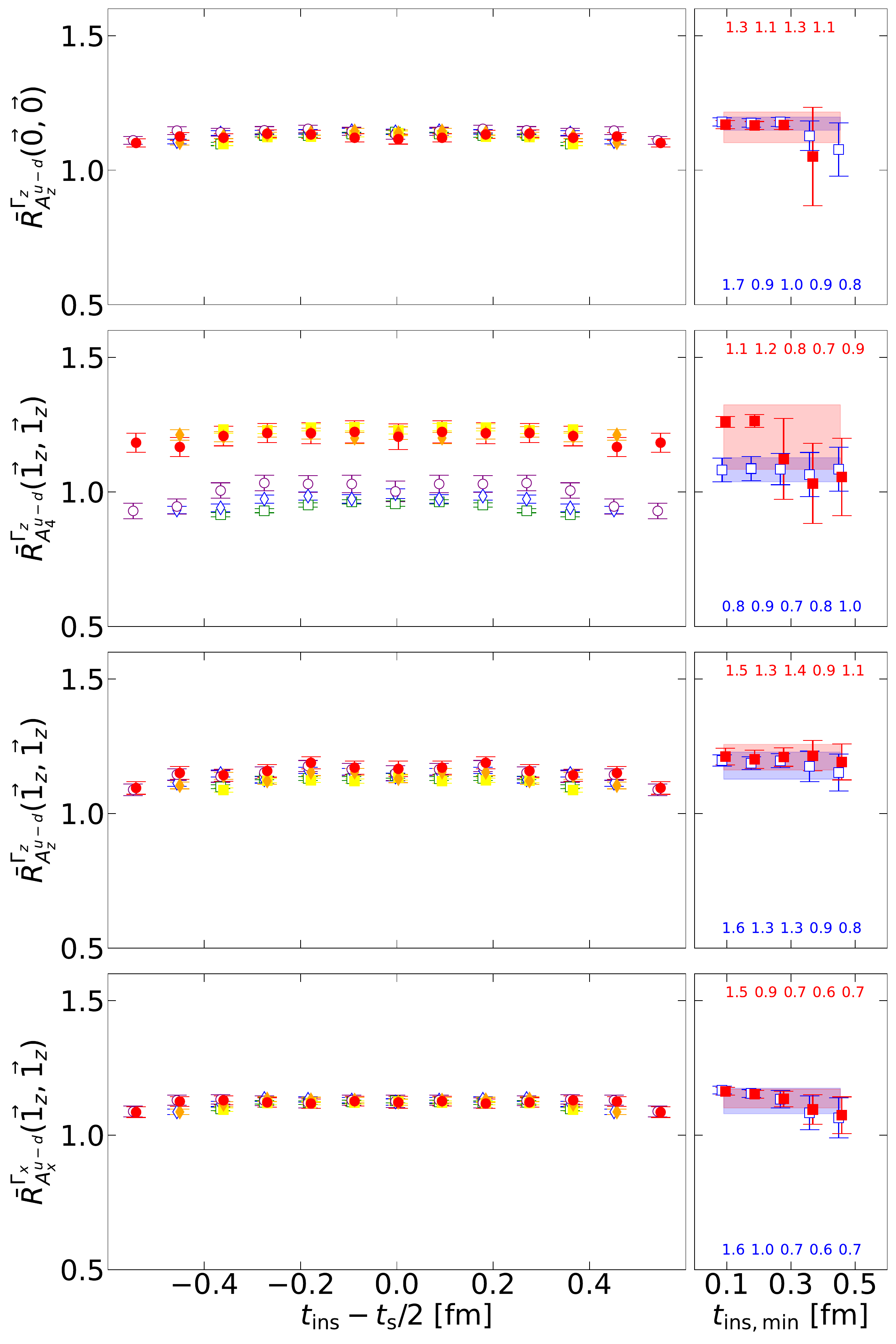}\\
    \includegraphics[width=\columnwidth]{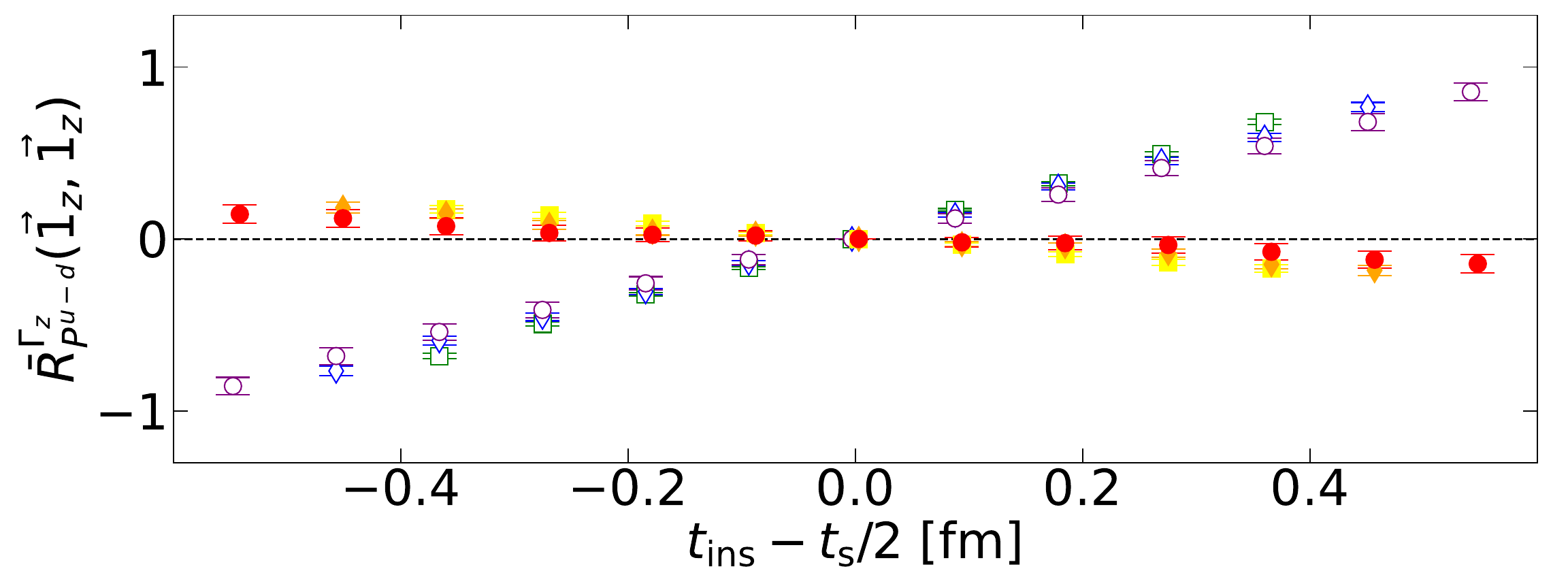}
    \caption{Ensemble  cA211.530.24. As in \cref{fig:3pt_raw_sep_S_24} for the isovector axial case of \cref{eq:ratio_AP0}. In the last row for $\bar{R}^{\Gamma_{z}}_P(\vec{1}_z,\vec{1}_z)$, we do not include fits since the data at the middle points $t_{\tc}-t_{\tf}/2=0$ are exactly zero after averaging ratios related by symmetries.}\label{fig:3pt_raw_sep_AJ_24}
\end{figure}
\begin{figure}[!ht]
    \centering
    \includegraphics[width=\columnwidth]{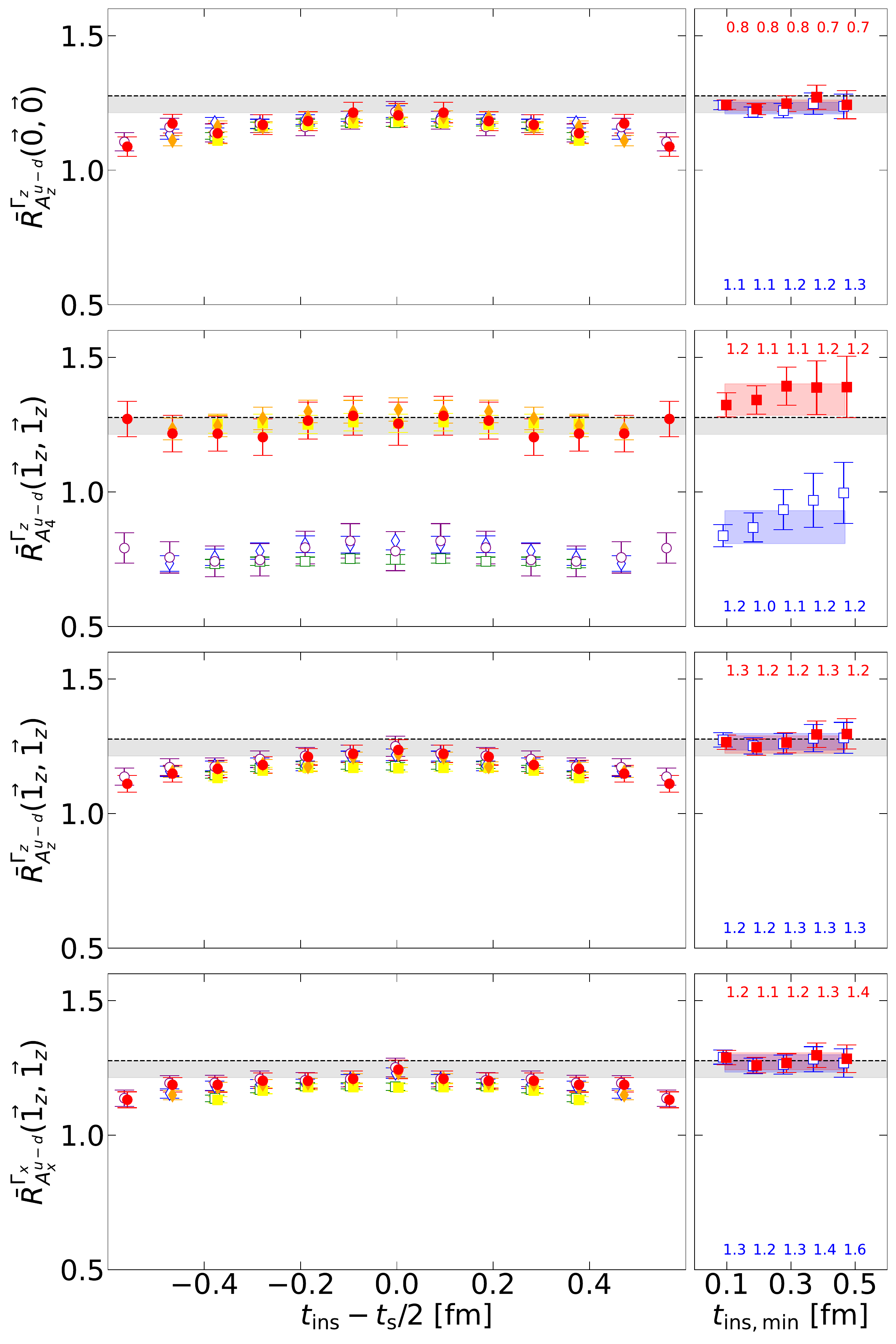}\\
    \includegraphics[width=\columnwidth]{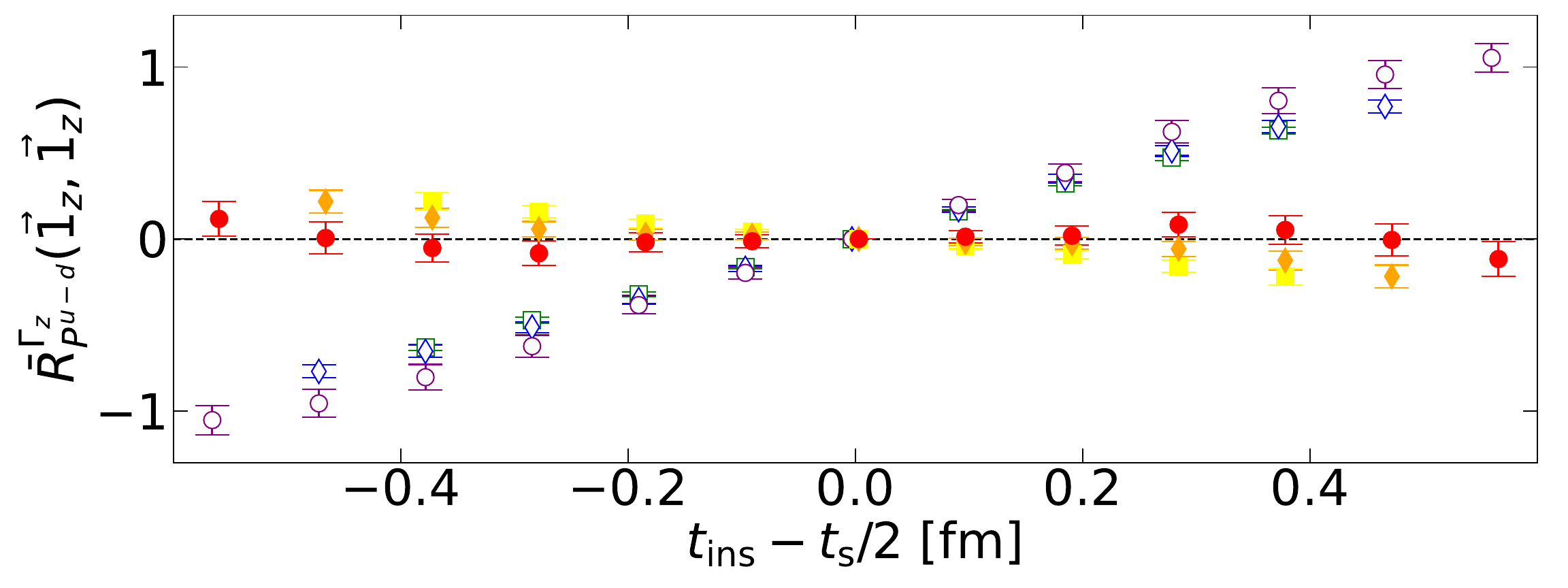}
    \caption{Ensemble  cA2.09.48. As in \cref{fig:3pt_raw_sep_AJ_24}. The grey bands are the results obtained in Ref.~\cite{Alexandrou:2023qbg}. The dashed lines indicate the experimental value $g^{u-d}_A= 1.27641(56)$ \cite{Markisch:2018ndu}.}\label{fig:3pt_raw_sep_AJ_48}
\end{figure}

For the isovector pseudoscalar and axial 3-point functions, the ratios are given in \cref{eq:ratio_AP0} for  $Q^2=0$ are shown in  \cref{fig:3pt_raw_sep_AJ_24,fig:3pt_raw_sep_AJ_48} accompanied with  the  two-state fits (to the 2-point and 3-point functions).
We have four different cases that yield  the axial charge $g^{u-d}_A$.
We note that the two-state fit results from three of these  are in agreement.
One yields  a smaller value for $g^{u-d}_A$.
After applying GEVP, significant improvement happens only on that specific case, and brings  agreement with the other three, demonstrating that for the specific case the contaminant state is the $N(1)\pi(0)$ state.
For the physical point ensemble, the common value is also compatible with both the the experimental value and the value from Ref.~\cite{Alexandrou:2023qbg}. 
The ratio $\bar{R}^{\Gamma_z}_{P^{u-d}}(\vec{1}_z,\vec{1}_z)$ presented in the last row of \cref{fig:3pt_raw_sep_AJ_24,fig:3pt_raw_sep_AJ_48} with the pseudoscalar current should be  $0$ asymptotically.
While it is  $0$ at the mid point  $\tc=\tf/2$ by symmetry, it is nonzero away from the mid point.
 Non-vanishing values come entirely from excited state contamination.
Indeed, after applying GEVP, most of these excited state contaminations are removed, and we get values much closer to $0$.
\begin{figure}[!ht]
    \centering
    \includegraphics[width=\columnwidth]{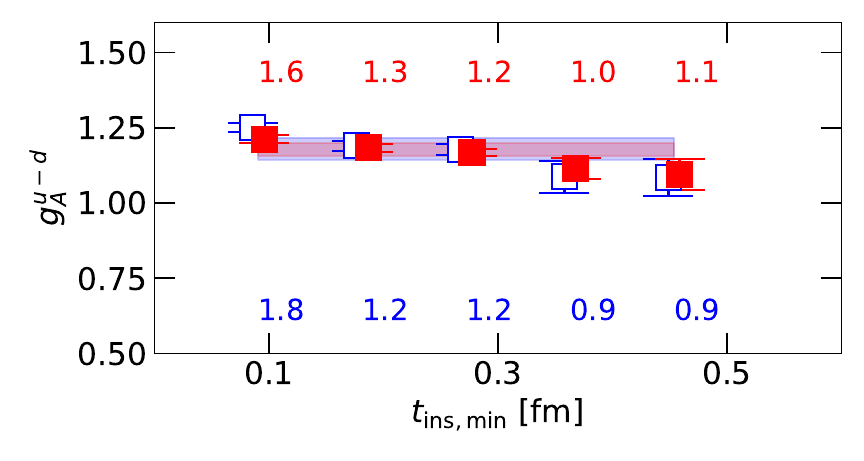}
    \caption{Ensemble  cA211.530.24: We show results on the extracted value of the  isovector axial charge $g_A^{u-d}$  as a function of $t_{\mathrm{ins,min}}$ used in the fit, that fits to two 2-point functions and five 3-point functions used to define the ratios in \cref{eq:ratio_AP0} for the isovector case.
    }\label{fig:3pt_A5_24}
\end{figure}
\begin{figure}[!ht]
    \centering
    \includegraphics[width=\columnwidth]{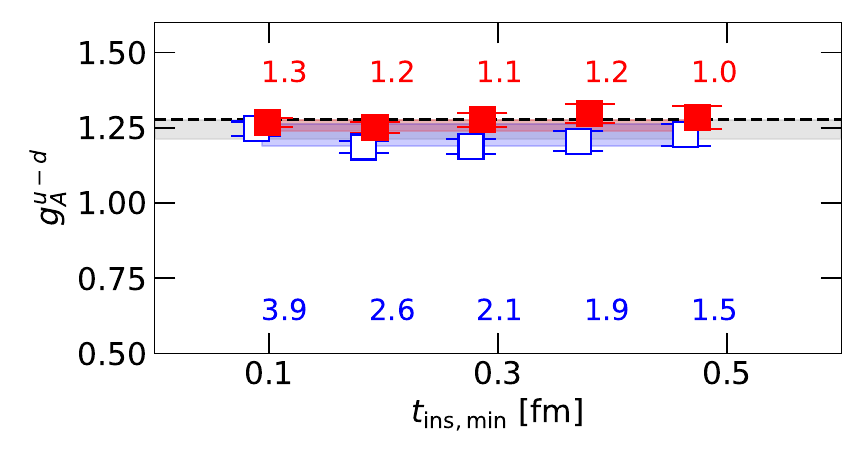}
    \caption{Ensemble  cA2.09.48. As in \cref{fig:3pt_A5_24}.}\label{fig:3pt_A5_48}
\end{figure}
In addition, we also perform simultaneous two-state  fits  to  the two 2-point and five 3-point functions used in the 5 ratios in \cref{eq:ratio_AP0} for the isovector case.
For the 3-point function corresponding to $\bar{R}^{\Gamma_z}_{P^{u-d}}(\vec{1}_z,\vec{1}_z)$, we set $a_{00}$ parameter to zero, and we do not include the values at the mid point $\tc=\tf/2$.
The results are presented in \cref{fig:3pt_A5_24,fig:3pt_A5_48}.
For the physical point ensemble, we get $g^{u-d}_A= 1.226(37)$ without GEVP and $g^{u-d}_A= 1.258(18)$ with GEVP. The result after GEVP agrees better with the experimental value, and also has a smaller error. This value is to be compared with the experimental value of $g^{u-d}_A= 1.27641(56)$ \cite{Markisch:2018ndu} and the  value of $g^{u-d}_A= 1.245(31)$ from Ref.~\cite{Alexandrou:2023qbg}. 

\begin{figure}[!ht]
    \centering
    \includegraphics[width=\columnwidth]{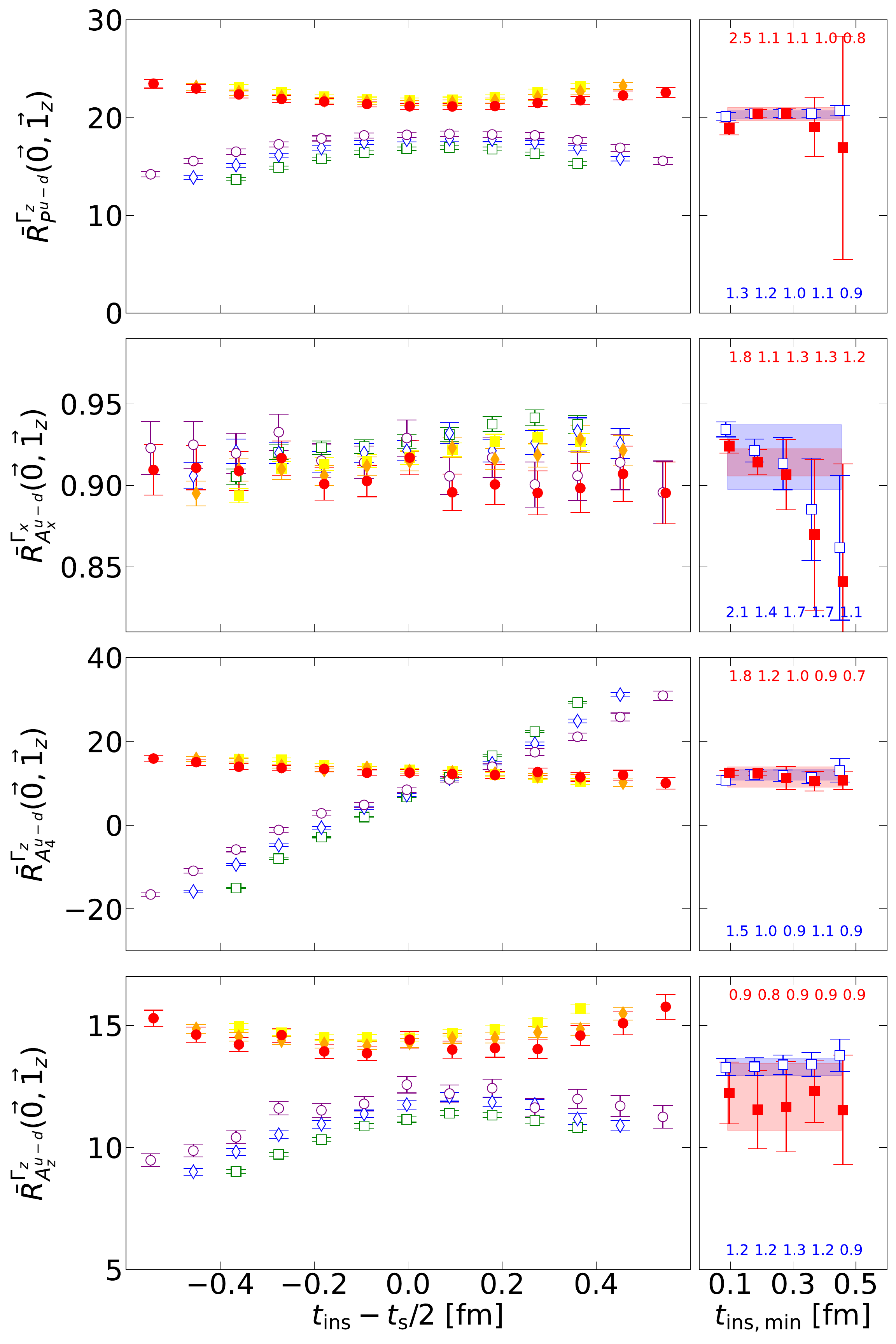}
    \includegraphics[width=\columnwidth]{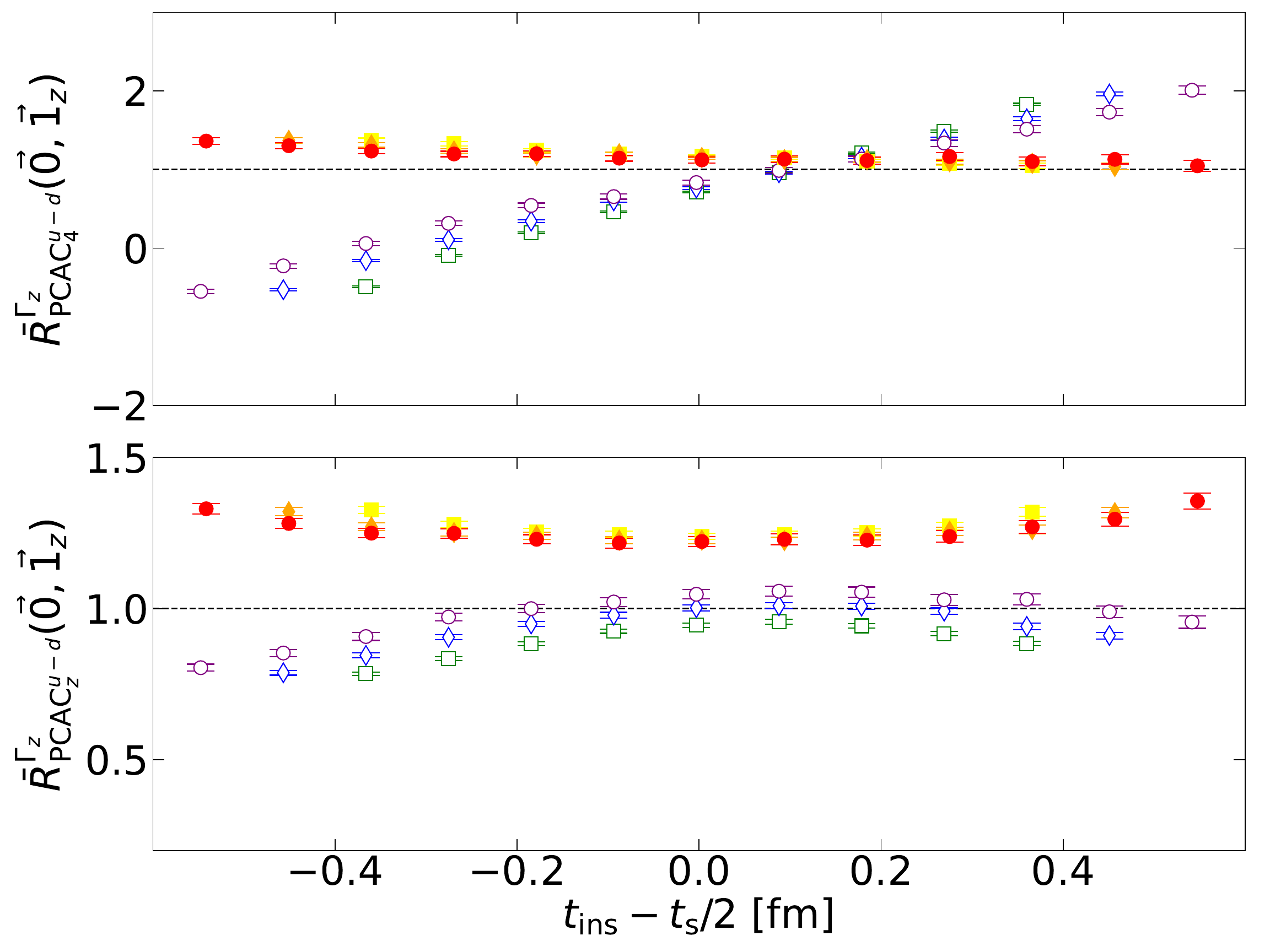}
    \caption{Ensemble  cA211.530.24. As in \cref{fig:3pt_raw_sep_S_24} for the isovector axial case given  in \cref{eq:ratio_AP1}. In the last two panels we show results for  $\bar{R}^{\Gamma_{k}}_{\mathrm{PCAC}_4^{u-d}}(\vec{0},\vec{1}_k)$ and $\bar{R}^{\Gamma_{k}}_{\mathrm{PCAC}_k^{u-d}}(\vec{0},\vec{1}_k)$, and there we do not include fits. The dashed lines  indicate $1$.}\label{fig:3pt_raw_sep_5APP_24}
\end{figure}
\begin{figure}[!ht]
    \centering
    \includegraphics[width=\columnwidth]{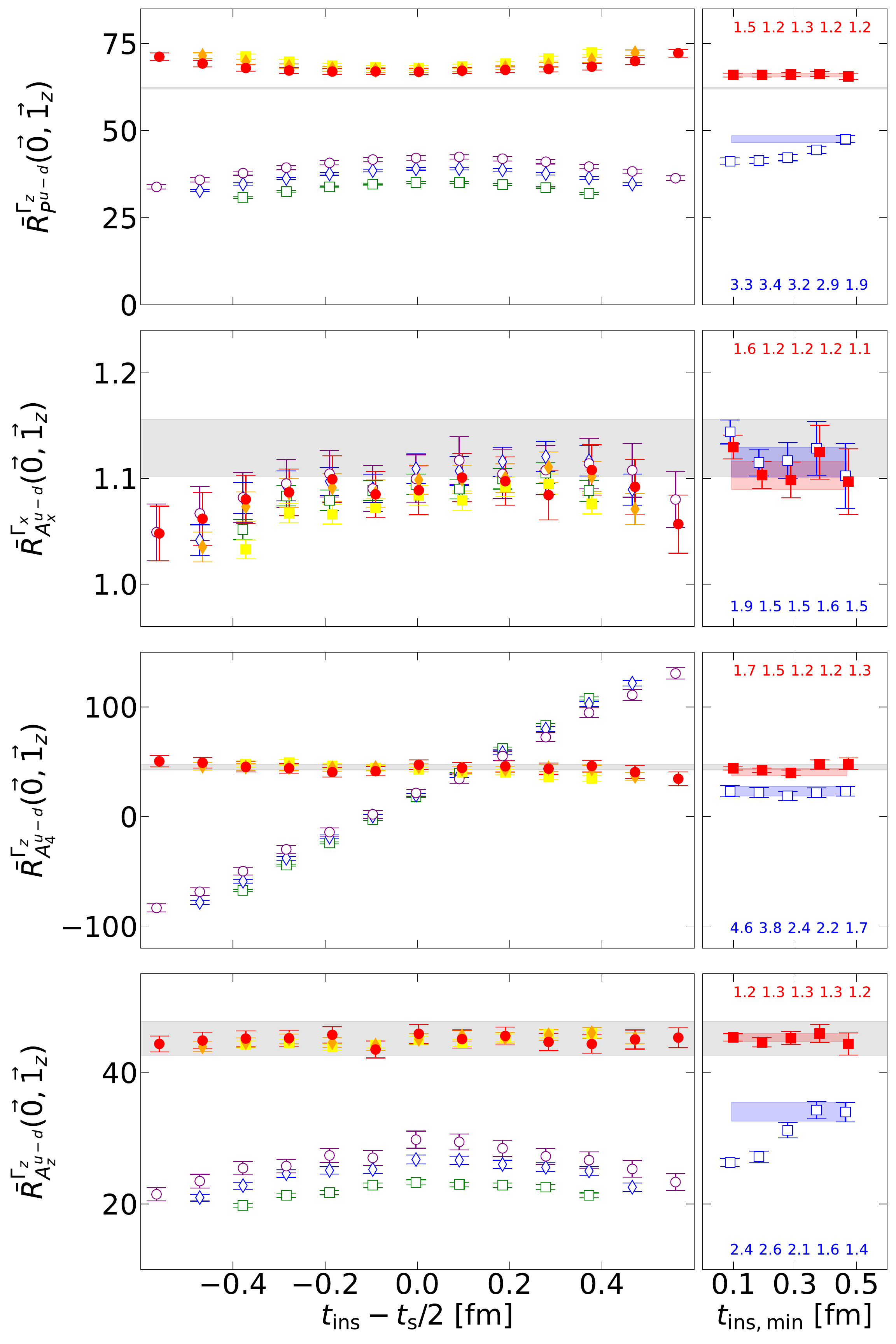}
    \includegraphics[width=\columnwidth]{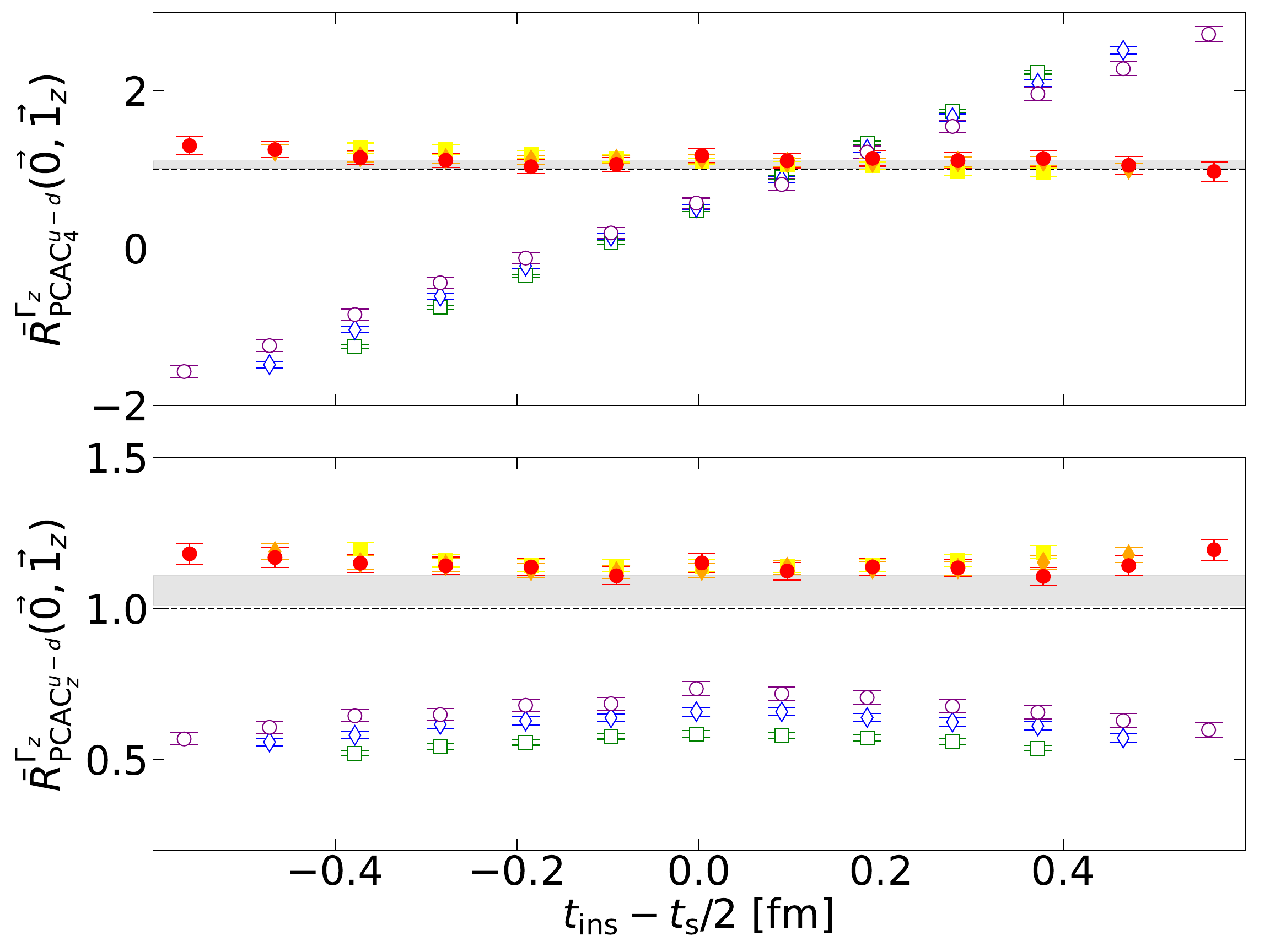}
    \caption{Ensemble  cA2.09.48. As in \cref{fig:3pt_raw_sep_5APP_24}. The grey bands are the results obtained in Ref.~\cite{Alexandrou:2023qbg} at the continuum limit.}\label{fig:3pt_raw_sep_5APP_48}
\end{figure}

For the ratios of the isovector pseudoscalar and axial 3-point functions at $Q^2=Q_1^2$ given in \cref{eq:ratio_AP1},
The results and the two-state fits to the 2-point and 3-point functions are presented in \cref{fig:3pt_raw_sep_5APP_24,fig:3pt_raw_sep_5APP_48}.
For the presented ratios, GEVP has a significant impact for all cases except for the ratio of the isovector axial 3-point function, $G^{u-d}_A(Q_1^2)$. We consider these changes as  improvements because the results with GEVP show milder time dependence.
For the two-state fits of the physical point ensemble, GEVP changes many results and stabilizes the $t_{\mathrm{ins,min}}$ dependence for several cases.
For the two-state fits of the $m_\pi=346\,$MeV ensemble, however, GEVP does not change the results within errors but worsens the precisions for several cases.
Our explanation is that the energies of $N\pi$ states are closer to that of the excited states observed in the 2-point function at the $m_\pi=346\,$MeV ensemble. 
For example, the non-interacting energy of the $N(0)\pi(1)$ state and the energy of the first excited state from 2-point function at one-unit of momentum are around $1848\,$MeV and $1934\,$MeV respectively, while they are $1236\,$MeV and $1637\,$MeV for the physical point ensemble.
For the $m_\pi=346\,$MeV ensemble, the excited states in the two-state fits can effectively be addressed  without GEVP. Furthermore, with $N\pi$ contaminations removed by GEVP, one gets less constraints on the excited states, which leads to the increases in errors.
\begin{figure}[!ht]
    \centering
    \includegraphics[width=\columnwidth]{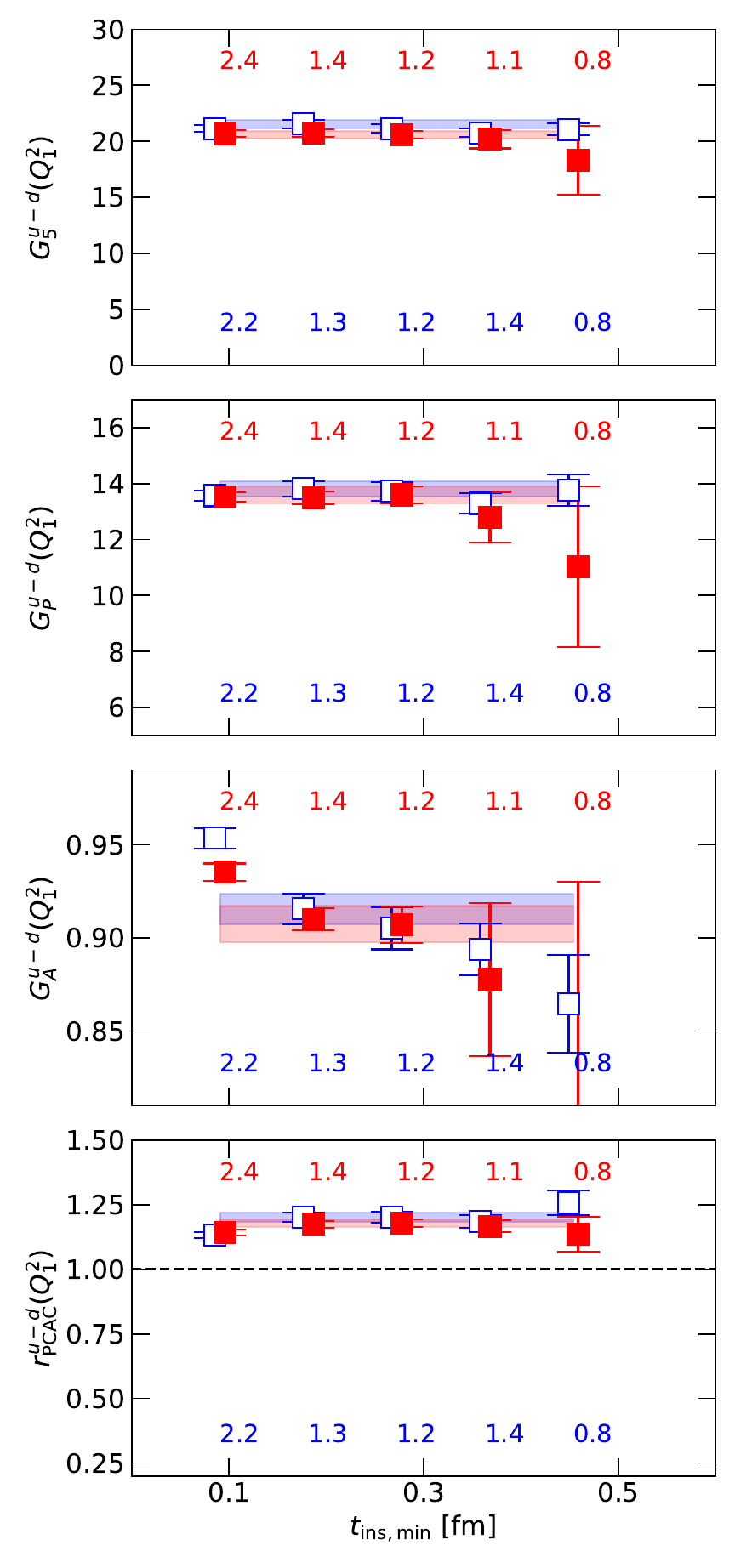}
    \caption{Ensemble  cA211.530.24. 
    From top to bottom, we show results on the extracted value of the isovector pseudoscalar, induced pseudoscalar, axial form factors and the PCAC ratio at one unit momentum transfer as a function of $t_{\mathrm{ins,min}}$ used in the fit, that fits to two 2-point functions and four 3-point functions used to define the ratios in \cref{eq:ratio_AP1} for the isovector case.
    }\label{fig:3pt_PCAC01_24}
\end{figure}
\begin{figure}[!ht]
    \centering
    \includegraphics[width=\columnwidth]{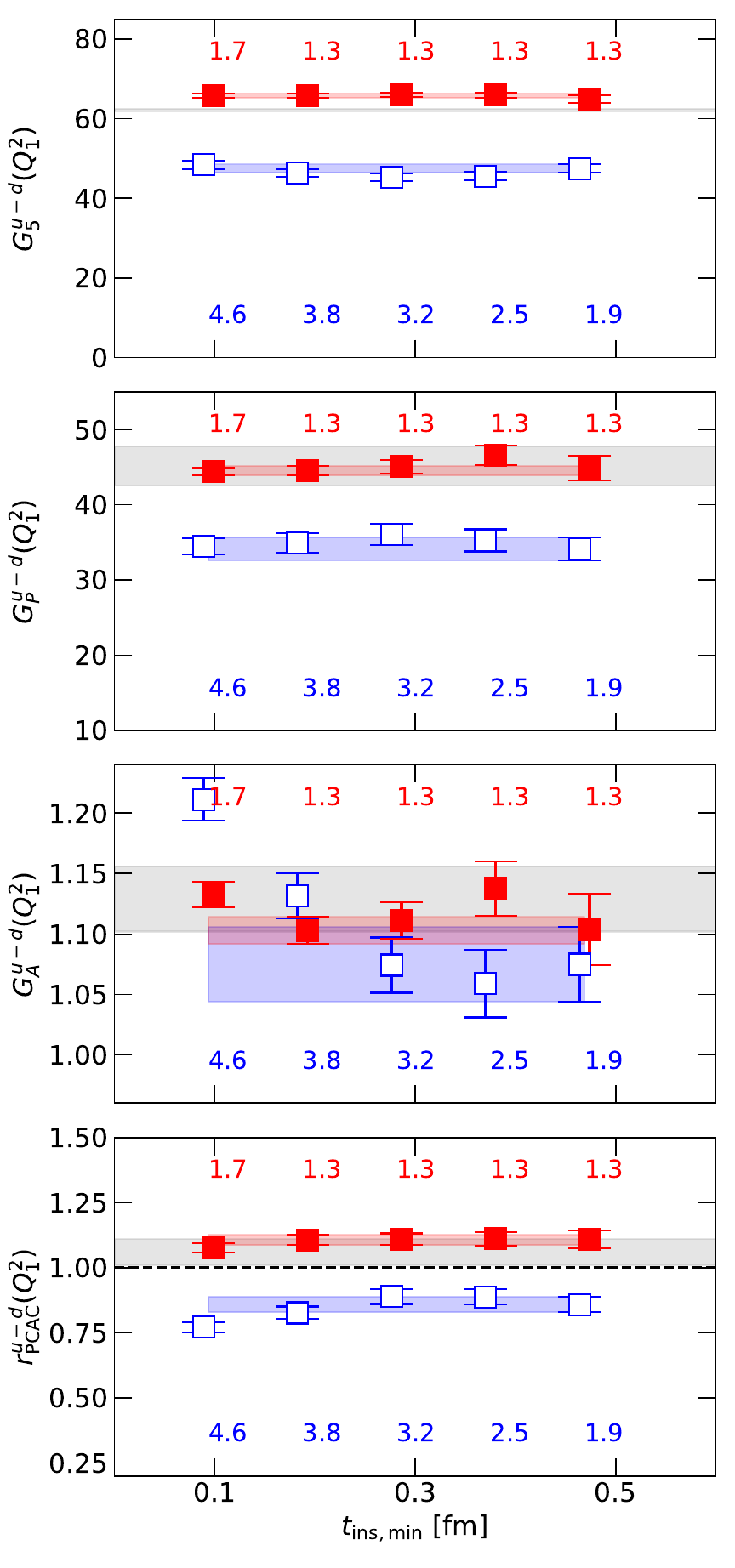}
    \caption{Ensemble  cA2.09.48. As in \cref{fig:3pt_PCAC01_24}.}\label{fig:3pt_PCAC01_48}
\end{figure}
In addition, we also perform simultaneous  fits to the two 2-point and four 3-point functions used in the ratios in \cref{eq:ratio_AP1} for the isovector case.
The results are presented in \cref{fig:3pt_PCAC01_24,fig:3pt_PCAC01_48}.
For the $m_\pi=346\,$MeV ensemble, GEVP does not really help, the reason of which has already been discussed previously.
For the physical point ensemble, for both cases without and with GEVP, $G^{u-d}_A$ agrees with the continuum-limit result (In Ref.~\cite{Alexandrou:2023qbg}, only small lattice discretization error is found for $G^{u-d}_A$), but GEVP stabilizes the $t_{\mathrm{ins,min}}$ dependence and reduces its error.
In contrast, in Ref.~\cite{Alexandrou:2023qbg}, large lattice artefacts were found for $G^{u-d}_P$. Including the  quark disconnected  contributions reduces these cut-off effects dramatically.
Using  GEVP and subtracting lattice artefacts using the quark disconnected at finite lattice spacing,  the results for $G^{u-d}_P$ approach  those in  Ref.~\cite{Alexandrou:2023qbg}, that did a very thorough analysis of both excited states and the took the continuum limit where these quark disconnected contributions vanish. 

Similarly, for 
 $G^{u-d}_5$, the results approach the ones Ref.~\cite{Alexandrou:2023qbg}, after GEVP. We note that in Ref.~\cite{Alexandrou:2023qbg}, we use several large source-sink time separations and included in the analysis the matrix element of the temporal axial current to eliminate excited states. Applying GVEP has a similar effect. Including the disconnected non-zero contributions to the isovector also reduces cut-off effects and the combining with GEVP bring the results close to the continuum extrapolated results of Ref.~\cite{Alexandrou:2023qbg} where excited state contamination were properly treated. A similar behavior is seen for the ratio, $r^{u-d}_{\mathrm{PCAC}}$. The fact, that is still consistently larger that unity indicates residual cut-off effects, since the ensemble used in this work is coarser than all the ensembles used in Ref.~\cite{Alexandrou:2023qbg}. 

More details on the effect of including the disconnected quark loop contributions to the isovector matrix elements can be found in \cref{app:iso}.

\section{Conclusions}\label{sec:conclusions}
We investigate contributions of excited states to nucleon matrix elements by studying the 2-point and 3-point functions using nucleon and pion-nucleon interpolating fields. We analyse two twisted mass fermion ensembles with pion masses $346$\,MeV and $131$\,MeV and lattice spacing $a\approx 0.09$~fm.
A variant of the application of the generalized eigenvalue problem (GEVP)  to nucleon matrix elements, which avoids the need of the most computationally demanding 3-point function, is developed and applied to the isoscalar and isovector scalar, vector, pseudoscalar, axial, and tensor nucleon matrix elements.
We include both connected and disconnected contributions to both isoscalar and isovector cases.
Since isospin symmetry is broken at finite lattice spacing in the twisted mass fermion formulation, 
the quark-disconnected contributions from topologies with insertion loops to isovector quantities 
are non-zero, and only vanish in the continuum limit.
When they are included, quantities that are known to have large cut-off effects,
such as those from the axial and pseudoscalar currents that couple to the pion field,
are shown to exhibit reduced cut-off effects. Thus, including these contributions leads to a better controlled extrapolation to the continuum.

 An important feature of our GEVP analysis is an analysis on the dependence on the reference time $t_0$.  While the effect on the eigenvalues is very mild, the eigenvectors show a dependence and only converge at large values of $t_0$. Following Ref.~\cite{Blossier:2009kd}, where it was shown that for $t_0\geq t/2$  higher states are suppressed with a  larger exponential energy gap, we  take $t-t_0=4a$. With this choice, we  find a faster convergence for the eigenvectors to their asymptotic large $t_0$ value.
We thus determine the eigenvectors using a plateau fit to the eigenvectors extracted for $t-t_0=4a$.

We next discuss the improvements on  the matrix elements of the  isovector pseudoscalar and axial matrix element.
\begin{itemize}
\item The isovector axial charge $g^{u-d}_A$ is extracted using four different combinations of the ratio that yield this quantity in the large time limit. 
  Before applying GEVP-improvement, the result extracted from the three-point function using moving-frame nucleon interpolator 
  and the temporal component of the axial current does not agree with results extracted from the other three cases. 
  However, after applying GEVP-improvement, it yields consistent results with the other three. 
Performing a joint fit using data for all four combinations included, we  obtain $g^{u-d}_A= 1.258(18)$ for the physical point ensemble, 
compatible with the experimental value $g^{u-d}_A= 1.27641(56)$~\cite{Markisch:2018ndu}.  Since this result is determined using four different matrix elements, which  suffer from different excited state contamination, it implies that we have reliable control on the elimination of excited state contributions.

\item The isovector pseudoscalar and induced pseudoscalar form factors with one lattice unit of momentum transfer 
  are significantly improved by applying the GEVP-improvement and, thus, belong to the third category. 
These form factors are used in defining the PCAC ratio $r^{u-d}_{\mathrm{PCAC}}$, which should be equal to $1$ as a consequence of the PCAC relation.
While significantly improved by GEVP, $r^{u-d}_{\mathrm{PCAC}}$ is still not fully compatible with $1$. 
Small residual lattice artifacts are still present and one has to extrapolate to the continuum limit as done  in Ref.~\cite{Alexandrou:2023qbg} to recover compatibility with the PCAC relation.
Therefore, we conclude that for these matrix elements, the contamination does not stem from the lowest $N\pi$ states included in our analysis.
\end{itemize}
 One quantity of interest is the nucleon sigma term $\sigma_{\pi N}$ due to its phenomenological importance and the tension between lattice determinations and phenomenology. The analysis of Ref.~\cite{Gupta:2021ahb,Gupta:2023cvo} suggests that there is significant contamination from $N\pi$ and $N\pi\pi$ states. Our findings do not support large contamination from the lowest  $N\pi$ states. In addition, all other matrix elements that we examine, namely matrix elements of the scalar, vector and tensor operators, do not show improvement  when applying the GEVP and, indicating that the dominant contaminant is not the $\pi N$ state.

\section*{Acknowledgements}
We thank all members of the ETM collaboration for a most conducive cooperation.
We would like to thank Lorenzo Barca, Marilena Panagiotou and Rainer Sommer for useful discussions and suggestions.
%
We acknowledge computing time granted on Piz Daint at Centro Svizzero di Calcolo Scientifico (CSCS) via the project with id s1174, JUWELS Booster at the J\"{u}lich Supercomputing Centre (JSC) via the project with id pines, and Cyclone at the Cyprus institute (CYI) via the project with ids P061, P146 and pro22a10951.
Y.L. is supported by the Excellence Hub project "Unraveling the 3D parton structure of the nucleon with lattice QCD (3D-nucleon)" id EXCELLENCE/0421/0043 co-financed by the European Regional Development Fund and the Republic of Cyprus through the Research and Innovation Foundation.
F.P. acknowledges financial support by the Cyprus Research and
Innovation foundation Excellence Hub project NiceQuarks under contract with number
 EXCELLENCE/0421/0195. 
C.A. and G. K. acknowledge partial support from the European Joint Doctorate AQTIVATE that received funding from the European Union’s research and innovation programme under the Marie Sklodowska-Curie Doctoral Networks action under the Grant Agreement No 101072344.
M.P. acknowledges support by the Sino-German collaborative research center CRC 110.

\appendix
\section{Conventions and useful relations}\label{app:con}
As customary we use Roman letters for spatial indices, and Greek letters for Euclidean spacetime indices.
The gamma matrices we use are given by
\begin{align}\label{eq:gms}
    &\gamma_i=\begin{bmatrix}
        & i\,\sigma_i \\
       -i\,\sigma_i &  \\
   \end{bmatrix} \,,\quad
   \gamma_4=\begin{bmatrix}
       \mathbf{1} & \\
        & -\mathbf{1} \\
   \end{bmatrix} \,, \nonumber\\
   &\gamma_5=\gamma_1\gamma_2\gamma_3\gamma_4=\begin{bmatrix}
     & \mathbf{1} \\
     \mathbf{1} & \\
   \end{bmatrix} \,,\quad
   \sigma_{\mu\nu}=\frac{1}{2}[\gamma_\mu,\gamma_\nu] \,.
\end{align}
For $\sigma_{\mu\nu}$, they have the following matrix form:
\begin{align}
    \sigma_{ij}=i\varepsilon_{ijk}\begin{bmatrix}
        \sigma_k & \\
         & \sigma_k \\
    \end{bmatrix} \,, \quad
    \sigma_{4k}=i\begin{bmatrix}
        & \sigma_k \\
       \sigma_k &  \\
   \end{bmatrix} \,.
\end{align}
The matrices for charge conjugation (C), parity (P), and time reversal (T) used in this work are given by
\begin{align}
    \Gamma_C&=i\gamma_2\gamma_4=\begin{bmatrix}
            & \sigma_2 \\
           \sigma_2 & \\
        \end{bmatrix} \,,\nonumber\\
    \Gamma_P&=\gamma_4=\begin{bmatrix}
        \mathbf{1} &  \\
         & -\mathbf{1} \\
     \end{bmatrix}\,, \nonumber\\
     \Gamma_T&=i\gamma_5\gamma_4=i\begin{bmatrix}
         & -\mathbf{1} \\
         \mathbf{1} &  \\
     \end{bmatrix}\,,
\end{align}
where all three matrices are Hermitian. 
Then we have
\begin{align}
    \Gamma_{PT}&=\Gamma_T\Gamma_P=i\gamma_5=i\begin{bmatrix}
        & \mathbf{1} \\
        \mathbf{1} &  \\
    \end{bmatrix}\,, \nonumber\\
    \Gamma_{CPT}&=\Gamma_T\Gamma_P\Gamma_C=\gamma_5\gamma_4\gamma_2=\sigma_{31}=\begin{bmatrix}
        i\,\sigma_2 &  \\
         & i\,\sigma_2 \\
    \end{bmatrix}\,.
\end{align}

The four-momentum $p$ of a nucleon satisfies $p^2=-m_N^2$, $p_4=i\,E_N$. For a nucleon at source with momentum $p$ and a nucleon at sink with momentum $p^\prime$, we define the transfer momentum $q=p-p^\prime$ and $Q^2=q^2>0$. We also define $P=p+p^\prime$.
The nucleon states are normalized as $\braket{N(p,s)|N(p,s)}=1$.
The Dirac spinor for the nucleon is taken as 
\begin{align}
    u_N(p,s)&=\frac{-i\slashed{p}+m_N}{\sqrt{2E_N(E_N+m_N)}}u^s \,, \nonumber\\
    \bar{u}_N(p,s)&=\bar{u}^s\frac{-i\slashed{p}+m_N}{\sqrt{2E_N(E_N+m_N)}} \,,
\end{align}
where $u^s=\begin{pmatrix} \phi^s \\ \vec{0} \\ \end{pmatrix}$. For spin-up along z-direction, we choose $\phi^{\uparrow_z}=\begin{pmatrix} 1 \\ 0 \\ \end{pmatrix}$, and for spin-down along z-direction, we choose $\phi^{\downarrow_z}=\begin{pmatrix} 0 \\ 1 \\ \end{pmatrix}$.

 $\varepsilon_{\mu\nu\rho\sigma}$ is the antisymmetric tensor with $\varepsilon_{1234}=1$. We will also use another antisymmetric tensor $\varepsilon_{ijk}=\varepsilon_{ijk4}$.
About the momenta $p^\prime=(P-q)/2$, $p=(P+q)/2$, $q=p-p^\prime$, $P=p+p^\prime$, we have the following relations:
\begin{align}
    &p^{\prime,2}=p^2=-m_N^2\,,\quad  q\cdot P=0\,, \nonumber\\
    &Q^2=q^2=-2m_N^2-2\,p^\prime\cdot p \,, \nonumber\\
    &p^\prime_{\{\mu}p_{\nu\}}=\frac{1}{2}(P_{\mu}P_{\nu}-q_{\mu}q_{\nu})  \,,\quad P_{[\mu}q_{\nu]}=2p^\prime_{[\mu}p_{\nu]}\,,
\end{align}
where
\begin{align}
    A_{\{\mu}B_{\nu\}} &= A_{\mu}B_{\nu} + A_{\nu}B_{\mu} \,,\nonumber\\
    A_{[\mu}B_{\nu]} &= A_{\mu}B_{\nu} - A_{\nu}B_{\mu} \,.
\end{align}
In addition, we have the Goldon identity:
\begin{align}
    \bar{u}(p^\prime)\gamma_\mu u(p) = \frac{-i}{2m_N} \bar{u}(p^\prime)\left[ P_\mu+\sigma_{\mu\nu}q_\nu \right]u(p) \,.
\end{align}

\section{Interpolating fields used in this work}\label{app:interfields}
\newcommand{\Jx}{\OJ}
\newcommand{\Jp}{\OJ}
We consider the nucleon $N$ and pion $\pi$ interpolating fields given by
\begin{align}
    \OJ_N(\vec{x};t)&=\epsilon_{abc} \left[\psi_1^{a}(\vec{x};t)^T \mathcal{C}\gamma_5\, \psi_2^{b}(\vec{x};t)\right] \psi_3^{c} \,, \nonumber\\
    \OJ_\pi(\vec{x};t)&=\bar{\psi}_1^a(\vec{x};t) \,i\gamma_5\, \psi_2^a(\vec{x};t) \,,\nonumber\\
\end{align}
where $\psi^a(\vec{x};t)$ is the quark field with the color index $a$ and the spin index suppressed.
Flavor of the quark fields depends on that of the $N$ and $\pi$.
For the proton $N=p$ (neutron $N=n$), we have $\psi_1\psi_2\psi_3=udu$ ($dud$). For the charged pion $\pi=\pi^+$ ($\pi^-$), we have $\bar{\psi_1}\psi_2=\bar{d}u$ ($\bar{u}d$). For the neutral pion $\pi=\pi^0$, we have $\bar{\psi_1}\psi_2=\frac{\bar{u}u-\bar{d}d}{\sqrt{2}}$.
The quark field has been smeared as explained in \cref{eq:smear}.
The spin index $\alpha$ of $\Jx_N$ is also suppressed, and we denote its full form by
\begin{align}
    [\OJ_N(\vec{x};t)]_\alpha \,, \quad \alpha=0,1,2,3\,.
\end{align}
Then the $N\pi$ interpolating field is given by
\begin{align}
\OJ_{N\pi}(\vec{x}_1,\vec{x}_2;t)&=\OJ_N(\vec{x}_1;t)
\OJ_\pi(\vec{x}_2;t) \,.
\end{align}
We perform the isospin projection to the $(I,I_3)=(\frac{1}{2},+\frac{1}{2})$ sector:
\begin{align}
    \OJ_N&=\OJ_p \,, \nonumber\\
    \OJ_{N\pi}&=\sqrt{2/3}\,\OJ_{n\pi^+} - \sqrt{1/3}\,\OJ_{p \pi^0} \,.
\end{align}
We perform the Fourier transformation:
\begin{align}
    \OJ_N(\vec{p};t)&=\sum_{\vec{x}} e^{i\vec{p}\cdot\vec{x}} \OJ_N(\vec{x};t) \,,\nonumber\\
    \OJ_\pi(\vec{p};t)&=\sum_{\vec{x}} e^{i\vec{p}\cdot\vec{x}} \OJ_\pi(\vec{x};t) \,, \nonumber\\
    \OJ_{N\pi}(\vec{p}_1,\vec{p}_2;t) &= \OJ_N(\vec{p}_1;t)
    \OJ_\pi(\vec{p}_2;t) \,.
\end{align}

On the lattice, the spherical symmetry group O$(3)$ is reduced to cubic symmetry group O$_h$. With a fixed momentum $\vec{p}$, the little group of O$_h$ is a subgroup with all elements that leave $\vec{p}$ invariant. One can construct the interpolating field that transforms as the $r$-th vector in the irreducible representation (irrep) $\Lambda$ of the little group.
In this work, we only consider the rest-frame nucleon and the nucleon with one unit of momentum.
The rest-frame nucleon lies in the $\Lambda=G_{1g}$ irrep of the little group O$_h$, and the nucleon with one unit of momentum lies in the $\Lambda=G_{1}$ of the little group Dih$_4$.
We restrict our discussion to the two irreps, and will suppress them in the following discussions. Since they are both two-dimensional, the irrep row can take the values $r=0,1$.
For the one unit of momentum case, we consider the representative direction $\vec{p}=\vec{1}_z=\frac{2\pi}{L}(0,0,1)$. 
We denote the group-projected interpolating fields by 
\begin{align}
    \OJ_k^r(\vec{p};t) \,,
\end{align}
where the index $r$ is for irrep row, and $k$ is for different interpolating fields with the same conserved quantum numbers as explained below.

The group-projected interpolating fields used in this work are given by
\begin{align}\label{eq:Ngrp2mom}
    \OJ_{N}^0(\vec{0};t)=[\OJ_{N}(\vec{0};t)]_0 \,, \nonumber\\
    \OJ_{N}^1(\vec{0};t)=[\OJ_{N}(\vec{0};t)]_1 \,, \nonumber\\
    \OJ_{N}^0(\vec{1}_z;t)=[\OJ_{N}(\vec{1}_z;t)]_0 \,, \nonumber\\
    \OJ_{N}^1(\vec{1}_z;t)=[\OJ_{N}(\vec{1}_z;t)]_1 \,,
\end{align}
where the left-hand-side interpolators are in group basis and the right-hand-side interpolators are in momentum basis,
and 
\begin{widetext}
    \begin{align}
        \OJ^0_{N(1)\pi(1)}(\vec{0};t)&=\frac{-i}{\sqrt{6}}\Big\{[\Jp_{N\pi}(-\vec{1}_x,\vec{1}_x;t)]_1 - [\Jp_{N\pi}(\vec{1}_x,-\vec{1}_x;t)]_1 -i\,[\Jp_{N\pi}(-\vec{1}_y,\vec{1}_y;t)]_1 \nonumber\\
        &\quad +i\,[\Jp_{N\pi}(\vec{1}_y,-\vec{1}_y;t)]_1 +[\Jp_{N\pi}(-\vec{1}_z,\vec{1}_z;t)]_0 - [\Jp_{N\pi}(\vec{1}_z,-\vec{1}_z;t)]_0 \Big\} \,,\nonumber\\
        \OJ^1_{N(1)\pi(1)}(\vec{0};t)&=\frac{-i}{\sqrt{6}}\Big\{[\Jp_{N\pi}(-\vec{1}_x,\vec{1}_x;t)]_0 - [\Jp_{N\pi}(\vec{1}_x,-\vec{1}_x;t)]_0 +i\,[\Jp_{N\pi}(-\vec{1}_y,\vec{1}_y;t)]_0 \nonumber\\
        &\quad -i\,[\Jp_{N\pi}(\vec{1}_y,-\vec{1}_y;t)]_0 -[\Jp_{N\pi}(-\vec{1}_z,\vec{1}_z;t)]_1 + [\Jp_{N\pi}(\vec{1}_z,-\vec{1}_z;t)]_1 \Big\} \,, \nonumber\\
        \OJ_{N(1)\pi(0)}^0(\vec{1}_z;t)&= -i\,[\Jp_{N\pi}(\vec{1}_z,\vec{0};t)]_0 \,,\qquad \OJ^1_{N(1)\pi(0)}(\vec{1}_z;t)= i\,[\Jp_{N\pi}(\vec{1}_z,\vec{0};t)]_1 \,, \nonumber\\
        \OJ^0_{N(0)\pi(1)}(\vec{1}_z;t)&= -i\,[\Jp_{N\pi}(\vec{0},\vec{1}_z;t)]_0 \,,\qquad \OJ^1_{N(0)\pi(1)}(\vec{1}_z;t)= i\,[\Jp_{N\pi}(\vec{0},\vec{1}_z;t)]_1 \,.
    \end{align}
\end{widetext}
To summarize, we have $k=N,N(1)\pi(1)$ for $\vec{p}=\vec{0}$, and $k=N,N(1)\pi(0),N(0)\pi(1)$ for $\vec{p}=\vec{1}_z$.

\section{Standard 2-point and 3-point functions in group basis}\label{app:23pt}
Because the unpolarized projection matrix $\Gamma_0$ and the polarized projection matrix $\Gamma_k$ given by
\begin{align}\label{eq:defG}
    \Gamma_0&= \frac{1+\gamma_4}{4} = \frac{1}{2}\begin{bmatrix}
        \mathbf{1} & \\
         & \mathbf{0}  \\
    \end{bmatrix}\,, \nonumber\\
    \Gamma_k&=i\gamma_5\gamma_k\,\Gamma_0 = \frac{1}{2}\begin{bmatrix}
        \sigma_k & \\
         & \mathbf{0} \\
    \end{bmatrix} \,,
\end{align}
have nonzero elements only in the left-upper $2\times2$ block, the single nucleon interpolators in momentum space $[\OJ_N(\vp;t)]_\alpha$ are only required for $\alpha=0,1$.
They can be linearly expanded in terms of the interpolators in the group basis as
\begin{align}
    [\OJ_N(\vp;t)]_\alpha=\sum_r c_\alpha^r(\vp)\,\OJ^r_N(\vp;t)\,.
\end{align}
The group basis, like in \cref{eq:Ngrp2mom}, can always be chosen to satisfy
\begin{align}\label{eq:gbcab}
    \sum_\alpha c_\alpha^{r^\prime}(\vp) [c_\alpha^r(\vp)]^*=\delta_{r^\prime r} \,.
\end{align}

The standard nucleon 2-point function can be expressed in the group basis as follows:
\begin{align}\label{eq:C2ptstd}
    &\quad \Cdpt_{NN}(\vp;t) \,,\nonumber\\
    &=\text{Tr}\left[ \braket{\OJ_N(\vp;t)\bar{\OJ}_N(\vp;0)}\Gamma_0 \right] \,,\nonumber\\
    &=\braket{[\OJ_N(\vp;t)]_\alpha[\OJ_N(\vp;0)]^\dagger_\beta}[\gamma_4\Gamma_0]_{\beta\alpha} \,,\nonumber\\
    &=\sum_{r^\prime r}c_\alpha^{r^\prime}(\vp)\braket{\OJ_N^{r^\prime}(\vp;t)[\OJ^{r}_N(\vp;0)]^\dagger}[c_\beta^{r}(\vp)]^*[\gamma_4\Gamma_0]_{\beta\alpha}  \,,\nonumber\\
    &=\frac{1}{2}\sum_r \braket{\OJ_N^{r}(\vp;t)[\OJ^{r}_N(\vp;0)]^\dagger} =\frac{1}{2}\sum_{r}\Cdpt^r_{NN}(\vp;t)\,,
\end{align}
where \cref{eq:gbcab} is used.

Similarly, the standard nucleon 3-point function in the group basis will be
\begin{align}\label{eq:C3ptstd}
    &\quad\Ctpt_{NN}(\Gamma,\OO;\vpp,\vp;\tf,\tc) \,,\nonumber\\
    &=\text{Tr}\left[ \braket{\OJ_N(\vpp;\tf)\OO(\vq;\tc)\bar{\OJ}_N(\vp;0)}\Gamma \right] \,,\nonumber\\
    &=\braket{[\OJ_N(\vpp;\tf)]_\alpha\OO(\vq;\tc)[\OJ_N(\vp;0)]_\beta^\dagger}[\gamma_4\Gamma]_{\beta\alpha} \,,\nonumber\\
    &=\sum_{r^\prime r} \braket{\OJ_N^{r^\prime}(\vpp;\tf)\OO(\vq;\tc)[\OJ^{r}_N(\vp;0)]^\dagger} \hat{\Gamma}_{r^\prime r}(\vpp,\vp) \,, \nonumber\\
    &= \sum_{r^\prime,r} \Ctpt_{NN}^{r^\prime r}(\OO;\vpp,\vp;\tf,\tc) \,\hat{\Gamma}_{r^\prime r}(\vpp,\vp) \,,
\end{align}
where
\begin{align}
    \hat{\Gamma}_{r^\prime r}(\vpp,\vp)= c_\alpha^{r^\prime}(\vpp) [\gamma_4\Gamma]_{\alpha\beta} [c_\beta^{r}(\vp)]^* \,.
\end{align}

\section{Data symmetrization}\label{app:dsbs}
In this work, data are random variables with their expectation values giving 2-point and 3-point functions.
Data related by symmetry share the same expectation value, and can be averaged to effectively increase the statistics. 
There are various transformations, including the flavor-exchanging ($u\leftrightarrow d$), translation, rotation, charge conjugation (C), parity (P), time reversal (T), and complex conjugation (c.c.).
While all of them are symmetries in the continuum, in the twisted-mass fermion formulation at a finite lattice spacing, $(u\leftrightarrow d)$, P, T are broken.
However, the twisted parity $\tilde{P}=(u\leftrightarrow d)\times P$, twisted time reversal $\tilde{T}=(u\leftrightarrow d)\times T$, and PT are still symmetries.
We symmetrize our data with these symmetry transformations.
Apart from the c.c. that will be explained in \cref{app:cc}, details of the other transformations can be found in, \eg, Ref.~\cite{Shindler:2007vp}.
Our data-symmetrization process are described below step by step.

In the first step, we average data at different source positions $\xsrc$ using translations. 

In the second step, we average data for different isospin sectors using the twisted parity $\tilde{P}$ transformation.
While we focus on data for the isospin $(I,I_3)=(\frac{1}{2},+\frac{1}{2})$ sector, data for the $(\frac{1}{2},-\frac{1}{2})$ sector also exist for the topologies including the connected piece `N', and will be averaged in this step.

In the third step, we average data for the forward and backward propagations using the PT transformation.
Data for backward propagation exist for all topologies including the connected pieces `N', `T', `Tf', `B2pt', `W2pt', `Z2pt'.

In the fourth step, we average data related by all 24 rotations of the octahedral group $\mathrm{O}$.
This step ensures that the data for the representative moving momentum $\vp=\vec{1}_z$ have effectively taken that for the other five directions into account.

In the fifth step, we perform the average based on c.c..
The c.c. transformation is to take complex conjugation on the 2-point and 3-point functions.
As explained in \cref{app:cc}, one can prove
\begin{align}\label{eq:ds_cc}
    \Cdpt_{jk}^r(\vp;t)&=[\Cdpt_{kj}^r(\vp;t)]^* \,,\nonumber\\
    \Ctpt_{jk}^{r^\prime r}(\OO;\vpp,\vp;\tf,\tc)&=s_\Gamma[\Ctpt_{kj}^{r r'}(\OO;\vpp,\vp;\tf,\tf-\tc)]^* \,,
\end{align}
where $s_\Gamma=\pm1$ is defined via
\begin{align}\label{eq:sO}
    \gamma_4\Gamma^\dagger\gamma_4=s_\Gamma\,\Gamma\,, \text{ for } \OO=\bar{\psi}\Gamma\psi \,.
\end{align}
In this step, we symmetrize the data to satisfy the above equalities.
As a result, the data of the 2-point function matrix is Hermitian. 

In the sixth step, we perform the average based on the composition of c.c. and CPT.
As explained in \cref{app:ccCPT}, for the operators used in this work, one can prove
\begin{align}\label{eq:ds_ccCPt}
    &\quad\Cdpt_{jk}^r(\vp;t)= [\Cdpt_{jk}^{\bar{r}}(\vp;t)]^* \,,\nonumber\\
    &\quad\Ctpt_{jk}^{r^\prime r}(\OO;\vpp,\vp;\tf,\tc) \nonumber\\
    &=(-1)^{r'+r}s_\Gamma s^\prime_\Gamma[\Ctpt_{jk}^{\bar{r}^{\prime} \bar{r}}(\OO;\vpp,\vp;\tf,\tc)]^* \,,
\end{align}
where $\bar{r}$ is defined as $\bar{0}=1$ and $\bar{1}=0$, and $s^\prime_\Gamma=\pm1$ is defined via
\begin{align}\label{eq:sO2}
    \Gamma_{CPT}^T\Gamma^T\Gamma_{CPT}^*=s^\prime_\Gamma\,\Gamma\,, \text{ for } \OO=\bar{\psi}\Gamma\psi \,,
\end{align}
where $\Gamma_{CPT}=\Gamma_{T}\Gamma_{P}\Gamma_{C}=\gamma_5\gamma_4\gamma_2$.
In this step, we symmetrize the data to satisfy the above equalities.
As a result, the data of the 2-point functions for the two irrep rows are complex conjugate to each other, and thus the GEVPs of the two irrep rows share the same eigenvalues and their eigenvectors can be chosen to be complex conjugate to each other.

In the seventh step, we remove the imaginary part of the 2-point functions at zero momentum.
The two irrep rows lie in the same irrep only for the zero total momentum when parity is broken as explained in \cref{app:pbreak}.
By Wigner-Eckart theorem, we have
\begin{align}\label{eq:WE0}
    \Cdpt_{jk}^0(\vec{0};t)=\Cdpt_{jk}^1(\vec{0};t) \,,
\end{align}
which, combined with \cref{eq:ds_ccCPt}, ensures the realness of 2-point functions.
As a result, the GEVP eigenvectors for zero total momentum can be made real, and thus equal between the two irrep rows.

\subsection{Complex conjugation}\label{app:cc}
The c.c. transformation is to take complex conjugation over the 2-point and 3-point functions. In Euclidean spacetime, since the time-dependence of an operator $O(t)$ (including both interpolating fields and insertion operators in this work) satisfies
\begin{align}
    O(t)=e^{Ht}O(0)e^{-Ht} \,.
\end{align} 
Because of such time dependence, the conjugation of the operator is $O^\dagger(-t)$ instead of $O(t)$.

For the 2-point function, we have
\begin{align}
  [\Cdpt_{jk}^{r}(\vp; t)]^*&=[\braket{\OJ^{r}_j(\vp; t) \,\, \OJ^{r;\dagger}_k(\vp;0)}]^* \nonumber\\
  &=\braket{\OJ^{r}_k(\vp; 0) \,\, \OJ^{r;\dagger}_j(\vp;-t)} \nonumber\\
  &=\braket{\OJ^{r}_k(\vp; t) \,\, \OJ^{r;\dagger}_j(\vp;0)} = \Cdpt_{kj}^{r}(\vp; t) \,,
\end{align}
where a temporal translation is used in the third equality.

For the 3-point function, we have
\begin{align}
  &\quad[\Ctpt^{r'r}_{jk}(\OO;\vpp,\vp;\tf,\tc)]^*  \nonumber\\
  &=[\braket{\OJ^{r'}_j(\vpp;\tf)\,\, \OO(\vec{q};\tc)\,\, \OJ^{r;\dagger}_k(\vp;0)}]^* \nonumber\\
  &=s_\Gamma\braket{\OJ^{r}_k(\vp;0)\,\, \OO(-\vec{q};-\tc)\,\, \OJ^{r';\dagger}_j(\vpp;-\tf)} \nonumber\\
  &=s_\Gamma\braket{\OJ^{r}_k(\vp;\tf)\,\, \OO(-\vec{q};\tf-\tc)\,\, \OJ^{r';\dagger}_j(\vpp;0)} \nonumber\\
  &=s_\Gamma\,\Ctpt^{r'r}_{kj}(\OO;\vp,\vpp;\tf,\tf-\tc) \,,
\end{align}
where a temporal translation is used in the third equality, and $s_\Gamma$ is defined in \cref{eq:sO}.

The c.c. transformation can also be derived in the path-integral formalism, where the conjugation of $O(t)$ in the path integral is just $O^\dagger(t)$. With the complex conjugation applied, one additionally applies the $\gamma_5$-hermiticity and time reversal transformation to reproduce the c.c. transformation derived above.

\subsection{Composition of complex conjugation and CPT}\label{app:ccCPT}
Applying the c.c. and CPT, the 2-point function transforms as
\begin{align}
  [\Cdpt_{jk}^{r}(\vp; t)]^*&=[\braket{\OJ^{r}_j(\vp; t) \; \OJ^{r;\dagger}_k(\vp;0)}]^* \nonumber\\
  &=\braket{\OJ^{r}_k(\vp; 0) \; \OJ^{r;\dagger}_j(\vp;-t)} \nonumber\\
  &=-\braket{\OJ^{r;\dagger}_j(\vp;-t) \; \OJ^{r}_k(\vp; 0)} \nonumber\\
  &=\braket{\OJ^{\bar{r}}_j(\vp;t) \; \OJ^{\bar{r};\dagger}_k(\vp; 0)} = \Cdpt_{jk}^{\bar{r}}(\vp; t) \,,
\end{align}
where in the third equality the minus sign comes from exchanging two fermionic interpolators ($N$ or $N\pi$), and in the fourth equality we apply the CPT transformation:
\begin{align}
    \OJ^{r;\dagger}_j(\vp;-t) &\xrightarrow{CPT} (-1)^{r+1}\OJ^{\bar{r}}_j(\vp;t) \,,\nonumber\\
    \OJ^{r}_k(\vp; 0) &\xrightarrow{CPT} (-1)^r [\OJ^{\bar{r}}_k(\vp; 0)]^\dagger \,,
\end{align}
where $\bar{r}$ is defined as $\bar{0}=1$ and $\bar{1}=0$. The above transformation rule holds for the interpolators defined in \cref{app:interfields} and the CPT transformation defined in Ref.~\cite{Shindler:2007vp}. We note a constant complex phase factor for the interpolators can spoil the above transformation rule because of the antilinearity of the c.c. transformation. 

Similarly, the 3-point function transforms as
\begin{align}
  &\quad[\Ctpt^{r'r}_{jk}(\OO;\vpp,\vp;\tf,\tc)]^*  \nonumber\\
  &=[\braket{\OJ^{r'}_j(\vpp;\tf)\,\, \OO(\vec{q};\tc)\,\, \OJ^{r;\dagger}_k(\vp;0)}]^* \nonumber\\
  &=s_\Gamma \braket{\OJ^{r}_k(\vp;0)\,\, \OO(-\vec{q};-\tc)\,\, \OJ^{r';\dagger}_j(\vpp;-\tf)} \nonumber\\
  &=-s_\Gamma \braket{\OJ^{r';\dagger}_j(\vpp;-\tf)]\,\, \OO(-\vec{q};-\tc)\,\, \OJ^{r}_k(\vp;0)} \nonumber\\
  &=(-1)^{r'+r}s_\Gamma s_\Gamma^\prime \braket{\OJ^{\bar{r}'}_j(\vpp;\tf)\,\, \OO(\vec{q};\tc)\,\, \OJ^{\bar{r};\dagger}_k(\vp;0)} \nonumber\\
  &=(-1)^{r'+r}s_\Gamma s_\Gamma^\prime \;\Ctpt^{\bar{r}'\bar{r}}_{jk}(\OO;\vpp,\vp;\tf,\tc) \,,
\end{align}
where the CPT transformation rule for the insertion operator
\begin{align}
    \OO(\vec{q};t) &\xrightarrow{CPT} s_\Gamma^\prime \OO(\vec{q};-t) 
\end{align}
is used in the fourth equality with $s_\Gamma^\prime$ defined in \cref{eq:sO2}.

\subsection{Overlapping factor}\label{app:of}
\newcommand{\ofz}{\mathcal{Z}}
The overlapping factor
\begin{align}
    \ofz_{\alpha,s}(\vp)&=\braket{N(p,s)|[\JN(\vp;0)]^\dagger_\alpha|\Omega} \,,
\end{align}
satisfies 
\begin{align}
    \ofz_{0,\downarrow_z}(\vp)&=\ofz_{1,\uparrow_z}(\vp)=0 \,. \nonumber\\
    \ofz_{0,\uparrow_z}(\vec{0})&=\ofz_{1,\downarrow_z}(\vec{0}) \,,
\end{align}
where $\uparrow_z$ ($\downarrow_z$) denoting the spin up (down) along z-direction.
One cannot, however, directly get $\ofz_{0,\uparrow_z}(\vp)=\ofz_{0,\downarrow_z}(\vp)$ for $\vp\neq\vec{0}$ because of the parity breaking in twisted-mass fermion formulation.
To prove that, we proceed as follows.
The factor appears in the asymptotic limit of the 2-point function:
\begin{align}
    &\quad\braket{[\OJ_N(\vp;t)]_\alpha[{\OJ}_N(\vp;0)]^\dagger_\beta}  \nonumber\\
    &\xrightarrow{t\to\infty} \sum_s [\ofz_{\alpha,s}(\vp)]^* \ofz_{\beta,s}(\vp) e^{-E_{N}(\vp)\,t} \,.
\end{align}
With the complex conjugation transformation explained in \cref{app:cc}, one has
\begin{align}
\braket{[\OJ_N(\vp;t)]_0[{\OJ}_N(\vp;0)]^\dagger_0}=\braket{[\OJ_N(\vp;t)]_0[{\OJ}_N(\vp;0)]^\dagger_0}^* \,.
\end{align}
With the composed transformation explained in \cref{app:ccCPT}, one has
\begin{align}
\braket{[\OJ_N(\vp;t)]_0[{\OJ}_N(\vp;0)]^\dagger_0}=\braket{[\OJ_N(\vp;t)]_1[{\OJ}_N(\vp;0)]^\dagger_1}^* \,.
\end{align}
Combine the above two equations, one gets
\begin{align}
\braket{[\OJ_N(\vp;t)]_0[{\OJ}_N(\vp;0)]^\dagger_0}=\braket{[\OJ_N(\vp;t)]_1[{\OJ}_N(\vp;0)]^\dagger_1} \,.
\end{align}
Then with the asymptotic limit, one gets
\begin{align}
    |\ofz_{0,\uparrow_z}(\vp)|=|\ofz_{1,\downarrow_z}(\vp)| \,.
\end{align}
In addition, apart from the convention $\braket{N(p,s)|N(p,s)}=1$, one has the freedom to choose the phase of $\ket{N(p,\uparrow_z)}$ and $\ket{N(p,\downarrow_z)}$ independently for $\vec{p}\neq\vec{0}$ since they are not connected by any symmetry transformation in the twisted-mass formulation.
With proper phase convention, one gets
\begin{align}
    \ofz_{0,\uparrow_z}(\vec{p})&=\ofz_{1,\downarrow_z}(\vec{p}) \,,
\end{align}
with both of them real numbers.

\section{Broken symmetries in twisted-mass fermion formulation}\label{app:bstma}

In the twisted-mass fermion formulation, isospin and parity symmetries are broken at a finite lattice spacing.
In Ref.~\cite{Frezzotti:2003ni,Frezzotti:2004wz}, it was proved that renormalized correlation functions suffer lattice discretization artefact of $\mathcal{O}(a^2)$ for twisted-mass actions at maximal twist.
Therefore, the breaking of these symmetries should only cause $\mathcal{O}(a^2)$ lattice artefacts for renormalized correlation functions.

However, for finite lattice spacing one could investigate the size of these cut-off effects. If these are large, including them can lead to a smoother continuum extrapolation. 
Indeed, in  Ref.~\cite{ETM:2008zte},  the inclusion of the disconnected contributions arising from  with pion loops to the  pion 2-point function, was shown to  significantly reduce the lattice artefacts, despite the fact the  such contributions was shown to be an $\mathcal{O}(a^2)$ lattice artefact \cite{Jansen:2005cg,Shindler:2007vp}.

In this appendix we investigate the size of disconnected quark loops to isovector form factors.

\subsection{Isospin symmetry: neutral pion and isovector insertion loops}\label{app:iso}
With an action where the isospin symmetry is exact, the $\pi^0$ loop would vanish.
The twisted-mass action, however, breaks isospin symmetry at a finite lattice spacing.
The isospin symmetry between 2-point functions of the neutral and charged pions has been studied long ago (see, \eg, Ref.~\cite{ETM:2008zte}).
Although it was proved \cite{Frezzotti:2003ni,Frezzotti:2004wz} that for the pion 2-point functions, the nonvanishing value of the $\pi^0$ loop is a lattice artefact at $\mathcal{O}(a^2)$, it still makes a large contribution numerically, and gives an important contribution that to the $\pi^0$ mass bringing closer  to the approximate-isospin-symmetric equality between the masses of the neural and charged pions.
Here we consider the $N \to N\pi$ correlation functions, and study the importance of  $\pi^0$-loop contributions to restoring the isospin symmetry.

The isospin $1/2$ and $3/2$ pion-nucleon interpolators $\OJ^{1/2}_{N\pi}$ and $\OJ^{3/2}_{N\pi}$ are related to the interpolators $\OJ_{n\,\pi^+}$ and $\OJ_{p\,\pi^0}$ via
\begin{align}
    \begin{bmatrix}
         \OJ^{1/2}_{N\pi} \\
         \OJ^{3/2}_{N\pi} \\
    \end{bmatrix} &=
    \begin{bmatrix}
        \sqrt{2/3}\;\; & -\sqrt{1/3} \\
        \sqrt{1/3}\;\; & \sqrt{2/3} \\
    \end{bmatrix}
    \begin{bmatrix}
        \OJ_{n\,\pi^+} \\
        \OJ_{p\,\pi^0} \\
   \end{bmatrix} \,,
\end{align}
or equivalently
\begin{align}
    \begin{bmatrix}
        \OJ_{n\,\pi^+} \\
        \OJ_{p\,\pi^0} \\
    \end{bmatrix}&=
    \begin{bmatrix}
        \sqrt{2/3}\;\; & \sqrt{1/3} \\
        -\sqrt{1/3}\;\; & \sqrt{2/3} \\
    \end{bmatrix}
    \begin{bmatrix}
        \OJ^{1/2}_{N\pi} \\
        \OJ^{3/2}_{N\pi} \\
   \end{bmatrix} \,.
\end{align}
Since the single proton interpolator $\OJ_{p}$ has the isospin $1/2$, assuming isospin symmetry, one would have
\begin{align}
    \braket{\OJ_p\;\OJ^\dagger_{n\,\pi^+}} &= \sqrt{2/3} \braket{\OJ_p\;\OJ^{1/2;\dagger}_{N\pi}} \,, \nonumber\\
    \braket{\OJ_p\;\OJ^\dagger_{p\,\pi^0}} &= -\sqrt{1/3} \braket{\OJ_p\;\OJ^{3/2;\dagger}_{N\pi}} \,, 
\end{align}
and therefore the isospin-symmetry relation
\begin{align}\label{eq:eqiso}
    \braket{\OJ_p\;\OJ^\dagger_{n\,\pi^+}}=-\sqrt{2} \braket{\OJ_p\;\OJ^\dagger_{p\,\pi^0}} \,.
\end{align}

\begin{figure}[!ht]
    \centering
    \includegraphics[width=\columnwidth]{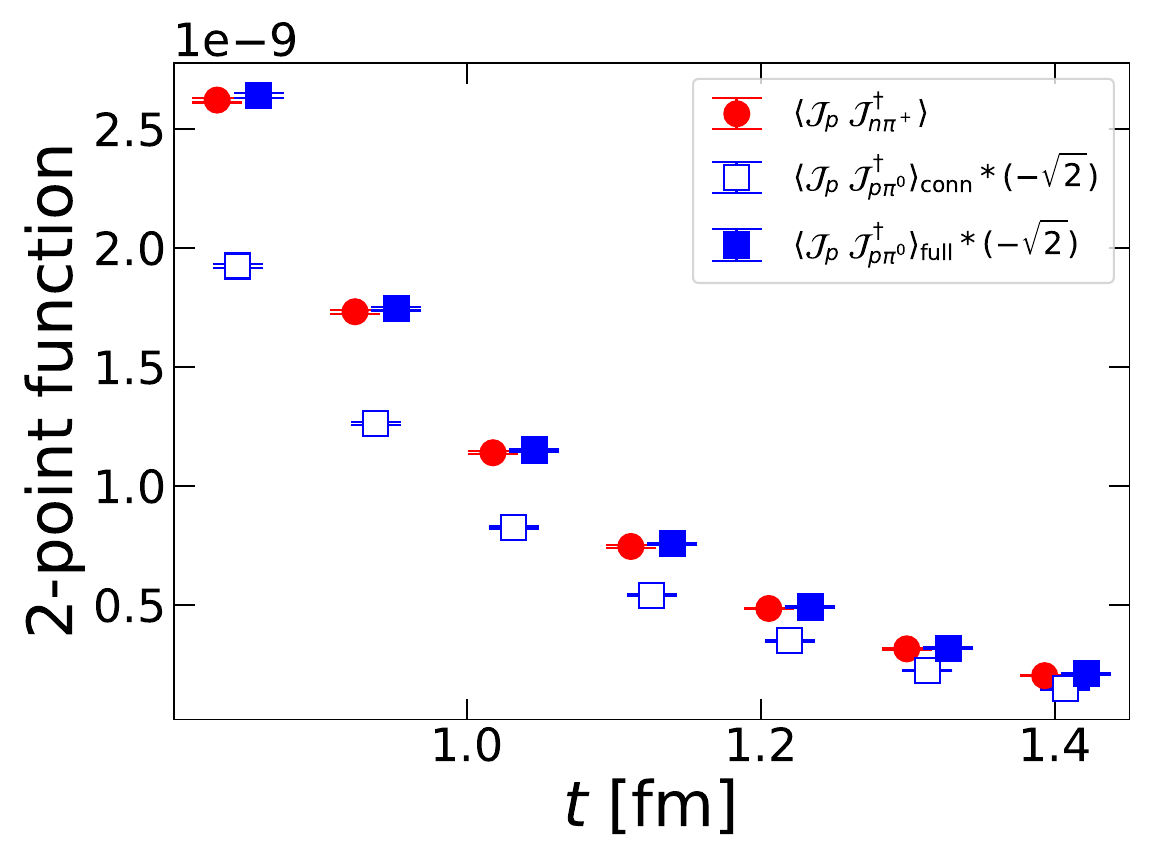}
    \caption{Ensemble  cA2.09.48. Comparison between 2-point functions related by isospin symmetry. In the $p\,\pi^0$ case, `conn' indicates only the `T' topology has been included, and `full' indicates both `T' and `N-\pii' topologies have been included. In the $n\,\pi^+$ case, there is only the `T' topology.}\label{fig:symmetry_isospin}
\end{figure}
In \cref{fig:symmetry_isospin}, we compare the numerical results between $ \braket{\OJ_p\;\OJ^\dagger_{n\,\pi^+}}$ and $\braket{\OJ_p\;\OJ^\dagger_{p\,\pi^0}}$. 
In the $n\,\pi^+$ case, there is only one topology `T', while in the other case, there are both `T' and `N-\pii'.
We present the results for both the connected $p\,\pi^0$ case (with `T'), and the full $p\,\pi^0$ case (with both `T' and `N-\pii').
One clearly sees that contribution from diagrams with the pion loop is important for \cref{eq:eqiso} to hold approximately.

\begin{figure*}[!ht]
    \centering
    \includegraphics[width=\textwidth]{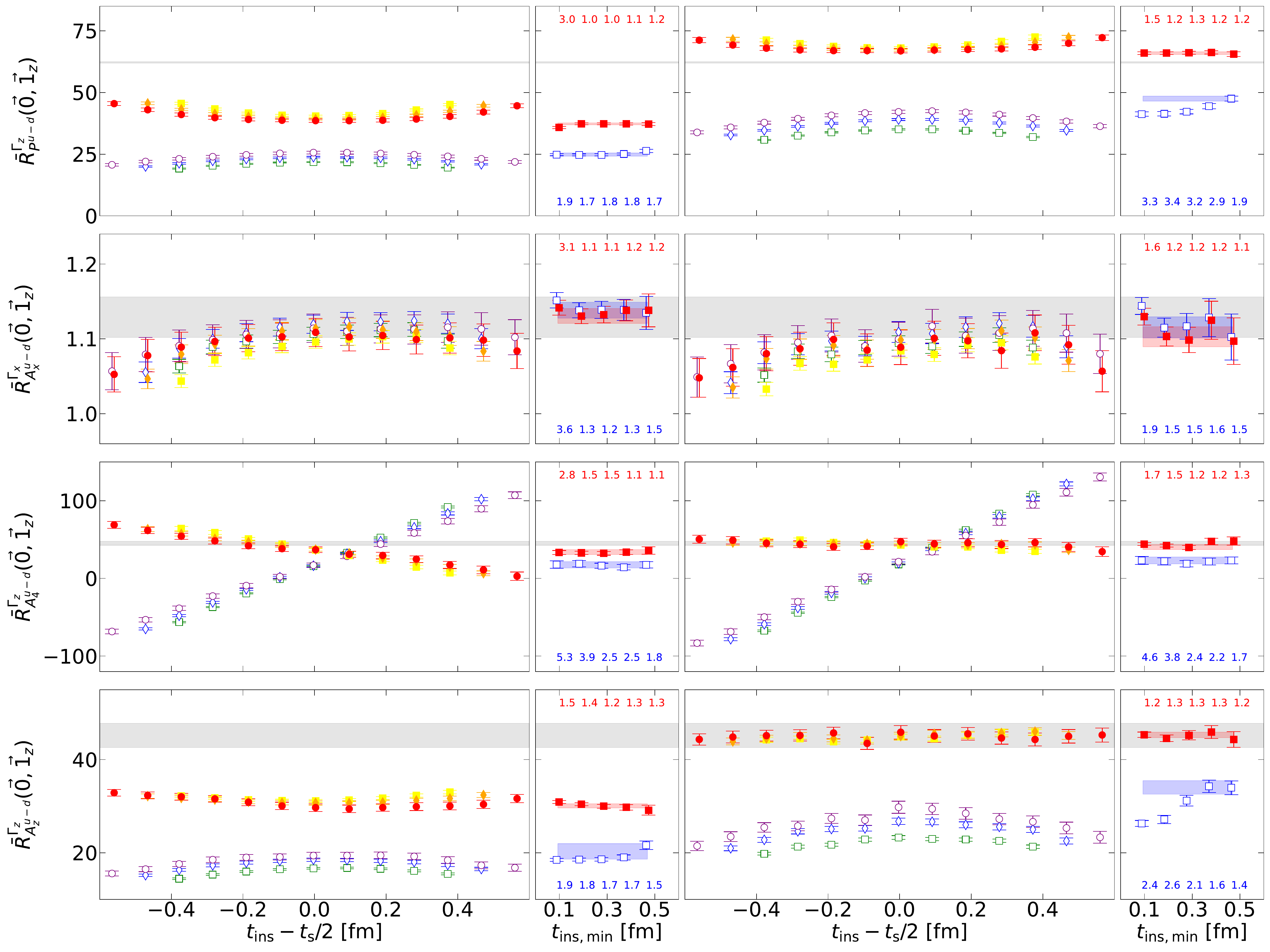}
    \includegraphics[width=\textwidth]{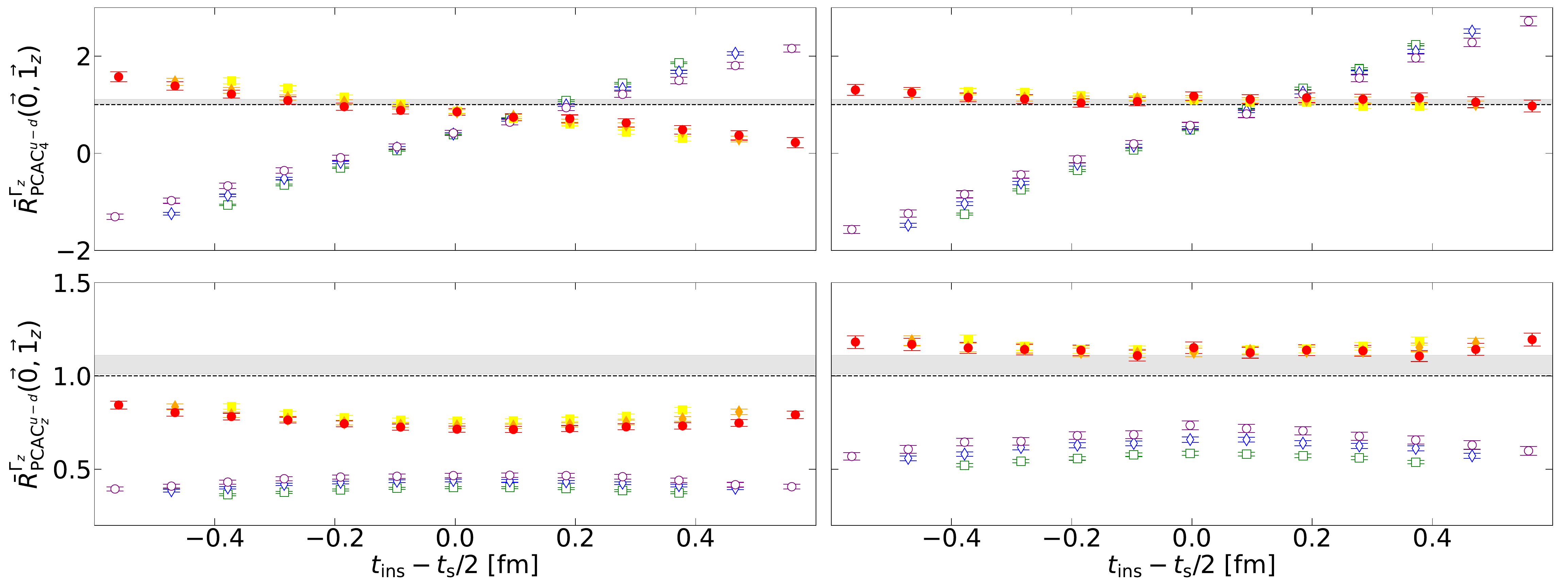}
    \caption{Ensemble  cA2.09.48. The right panels are exactly the same as in \cref{fig:3pt_raw_sep_5APP_48}. The left panels are the same plots without including any topologies containing the insertion loop `J'.}\label{fig:3pt_raw_sep_5APP_48_noJ_all}
\end{figure*}

Like the $\pi^0$ loop, the insertion loop in the isovector case is nonvanishing on a twisted mass lattice.
In \cref{fig:3pt_raw_sep_5APP_48_noJ_all}, we show a comparison for all PCAC-relevant cases between results without (left) and with (right) the insertion loop. 
We first look at those blue open points that are without GEVP improvement. The 3-point functions used in these cases include only $\NJN$. From \cref{fig:diags_3pt_Id}, it has only two topologies: `NJN' and `N-j', and the latter one involves an insertion loop.
The second row in \cref{fig:3pt_raw_sep_5APP_48_noJ_all} for $G_A$ does not receive significant contribution from the insertion loop, and gets only noisier after including it. 
The first for $G_5$, the third and fourth for $G_P$, and the last two for $r_{\mathrm{PCAC}}$, do receive significant contribution from the insertion loop. For the $G_5$ case, its values get almost doubled after including the insertion loop.
It is also clear that the first, fourth and last rows get closer to the grey bands, which are results from continuum extrapolation. 

Then we look at those filled points that are after GEVP improvement.
As shown in \cref{fig:diags_3pt_Id}, we have a lot more topologies with an insertion loop for $\NJNpi$ and $\NpiJN$.
In the second row for $G^{u-d}_A$, even before including the insertion loop, it agrees with the continuum-limit result already. So in this case, the insertion loop does not help.
In the third and fourth rows for $G^{u-d}_P$, only with the insertion loop, it agrees with the continuum-limit result. So in this case, the insertion loop removes almost all the lattice artefact.
In the first row for $G^{u-d}_5$, with the insertion loop, it gets much closer to but still disagrees with the continuum-limit result. So in this case, the insertion loop removes most of the lattice artefact, but there still seems to be a much smaller residual lattice artefact.
We emphasize that the ensembles used in this work are coarser than all the ensembles used in Ref.~\cite{Alexandrou:2023qbg} for the continuum extrapolation.
Furthermore, in the third, fourth and fifth rows, although the GEVP reduces a lot of the contaminations for both cases without and with the insertion loop, there are relatively larger contaminations left in the case without the insertion loop.
This indicates that, the removal of the contaminations does not happen separately for topologies without and with the insertion loop.

\subsection{Parity: asymmetry between irrep rows}\label{app:pbreak}
The spherical symmetry group $\mathrm{O(3)}$ is isomorphic to $\mathrm{SO(3)}\times\mathrm{C}^P_2$ where the $\mathrm{C}^P_2$ is the cyclic group generated by the parity transformation. The relation can be formally written as
\begin{align}
    \begin{bmatrix} 
        &&\\
        &\text{ O(3)}&\\
        &&
    \end{bmatrix}
    &=
    \begin{bmatrix}
        &&\\
        &\text{ SO(3)}&\\
        && 
    \end{bmatrix} \nonumber\\
    &\hspace{0.2cm}\text{\Large$\times$}
    \left\{
        \begin{bmatrix} 1&&\\ &1&\\ &&1 \end{bmatrix}
        \,,
        \begin{bmatrix} -1&&\\ &-1&\\ &&-1 \end{bmatrix}
    \right\}
    \,.
\end{align}
Most of the elements in O(3) do not leave lattice sites invariant, and are not symmetries on the lattice.
The subgroup of $\mathrm{O(3)}$ with elements leaving the lattice invariant is labelled as $\mathrm{O}_h=\mathrm{O}\times\mathrm{C}^P_2$, where $\mathrm{O}$ is the octahedral group of dimension $24$.

Because we work with interpolating fields with a fixed total momentum $\vp$.
We also consider the subgroup of $\mathrm{O}_h$ with elements leaving $\vp$ invariant.
Such subgroup is called a little group of $\mathrm{O}_h$.
For $\vp=\vec{0}$, the little group is the entire $\mathrm{O}_h$ group.
For $\vp\neq\vec{0}$, it is useful to consider the $\mathrm{O(2)}$ group since the little group is also always a subgroup of $\mathrm{O(2)}$. 
The group $\mathrm{O(2)}$ is isomorphic to the semidirect product $\mathrm{SO(2)}\rtimes\mathrm{C}^r_2$ formally written as
\begin{align}
    \begin{bmatrix}
        \text{ O(2)} & \begin{matrix}\\\\\end{matrix} \\
        \begin{matrix}&&&&\end{matrix} &	1	\\
    \end{bmatrix}
    &=
    \begin{bmatrix}
        \text{ SO(2)} & \begin{matrix}\\\\\end{matrix} \\
        \begin{matrix}&&&&\end{matrix} &	1	\\
    \end{bmatrix} \nonumber\\
    &\hspace{0.5cm}
    \text{\Large$\rtimes$}\left\{
    \begin{bmatrix}
        1&&\\
        &1&\\
        &&1
    \end{bmatrix}
    ,
    \begin{bmatrix}
        1&&\\
        &-1&\\
        &&1
    \end{bmatrix}
    \right\}
        \,,
\end{align}
where we choose the third dimension of the above matrices to be the direction of $\vp$ rather than the lattice $z$-direction, and the $\mathrm{C}^r_2$ is generated by a reflection that flips a direction perpendicular to $\vp$. 
For $\vp=\vec{1}_z$, the little group is $\mathrm{Dih}_{4}=\mathrm{C}_4\rtimes\mathrm{C}^r_2$. 

As explained in \cref{app:interfields}, in this work, for $\vp=\vec{0}$ we consider the irrep $\Lambda=G_{1g}$ of the little group $\mathrm{O}_h$, while for $\vp=\vec{1}_z$ we consider the irrep $\Lambda=G_1$ of the little group $\mathrm{Dih}_{4}$.
Both irreps are two-dimentional with irrep rows $r=0,1$.
In the twisted-mass fermion formulation, the party symmetry is broken. The little group $\mathrm{O}_h$ reduces to $\mathrm{O}$, while $\mathrm{Dih}_{4}$ reduces to $\mathrm{C}_4$.
A major difference for $\vp=\vec{0}$ and $\vec{1}_z$ is that, $r=0$ and $1$ still lie in a single two-dimensional irrep of $\mathrm{O}$, while they split into two one-dimensional irreps of $\mathrm{C}_4$.
In the latter case, the symmetry between two different irrep rows are no longer protected by Wigner-Eckart theory, \eg, a similar equality with \cref{eq:WE0} for $\vp=\vec{1}_z$ does not hold.
As a result, one cannot use \cref{eq:ds_ccCPt} to further prove that the 2-point function is real.

\begin{figure}[!ht]
    \centering
    \includegraphics[width=\columnwidth]{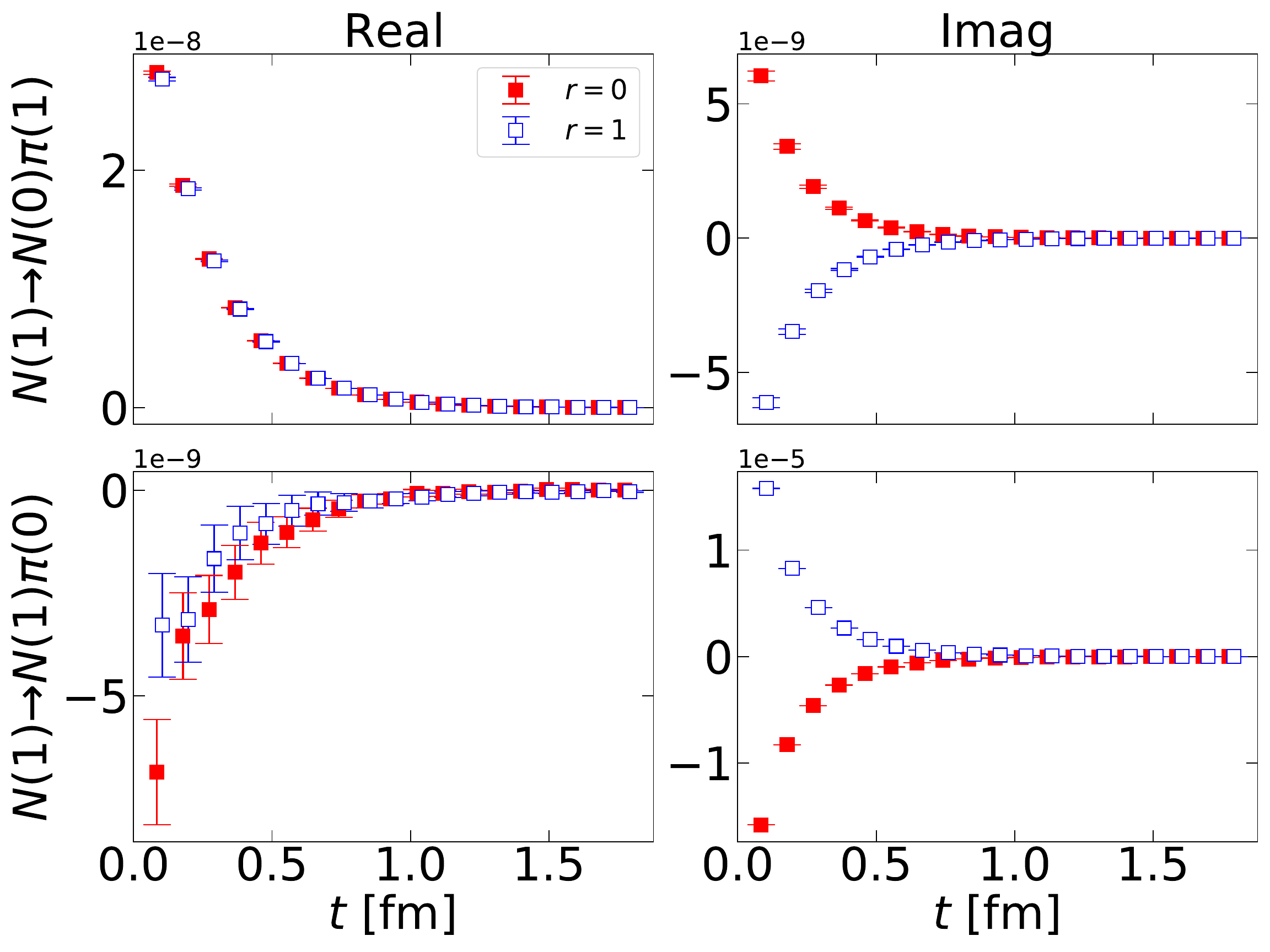}
    \caption{Ensemble  cA2.09.48. Comparison of both real (left) and imaginary (right) parts of the 2-point functions $N(1)\to N(0)\pi(1)$ (top) and $N(1)\to N(1)\pi(0)$ (bottom) between $\lambda=1$ (filled points) and $\lambda=2$ (open points) irrep vectors. While having equal real parts, their imaginary parts are nonzero and opposite to each other.}\label{fig:ccl1l2}
\end{figure}

In fact, moving-frame 2-point functions do acquire nonzero imaginary part in this way as illustrated in \cref{fig:ccl1l2}.
In the $N(1)\to N(0)\pi(1)$ case, the imaginary part is smaller than but comparable with the real part.
In the $N(1)\to N(1)\pi(0)$ case, the imaginary part receives huge contribution from the expectation value of the $\pi^0$ loop, and is few orders of magnitude larger than the real part.
In either case, it is important to include the imaginary part into the analysis, and treat the two irrep rows differently.

\section{Proof of \cref{eq:Id-2}}\label{app:Idproof}
\newcommand{\cpl}{\sum_k}

As in the main text, we assume the model situation, that the operator set entering the GEVP is a basis, i.e. that there are
as many independent operators as there are states, which contribute significantly.
In this work, we include $2$ interpolators for $\vp=\vec{0}$ and $3$ interpolators for $\vp=\vec{1}_z$, so we assume that there are only $2$ or $3$ states for each momentum sector that contributes significantly.

To be clear, we note that we use the letters $m,n$ for indices of eigenstates of the system with $n=0$ for the ground nucleon state 
(with any momentum), and we use the letters $j,k$ for indices of interpolators used in this work with $k=N$ for the single nucleon interpolator.
We first note
\begin{align}
    W&=\frac{1}{v_{0N}\,[v^{-1}]_{N0}}-1 = \cpl \frac{v_{0k}\,[v^{-1}]_{k0}}{v_{0N}\,[v^{-1}]_{N0}}-1 \nonumber\\
    &= \cpl \frac{v_{0k}\,Z_{k0}}{v_{0N}\,Z_{N0}}-1 = \frac{A_0}{v_{0N}\,Z_{N0}}-1 \,,
\end{align}
where we have suppressed all the $\vp$ arguments for simplicity, and the third equality follows from \cref{eq:viz}.
Then we have
\begin{align}
    1-W^*(\vpp)\,W(\vp) &= \frac{A_{0}^*(\vpp)}{v_{0N}^*(\vpp)\,Z^*_{N0}(\vpp)} + \frac{A_{0}(\vp)}{v_{0N}(\vp)\,Z_{N0}(\vp)} \nonumber\\
    & - \frac{A_{0}^*(\vpp)}{v_{0N}^*(\vpp)\,Z^*_{N0}(\vpp)}\frac{A_{0}(\vp)}{v_{0N}(\vp)\,Z_{N0}(\vp)} \nonumber\\
    1+W^*(\vpp)&=\frac{A_{0}^*(\vpp)}{v_{0N}^*(\vpp)\,Z^*_{N0}(\vpp)} \nonumber\\
    1+W(\vp)&=\frac{A_{0}(\vp)}{v_{0N}(\vp)\,Z_{N0}(\vp)} \,.
\end{align}
Expressing $I_d$ in the following sum:
\begin{align}
    I_d&=\sum_{j,k}\sum_{m,n}d_{jk}\,v_{0j}^*(\vpp)\,v_{0k}(\vp)\,Z_{jm}^*(\vpp)\,Z_{kn}(\vp)  \nonumber \\
    &\quad \times \braket{m(\vpp)|\OO|n(\vp)} e^{-E_{m}(\vpp)(\tf-\tc)}e^{-E_{n}(\vp)\tc} \nonumber\\
    &\equiv \sum_{m,n} I_{d;mn} \braket{m(\vpp)|\OO|n(\vp)} e^{-E_{m}(\vpp)(\tf-\tc)}e^{-E_{n}(\vp)\tc} \,,
\end{align}
the proof proceeds as follows:
{\allowdisplaybreaks
\begin{widetext}
\begin{align}
    I_{d;mn}=&\sum_{j,k}d_{jk}\,v_{0j}^*(\vpp)\,v_{0k}(\vp)\,Z_{jm}^*(\vpp)\,Z_{kn}(\vp) \nonumber\\
    =&\left[1-W^*(\vpp)\,W(\vp)\right]\,v_{0N}^*(\vpp)\,v_{0N}(\vp)\,Z_{Nm}^*(\vpp)\,Z_{Nn}(\vp) \nonumber\\
    &\hspace{1cm}+\left[1+W^*(\vpp)\right]\sum_{k\neq N}^{}v_{0N}^*(\vpp)\,v_{0k}(\vp)\,Z_{Nm}^*(\vpp)\,Z_{kn}(\vp) +\left[1+W(\vp)\right]\sum_{j\neq N}^{}v_{0j}^*(\vpp)\,v_{0N}(\vp)\,Z_{jm}^*(\vpp)\,Z_{Nn}(\vp) \nonumber\\
    =&\left[1-W^*(\vpp)\,W(\vp)\right]\,v_{0N}^*(\vpp)\,v_{0N}(\vp)\,Z_{Nm}^*(\vpp)\,Z_{Nn}(\vp) \nonumber\\
    &\hspace{2cm}+\left[1+W^*(\vpp)\right]v_{0N}^*(\vpp)\,Z_{Nm}^*(\vpp)\left[A_{0}(\vp)\,\delta_{0n}-v_{0N}(\vp)\,Z_{Nn}(\vp)\right] \nonumber\\
    &\hspace{2cm}+\left[1+W(\vp)\right]\,v_{0N}(\vp)\,Z_{Nn}(\vp)\left[A_{0}^*(\vpp)\,\delta_{0m}-v_{0N}^*(\vpp)\,Z_{Nm}^*(\vpp)\right] \nonumber\\
    =& \left[{\frac{A_{0}^*(\vpp)}{Z^*_{N0}(\vpp)}v_{0N}(\vp)\,Z_{Nm}^*(\vpp)\,Z_{Nn}(\vp)} + {\frac{A_{0}(\vp)}{Z_{N0}(\vp)}v_{0N}^*(\vpp)\,Z_{Nm}^*(\vpp)\,Z_{Nn}(\vp)} -\frac{A_{0}^*(\vpp)}{Z^*_{N0}(\vpp)}\frac{A_{0}(\vp)}{Z_{N0}(\vp)}Z_{Nm}^*(\vpp)\,Z_{Nn}(\vp)\right] \nonumber\\
    &\hspace{2cm} +\left[{\frac{A_{0}^*(\vpp)}{Z^*_{N0}(\vpp)}Z_{Nm}^*(\vpp)A_{0}(\vp)\delta_{0n}}-{\frac{A_{0}^*(\vpp)}{Z^*_{N0}(\vpp)}Z_{Nm}^*(\vpp)v_{0N}(\vp)\,Z_{Nn}(\vp)}\right]\nonumber\\
    &\hspace{2cm}+\left[\frac{A_{0}(\vp)}{Z_{N0}(\vp)}Z_{Nn}(\vp)\,A_{0}^*(\vpp)\delta_{0m} - {\frac{A_{0}(\vp)}{Z_{N0}(\vp)}Z_{Nn}(\vp)\,v_{0N}^*(\vpp)\,Z_{Nm}^*(\vpp)}\right] \nonumber\\
    =&\frac{A_{0}^*(\vpp)}{Z^*_{N0}(\vpp)}\frac{A_{0}(\vp)}{Z_{N0}(\vp)}\left[Z_{Nm}^*(\vpp)\,Z_{N0}(\vp)\,\delta_{0n}+Z_{N0}^*(\vpp)\,Z_{Nn}(\vp)\,\delta_{0m}-Z_{Nm}^*(\vpp)\,Z_{Nn}(\vp)\right] \nonumber\\
    =&\frac{A_{0}^*(\vpp)}{Z^*_{N0}(\vpp)}\frac{A_{0}(\vp)}{Z_{Nn}(\vp)}Z_{Nm}^*(\vpp)\,Z_{Nn}(\vp)\,(\delta_{0n}+\delta_{0m}-1) = Z_{Nm}^*(\vpp)\,Z_{Nn}(\vp)\,(\delta_{0n}+\delta_{0m}-1) \,,
\end{align}
\end{widetext}
where the last step is the only place where the normalization convention \cref{eq:nmlz} is used.
This directly leads to \cref{eq:Id-2}.
}

\section{Decomposition of nucleon matrix elements}\label{app:deNME}
We first show that the ratio of 3-point to 2-point function converges to the nucleon matrix element as described by \cref{eq:ratio0}.
With a proper phase convention as explained in \cref{app:of}, we have the overlapping factor
\begin{align}
    \ofz_{\alpha,s}(\vp)&=\braket{N(p,s)|[\JN(\vp;0)]^\dagger_\alpha|\Omega} \,, \nonumber\\
    \ofz_{0,\downarrow_z}(\vp)&=\ofz_{1,\uparrow_z}(\vp)=0 \,. \nonumber\\
    \ofz_{0,\uparrow_z}(\vp)&=\ofz_{1,\downarrow_z}(\vp)\equiv \ofz(\vp) \,, \nonumber\\
    \ofz(\vp)&=\ofz(\vp)^*
\end{align}
where $\ket{\Omega}$ denotes the vacuum, $\uparrow_z$ ($\downarrow_z$) denotes the spin up (down) along z-direction.
Then the asymptotic limit for the standard 2-point ($t\to\infty$) and 3-point functions ($\tc,\tf-\tc\to\infty$) defined in \cref{eq:C2ptstd,eq:C3ptstd} are given by 
\begin{widetext}
\begin{align}
    &\quad \Cdpt_{NN}(\vp;t)=\text{Tr}\left[ \braket{\OJ_N(\vp;t)\bar{\OJ}_N(\vp;0)}\Gamma_0 \right]  \nonumber\\
    &\to \sum_{\alpha,\beta}\sum_s [\gamma_4\Gamma_{0}]_{\beta\alpha} \braket{\Omega|[\OJ_{N}(\vp;0)]_\alpha|N(p,s)}\braket{N(p,s)|[{\OJ}_{N}(\vp;0)]^\dagger_\beta|\Omega} e^{-E_{N}(\vp)\,t} \nonumber\\
    &= \sum_{\alpha,\beta} [\gamma_4\Gamma_{0}]_{\beta\alpha} \;\delta_{\alpha\beta} \;\ofz(\vp)^2 e^{-E_{N}(\vp)\,t} = \ofz(\vp)^2 e^{-E_{N}(\vp)\,t} \,,\nonumber\\
    &\quad \Ctpt_{NN}(\Gamma,\OO;\vpp,\vp;\tf,\tc)=\text{Tr}\left[ \braket{\OJ_N(\vpp;\tf)\OO(\vq;\tc)\bar{\OJ}_N(\vp;0)}\Gamma \right] \nonumber\\
    &\to \sum_{\alpha,\beta;s^\prime,s} [\gamma_4\Gamma]_{\beta\alpha} \braket{\Omega|[\OJ_{N}(\vpp;0)]_\alpha|N(p^\prime,s^\prime)} \braket{N(p^\prime,s^\prime)|\OO(\vec{q};0)|N(p,s)} \braket{N(p,s)|[{\OJ}_{N}(\vp;0)]^\dagger_\beta|\Omega} e^{-E_{N}(\vpp)(\tf-\tc)}e^{-E_{N}(\vp)\,\tc} \nonumber\\
    &= \sum_{\alpha,\beta;s^\prime,s} [\gamma_4\Gamma]_{\beta\alpha} \ofz_{\alpha,s^\prime}(\vpp) \ofz_{\beta,s}(\vp) \braket{N(p^\prime,s^\prime)|\OO(\vec{q};0)|N(p,s)} e^{-E_{N}(\vpp)(\tf-\tc)}e^{-E_{N}(\vp)\,\tc} \nonumber\\
    &= \sum_{s^\prime, s} \ofz(\vpp)\ofz(\vp)\; \tilde{\Gamma}_{s s^\prime} \braket{N(p^\prime,s^\prime)|\OO(\vq;0)|N(p,s)} e^{-E_{N}(\vpp)(\tf-\tc)}e^{-E_{N}(\vp)\,\tc} \,,
\end{align}
\end{widetext}
where
\begin{align}
    \tilde{\Gamma}_{s s^\prime}&=\sum_{\alpha,\beta} [\gamma_4\Gamma]_{\beta\alpha} \frac{\ofz_{\alpha,s^\prime}(\vpp) \ofz_{\beta,s}(\vp)}{\ofz(\vpp)\ofz(\vp)}   \,.
\end{align}
For $\Gamma=\Gamma_0$ and $\Gamma_k$, we have
\begin{align}
    \tilde{\Gamma}_0=\frac{1}{2}\mathbf{1}\,, \tilde{\Gamma}_k=\frac{1}{2}\sigma_k \,.
\end{align}
Then it is straightforward to get \cref{eq:ratio0}:
\begin{align}
    &\quad R^{\Gamma}_\OO(\vpp,\vp;\tf,\tc) \nonumber\\
    &\to \sum_{s^\prime, s} \tilde{\Gamma}_{s s^\prime} \braket{N(p^\prime,s^\prime)|\OO(\vq;0)|N(p,s)} \equiv \Pi^{\Gamma}_\OO(p^\prime,p) \,,
\end{align}

The nucleon matrix element can be written in terms of the Dirac spinor
\begin{align}
    \braket{N(p^\prime,s^\prime)|\OO(\vq;0)|N(p,s)}=\bar{u}_N(p^\prime,s^\prime)\Lambda_\OO(Q^2) u_N(p,s) \,,
\end{align}
where the convention for the Dirac spinor can be found in \cref{app:con}.
Then we have
\begin{align}
    &\quad\Pi^\Gamma_J(p^\prime,p) = \sum_{s^\prime, s} \tilde{\Gamma}_{s s^\prime} \braket{N(p^\prime,s^\prime)|\OO(\vq;0)|N(p,s)}\nonumber\\
    &=\text{Tr}\left[\Gamma \frac{-i\slashed{p}^\prime+m_N}{\sqrt{2E_N^\prime(E_N^\prime+m_N)}}  \Lambda_\OO(Q^2) \frac{-i\slashed{p}+m_N}{\sqrt{2E_N(E_N+m_N)}}  \right] \nonumber\\
    &\equiv\frac{1}{\sqrt{2E_N^\prime(E_N^\prime+m_N)}}\frac{1}{\sqrt{2E_N(E_N+m_N)}} M^\Gamma_J(p^\prime,p) \,,
\end{align}
with
\begin{align}
    M^\Gamma_\OO(p^\prime,p)&=\text{Tr}\left[\Gamma^\prime (-i\slashed{p}^\prime+m_N) \Lambda_\OO(Q^2) (-i\slashed{p}+m_N)  \right]\,.
\end{align}
Here $\Gamma^\prime=\sum_{s, s^\prime}\tilde{\Gamma}_{s s^\prime}\,u^s \bar{u}^{s^\prime}$ coincides with the $\Gamma$ defined previously in \cref{eq:defG}:
\begin{align}
    \Gamma_0^\prime&=\frac{1}{2}(u^{\uparrow_k}\bar{u}^{\uparrow_k}+u^{\downarrow_k}\bar{u}^{\downarrow_k}) = \frac{1+\gamma_4}{4} = \frac{1}{2}\begin{bmatrix}
        \mathbf{1} & \\
         & \mathbf{0}  \\
    \end{bmatrix}\,, \nonumber\\
    \Gamma_k^\prime&=\frac{1}{2}(u^{\uparrow_k}\bar{u}^{\uparrow_k}-u^{\downarrow_k}\bar{u}^{\downarrow_k}) = i\gamma_5\gamma_k\,\Gamma_0 = \frac{1}{2}\begin{bmatrix}
        \sigma_k & \\
         & \mathbf{0} \\
    \end{bmatrix} \,.
\end{align}
Therefore, we will suppress the prime for this new $\Gamma^\prime$ hereafter.

In the following subsections, we will consider the decompositions for different insertion operators $\OO$. 

\subsection{Scalar: \texorpdfstring{$S=\bar{\psi}\psi$}{Lg}}
For the scalar insertion $S=\bar{\psi}\psi$, we have
\begin{align}
    \Lambda_S(Q^2) &= G_S(Q^2) \,.
\end{align}
Then we get
\begin{align}
    M^{\Gamma_0}_{S}(p^\prime,p)&=\left[2m_N^2+\frac{Q^2}{2}-i\,m_NP_4\right] G_S(Q^2) \,,\nonumber\\
    M^{\Gamma_k}_{S}(p^\prime,p)&=\left[-i\,\varepsilon_{ijk4}\,p^\prime_i\,p_j\right] G_S(Q^2) \,.
\end{align}
When $\vpp=\vp=\vec{0}$, it reduces to
\begin{align}
    \Pi^{\Gamma_0}_{S} &= G_S(0) \equiv g_S \,, \nonumber\\
    \text{Others}&=0 \,.
\end{align}
When $\vpp=\vp$, it reduces to
\begin{align}
    \Pi^{\Gamma_0}_{S} &= \frac{m_N}{E_N} g_S \,, \nonumber\\
    \text{Others}&=0 \,.
\end{align}
When $\vpp=\vec{0}$, it reduces to
\begin{align}
    \Pi^{\Gamma_0}_{S} &= \mathcal{C}^{-1}(E_N+m_N)G_S(Q^2) \,, \nonumber\\
    \text{Others}&=0 \,,
\end{align}
where $\mathcal{C}=\sqrt{2E_N(E_N+m_N)}$.

\subsection{Vector: \texorpdfstring{$V_{\mu}=\bar{\psi}\gamma_\mu \psi$}{Lg}}
For the vector insertion $V_{\mu}=\bar{\psi}\gamma_\mu \psi$, we have
\begin{align}
    \Lambda_{V_\mu}(Q^2) &= \gamma_\mu F_1(Q^2) - \frac{i\sigma_{\mu\nu}q_\nu}{2m_N}F_2(Q^2) \,,
\end{align}
where a possible $q_\mu$ term is absent because it violates the Ward identity $q_\mu\,\Lambda_{V_\mu}=0$.
The electric and magnetic form factors $G_E$ and $G_M$ are linear combinations of $F_1$ and $F_2$:
\begin{align}
    G_E(Q^2)&=F_1(Q^2)-\frac{Q^2}{(2m_N)^2}F_2(Q^2) \,, \nonumber\\
    G_M(Q^2)&=F_1(Q^2)+F_2(Q^2) \,.
\end{align}
Then we get
\begin{widetext}
\begin{align}
    M^{\Gamma_0}_{V_\mu}(p^\prime,p) &=  \left[ -i m_N P_\mu -\frac{Q^2}{2} \delta_{4\mu}-(p^\prime_4\,p_\mu + p^\prime_\mu\,p_4) \right] F_1(Q^2) +\left[ \frac{i Q^2}{4m_N}P_\mu+\frac{1}{2}q_4\, q_\mu-\frac{Q^2}{2}\delta_{4\mu} \right] F_2(Q^2) \nonumber\\
    M^{\Gamma_k}_{V_\mu}(p^\prime,p) &=  \left[ i\varepsilon_{\rho\sigma k \mu}\,p^\prime_{\rho}\,p_\sigma-m_N\varepsilon_{i\mu k4}\,q_i \right] F_1(Q^2) +\left[ i\varepsilon_{\rho\sigma k\mu}\,p^\prime_\rho\,p_\sigma -m_N\varepsilon_{i\mu k 4}\,q_i + \frac{1}{2m_N}P_\mu\,\varepsilon_{ijk4}\,p^\prime_i\,p_j \right] F_2(Q^2) \,.
\end{align}
\end{widetext}
When $\vpp=\vp=\vec{0}$, it reduces to
\begin{align}
    \Pi^{\Gamma_0}_{V_4} &= F_1(0) = G_E(0) = N_q \,, \nonumber\\
    \text{Others}&=0 \,,
\end{align}
where $N_q$ is the number of quarks of flavor $q$ in the nucleon.
When $\vpp=\vp$, it reduces to
\begin{align}
    \Pi^{\Gamma_0}_{V_4} &= N_q \,, \nonumber\\
    \Pi^{\Gamma_0}_{V_k} &= -i\frac{p_k}{E_N}N_q \,, \nonumber\\
    \text{Others}&=0 \,.
\end{align}
When $\vpp=\vec{0}$, it reduces to
\begin{align}
    \Pi^{\Gamma_0}_{V_4} &=\mathcal{C}^{-1} [E_N+m_N]G_E(Q^2)\,, \nonumber\\
    \Pi^{\Gamma_0}_{V_k} &=\mathcal{C}^{-1} [-i\,q_k]G_E(Q^2) \,, \nonumber\\
    \Pi^{\Gamma_k}_{V_{i\perp k}} &=\mathcal{C}^{-1} [\varepsilon_{ijk4}\,q_j]G_M(Q^2)\,, \nonumber\\
    \text{Others}&=0 \,.
\end{align}

\subsection{Pseudoscalar: \texorpdfstring{$P=\bar{\psi}\gamma_5 \psi$}{Lg}}
For the pseudoscalar insertion $P=\bar{\psi}\gamma_5 \psi$, we have
\begin{align}
    \Lambda_P(Q^2) &= \gamma_5\,G_5(Q^2) \,.
\end{align}
Then we get
\begin{align}
    M^{\Gamma_0}_{P}(p^\prime,p)&=0 \,,\nonumber\\
    M^{\Gamma_k}_{P}(p^\prime,p)&=\left[-m_N\,q_k+i(p^\prime_4 p_k -p^\prime_k p_4)\right] G_5(Q^2) \,.
\end{align}
When $\vpp=\vp$, it reduces to
\begin{align}
    \text{All}&=0 \,.
\end{align}
When $\vpp=\vec{0}$, it reduces to
\begin{align}
    \Pi^{\Gamma_k}_{P} &= \mathcal{C}^{-1}[-q_k] G_5(Q^2) \,, \nonumber\\
    \text{Others}&=0 \,.
\end{align}

\subsection{Axial: \texorpdfstring{$A_\mu=\bar{\psi}\gamma_\mu\gamma_5 \psi$}{Lg}}
For the axial insertion $A_\mu=\bar{\psi}\gamma_\mu\gamma_5 \psi$, we have
\begin{align}
    \Lambda_{A_\mu}(Q^2) &= \gamma_\mu\gamma_5 \,G_A(Q^2) +\frac{i\,q_\mu}{2m_N} \gamma_5 \,G_P(Q^2) \,.
\end{align}
Then we get
\begin{widetext}
\begin{align}
    M^{\Gamma_0}_{A_\mu}(p^\prime,p)&=\left[\varepsilon_{ij\mu4}\,p^\prime_i\,p_j\right] G_A(Q^2) \,,\nonumber\\
    M^{\Gamma_k}_{A_\mu}(p^\prime,p)&=\left[ (2i\,m_N^2 + i\frac{Q^2}{2}+m_N\,P_4)\delta_{\mu k} -m_N\,P_k\,\delta_{4\mu} + i(p^\prime_k\,p_\mu +p^\prime_\mu\,p_k)\right]G_A(Q^2)  \nonumber\\
    &\quad +\frac{i\,q_\mu}{2m_N}\left[-m_N\,q_k+i(p^\prime_4 p_k -p^\prime_k p_4)\right] G_P(Q^2) \,.
\end{align}
\end{widetext}
When $\vpp=\vp=\vec{0}$, it reduces to
\begin{align}
    \Pi^{\Gamma_k}_{A_k} &= i\,G_A(0) \equiv i\,g_A \,, \nonumber\\
    \text{Others}&=0 \,.
\end{align}
When $\vpp=\vp$, it reduces to
\begin{align}
    \Pi^{\Gamma_k}_{A_4} &= -\frac{p_k}{E_N} g_A \,, \nonumber\\
    \Pi^{\Gamma_k}_{A_k} &= i\left( \frac{m_N}{E_N}+\frac{p_k^2}{E_N(E_N+m_N)} \right)g_A \,, \nonumber\\
    \Pi^{\Gamma_k}_{A_{i\perp k}} &= i\frac{p_k\,p_i+p_i\,p_k}{2E_N(E_N+m_N)} g_A \,, \nonumber\\
    \text{Others}&=0 \,.
\end{align}
When $\vpp=\vec{0}$, it reduces to
\begin{align}
    \Pi^{\Gamma_k}_{A_4} &=\mathcal{C}^{-1}\left[-q_k\,G_A(Q^2)+\frac{q_k(E_N-m_N)}{2m_N}G_P(Q^2)\right] \,, \nonumber\\
    \Pi^{\Gamma_k}_{A_k} &= \mathcal{C}^{-1}\left[i(E_N+m_N)G_A(Q^2)-i\frac{q_k^2}{2m_N}G_P(Q^2)\right] \,, \nonumber\\
    \Pi^{\Gamma_k}_{A_{i\perp k}} &=\mathcal{C}^{-1}\left[0-i\frac{q_i\,q_k}{2m_N}G_P(Q^2)\right] \,, \nonumber\\
    \text{Others}&=0 \,.
\end{align}

\subsection{Tensor: \texorpdfstring{$T_{\mu\nu}=\bar{\psi}\sigma_{\mu\nu}\psi$}{Lg}}
For the tensor insertion $T_{\mu\nu}=\bar{\psi}\sigma_{\mu\nu}\psi$, we have
\begin{align}
    \Lambda_{T_{\mu\nu}}(Q^2) &= \sigma_{\mu\nu}A_{T10}(Q^2) - i\frac{\gamma_{[\mu}q_{\nu]}}{2m_N}B_{T10}(Q^2) \nonumber\\
    &\hspace{2cm}- \frac{P_{[\mu}q_{\nu]}}{2m_N^2}\tilde{A}_{T10}(Q^2)  \,.
\end{align}
Then we get
\begin{widetext}
\begin{align}
    M^{\Gamma_0}_{T_{\mu\nu}}(p^\prime,p) &=  \left[-\frac{1}{2}P_{[\mu} q_{\nu]}  - i m_N\delta_{4[\mu}\,q_{\nu]}  \right] A_{T10}(Q^2) - \frac{i}{2 m_N}\left[ \left(-i m_N-\frac{1}{2}P_4\right) P_{[\mu}q_{\nu]} -\frac{Q^2}{2} \delta_{4[\mu}q_{\nu]} \right] B_{T10}(Q^2)  \nonumber\\
    & - \frac{1}{2m_N^2}\left[ 2m_N^2+\frac{Q^2}{2}-i\,m_N P_4 \right]P_{[\mu}q_{\nu]}\tilde{A}_{T10}(Q^2)\,, \nonumber\\
    M^{\Gamma_k}_{T_{\mu\nu}}(p^\prime,p) &= \left[i m_N^2 \varepsilon_{\mu\nu k4}+m_N\varepsilon_{\mu\nu k\rho}P_\rho +i\left( \varepsilon_{i\nu k4}p^\prime_{\mu}p_{i} - \varepsilon_{i\mu k4}p^\prime_{\nu}p_{i} - \varepsilon_{i\mu\nu 4}p^\prime_{i}p_{k} - \varepsilon_{\mu\nu k\rho}p^\prime_{\rho}p_{4} \right)  \right] A_{T10}(Q^2) \nonumber\\
    & - \frac{i}{2 m_N} \left[( i\varepsilon_{\rho\sigma k \mu}\,p^\prime_{\rho}\,p_\sigma-m_N\varepsilon_{i\mu k4}\,q_i)q_\nu - (\mu \leftrightarrow \nu) \right] B_{T10}(Q^2)  - \frac{1}{2m_N^2}\left[ -i\,\varepsilon_{ijk4}\,p^\prime_i\,p_j \right]P_{[\mu}q_{\nu]}\tilde{A}_{T10}(Q^2) \,.
\end{align}
\end{widetext}
When $\vpp=\vp=\vec{0}$, it reduces to
\begin{align}
    \Pi^{\Gamma_k}_{T_{ij}} &= i\,\varepsilon_{ijk4}\, A_{10}(0) \equiv i\,\varepsilon_{ijk4}\,g_T \,, \nonumber\\
    \text{Others}&=0 \,.
\end{align}
When $\vpp=\vp$, it reduces to
\begin{align}
    \Pi^{\Gamma_k}_{T_{ij}} &=  i\,\varepsilon_{ijk4}\left(1-\frac{p_k^2}{E_N(E_N+m_N)} \right)\,g_T \,, \nonumber\\
    \Pi^{\Gamma_k}_{T_{4j}}&=-\Pi^{\Gamma_k}_{T_{j4}} =  -\frac{\varepsilon_{ijk4}\,p_i}{E_N}\,g_T \,, \nonumber\\
    \text{Others}&=0 \,.
\end{align}
When $\vpp=\vec{0}$, it reduces to
\begin{widetext}
\begin{align}
    \Pi^{\Gamma_{0}}_{T_{4j}} &=-\Pi^{\Gamma_{0}}_{T_{j4}}= \mathcal{C}^{-1}p_j\left\{ -2im_N\left[A_{T10}(Q^2)+B_{T10}(Q^2)\right] - 2i(E_N+m_N)\tilde{A}_{T10}(Q^2)\right\}  \,, \nonumber\\
    \Pi^{\Gamma_k}_{T_{ik}} &= -\Pi^{\Gamma_k}_{T_{ki}}= \mathcal{C}^{-1}\varepsilon_{ijk4}\,p_j\,p_k\left\{ -i B_{T10}(Q^2) \right\}  \,, \nonumber\\
    \Pi^{\Gamma_{k\neq i,j}}_{T_{ij}} &= \mathcal{C}^{-1}\varepsilon_{ijk4} \left\{ 2im_N(E_N+m_N)A_{T10}(Q^2) -i(p_i^2+p_j^2)B_{T10}(Q^2) \right\}   \,, \nonumber\\
    \Pi^{\Gamma_k}_{T_{4j}}&=-\Pi^{\Gamma_k}_{T_{j4}}=\mathcal{C}^{-1}\varepsilon_{ijk4}\,p_i \left\{ -2m_N A_{T10}(Q^2)+(E_N-m_N)B_{T10}(Q^2) \right\}  \,, \nonumber\\
    \text{Others}&=0 \,. 
\end{align}
\end{widetext}


\bibliography{refs}

\end{document}